\documentclass[reprint, nofootinbib, aps,physrev]{revtex4-2}
\usepackage[english]{babel} 
\usepackage[T1]{fontenc}
\usepackage[utf8]{inputenc}
\usepackage{amsfonts,amsmath,amssymb,mathbbol,dsfont}
\usepackage{graphicx,psfrag,xcolor}
\usepackage[normalem]{ulem}
\usepackage{dcolumn}
\usepackage{bm} 
\usepackage{hyperref}
\definecolor{linkcolor}{HTML}{0176ba}
\definecolor{urlcolor}{HTML}{0176ba} 
\definecolor{citecolor}{HTML}{900020}
\hypersetup{pdfstartview=FitH,  linkcolor=linkcolor,urlcolor=urlcolor, citecolor=citecolor, colorlinks=true}
\usepackage[export]{adjustbox}
\usepackage{multirow}
\usepackage{array}
\usepackage{tikz}
\usepackage{mathrsfs}
\usepackage{mathtools}
\DeclarePairedDelimiter\bra{\langle}{\rvert}
\DeclarePairedDelimiter\ket{\lvert}{\rangle}
\DeclarePairedDelimiterX\braket[2]{\langle}{\rangle}{#1\,\delimsize\vert\,\mathopen{}#2}
\usepackage{array}

\usepackage{listings}
\usepackage{color} 
\definecolor{mygreen}{RGB}{28,172,0} 
\definecolor{mylilas}{RGB}{170,55,241}
\definecolor{burntorange}{RGB}{204,85,0}
\usepackage{bbold}
\usepackage{mathtools}

\renewcommand{\Im}{\operatorname{Im}}

\usepackage{comment} 
\excludecomment{hide}
\includecomment{keep}
\usepackage{orcidlink}

\begin{document}

\title{Singular value decomposition to describe bound states in the continuum in periodic metasurfaces}

\author{Nikita Ustimenko\orcidlink{https://orcid.org/0000-0002-5137-493X}$^{1}$}
\email{nikita.ustimenko@kit.edu}
\author{Ivan Fernandez-Corbaton\orcidlink{https://orcid.org/0000-0003-2834-5572}$^{2}$}
 \author{Carsten Rockstuhl\orcidlink{https://orcid.org/0000-0002-5868-0526}$^{1,2,3}$}
\affiliation{$^1$Institute of Theoretical Solid State Physics, Karlsruhe Institute of Technology, Kaiserstr. 12, 76131 Karlsruhe, Germany}
\affiliation{$^2$Institute of Nanotechnology, Karlsruhe Institute of Technology, Kaiserstr. 12, 76131 Karlsruhe, Germany}
\affiliation{$^3$Center for Integrated Quantum Science and Technology (IQST), Karlsruhe Institute of Technology, Wolfgang-Gaede-Str. 1, 76131 Karlsruhe, Germany}

\begin{abstract}
Understanding how bound states in the continuum (BICs) emerge in periodic metasurfaces is essential for the controlled design of high-Q resonances and their systematic manipulation. Here, we investigate the singular value decomposition (SVD) of the effective transition matrix and the scattering matrix of periodic metasurfaces within a parameter range where the metasurface sustains a BIC. Our analysis yields general and practically applicable conditions on the singular values and singular vectors that enable BIC formation. At the BIC eigenfrequency, the inverse of the largest singular value of both matrices vanishes, and the corresponding left (right) singular vector is orthogonal to outgoing (incoming) plane waves that propagate in the directions of open diffraction orders. Our SVD-based approach predicts the spectral position of the BIC and provides detailed information about its properties, including the expansion coefficients in the multipole and plane-wave bases, as well as its behavior under perturbations that transform the BIC into a quasi-BIC. The approach is numerically validated by considering both symmetry-protected and accidental BICs in arrays of scatterers supporting electromagnetic or acoustic multipole resonances. The presented SVD framework offers a broadly applicable foundation for engineering BICs and quasi-BICs in complex metasurfaces, potentially enabling new routes for wave-based devices with tailored radiative properties.
\end{abstract}
 \maketitle

 \section{Introduction}

Bound states in the continuum (BICs) are nonradiating eigenmodes of periodic structures that are embedded within the radiation continuum yet remain completely decoupled from it. In this context, one can distinguish between symmetry-protected BICs (S-BICs), whose symmetry is incompatible with that of the continuum modes, and accidental BICs (A-BICs), which arise from destructive interference of radiating channels~\cite{Koshelev2023May}. In the literature, BICs are also commonly referred to as trapped modes~\cite{Evans1994Feb,Prokhorov2022Dec,Hsu2013Jul} because their energy is fully confined within the structure or its near field and cannot escape due to an infinite radiative quality factor ($Q$ factor). Moreover, in the absence of material losses or absorption, a BIC occurs at a purely real-valued eigenfrequency. 
 
BICs have been experimentally and theoretically demonstrated in nanophotonic~\cite{Hsu2016Jul} and acoustic~\cite{He2025Jul} systems. A broad and practically important class of systems supporting BICs comprises periodic arrangements of resonant scatterers, commonly referred to as metasurfaces. Upon scattering of an external wave by a metasurface, a BIC manifests itself as a dark mode~\cite{Bulgakov2011Jun}. In realistic finite-size arrays, translational symmetry is broken, and a genuine BIC is transformed into a quasi-BIC with a large but finite $Q$ factor due to edge-induced diffraction losses~\cite{Taghizadeh2017Jul,Sadrieva2019May}. However, even in infinite arrays, a genuine BIC can be converted into a quasi-BIC under certain perturbations, enabling its excitation by an external plane wave. Such a conversion can be achieved, for example, by a slight variation of the angle of incidence~\cite{Maksimov2020Dec,Joseph2021Apr}. Alternatively, only structural symmetry breaking~\cite{Fedotov2007Oct,Koshelev2018Nov} or parameter tuning~\cite{Joseph2021Apr,Allayarov2024Jun} can be employed for S-BICs and A-BICs, respectively. The resulting quasi-BIC response has been exploited in a wide range of applications~\cite{Joseph2021Dec}, including field enhancement~\cite{Bulgakov2017Jun}, nonlinear generation~\cite{Zograf2022Feb}, and lasing~\cite{Kodigala2017Jan}.

Several theoretical approaches have been proposed to diagnose and analyze the formation of BICs. One such approach relies on expanding the electromagnetic fields into quasinormal modes~\cite{Lalanne2018May,Sauvan2013Jun} or resonant states~\cite{Doost2014Jul,Muljarov2016Feb,Both2021Dec}, which are solutions to Maxwell’s equations with complex eigenfrequencies for an open, non-Hermitian system, satisfying the radiation boundary conditions at $|\mathbf{r}| \to +\infty$. Recently, Laude and Wang adopted this method for describing the elastodynamics of open phononic systems~\cite{Laude2023Apr}. Neale and Muljarov demonstrated that, in photonic-crystal slabs, BICs can arise from the hybridization of two or more resonant states~\cite{Neale2021Apr}. This phenomenon is commonly described using non-Hermitian effective Hamiltonians~\cite{Sadreev2006Jun,Bulgakov2008Aug,CanosValero2025Mar} or temporal coupled-mode theory~\cite{Maksimov2020Sep,Maksimov2025Dec,Manez-Espina2025Dec}, which are equivalent to a certain extent~\cite{Sadreev2021Apr,Maksimov2025Jul} and, in general, do not provide an exact description of the dynamics of a resonant system~\cite{Wu2024Jun}. Additional drawbacks of the resonant-state expansion include the divergence of resonant states in the limit $|\mathbf{r}| \to +\infty$ and the need to compute resonant states, complex eigenfrequencies, and coupling constants for all parameters of a system that can be deterministically controlled.

One established technique for computing resonant states is the pole expansion of the scattering matrix (\textit{S}-matrix), which relates incoming and outgoing scattering channels in stratified systems~\cite{Neviere1995Mar,Alpeggiani2017Jun,Colom2023Jun}. Blanchard, Hugonin, and Sauvan showed that a BIC corresponds to the coalescence of a pole and a zero of the \textit{S}-matrix at a real frequency~\cite{Blanchard2016Oct}. Moreover, Liu \textit{et al.} demonstrated that S-BICs can act as chain points that connect two nodal lines of the \textit{S}-matrix at the $\Gamma$-point in three-dimensional frequency-momentum space~\cite{Liu2025Dec}. In our contribution here, we focus on another aspect of the \textit{S}-matrix behavior near a BIC resonance, namely its singular value decomposition (SVD)~\cite{Horn1985Dec}. Singular values and singular vectors of the \textit{S}-matrix encode intrinsic scattering properties that govern transmission and reflection~\cite{Guo2022Jun,Zhang2025Sep}, absorption and emission~\cite{Miller2017Apr,Guo2023Apr}, and other fundamental physical characteristics~\cite{Miller2019Sep}. In particular, Guo \textit{et al.} demonstrated how coherent perfect absorption~\cite{Baranov2017Oct} and coherent perfect extinction manifest in the topology of the singular values and vectors of the \textit{S}-matrix~\cite{Guo2023Oct}.

Beyond resonant-state expansions, the decomposition of fields into spherical and cylindrical multipole waves, or simply multipoles, can also shed light on the origin of BICs in metasurfaces. Sadrieva \textit{et al.} showed that the multipole content of an S-BIC with Bloch (in-plane) wave vector $\mathbf{k}^{\rm BIC}_{\parallel}$ consists exclusively of multipoles that do not radiate along the direction specified by $\mathbf{k}^{\rm BIC}_{\parallel}$. In contrast, an A-BIC contains radiating multipoles whose far-field contributions destructively interfere, thereby suppressing radiation~\cite{Sadrieva2019Sep}. In the multipole basis, the transition matrix (\textit{T}-matrix) formalism, introduced by Waterman, provides a rigorous description of the electromagnetic and acoustic response of a single resonator of arbitrary shape~\cite{Waterman1965Aug,Waterman1969Jun,Asadova2025Mar}, as well as of their arrays, including those with periodic boundary conditions (lattices)~\cite{Xu2013Jun,Li2024Aug}.  In the latter case, one commonly refers to the effective \textit{T}-matrix of a unit cell~\cite{Zerulla2023Feb}, which considers the mutual interaction among infinitely many unit cells via analytically known lattice sums~\cite{Beutel2021Jun}. While lattice interactions do not alter the definition of multipoles or their radiation patterns, they can substantially modify the expansion coefficients through linear relations involving the effective \textit{T}-matrix, analogous to single-particle scattering. Consequently, the effective \textit{T}-matrix contains information about collective lattice resonances, including BICs~\cite{Rahimzadegan2022May}.

The eigenvalue decomposition (EVD) of the inverse effective \textit{T}-matrix, also known as the interaction matrix, has been used to extract the multipole content of BICs in periodic arrays of particles~\cite{Bulgakov2014Nov,Bulgakov2018Oct}. However, the EVD has a few limitations: (1) It does not necessarily exist, and (2) the interaction matrix is non-Hermitian (as well as the \textit{S}-matrix); therefore, its left and right eigenvectors do not form orthonormal basis sets for incident and scattered waves. In contrast, the SVD of a matrix, even complex valued, is always well defined and free from these limitations, making it a more robust descriptor of scattering processes~\cite{Suryadharma2017May} and lattice resonances~\cite{Wang2025Feb}. This motivates us to investigate the SVDs of the effective \textit{T}-matrix, expressed in a multipolar basis, and of the \textit{S}-matrix, expressed in a plane-wave (diffraction-order) basis, as a unified framework for diagnosing and characterizing BICs in metasurface-based systems. Although this work focuses on applying this approach to describe electromagnetic and acoustic BICs, it is not restricted to these cases and can be extended to other wave domains, such as elastic-wave BICs~\cite{Cao2025Dec, Liu2025Oct}. 

To this end, in Sec.~\ref{sec:theory}, we introduce the relevant matrix formulations and derive general SVD-based criteria for BIC formation, including the vanishing of the inverse of the largest singular values and the orthogonality of the associated singular vectors to propagating plane waves (typically corresponding to the zeroth diffraction order). In Sec.~\ref{sec:numerics}, we validate these theoretical predictions through numerical simulations of BICs in various systems and compare the different approaches. The numerical results are reproducible using publicly available codes~\cite{svd}. Section~\ref{sec:discussion} discusses how to distinguish between genuine BICs and quasi-BICs in numerical schemes based on the proposed SVD criteria. Finally, Sec.~\ref{sec:conclusion} concludes the work.

\section{Theory}\label{sec:theory}
\begin{figure}
    \centering
    \includegraphics[scale=0.27]{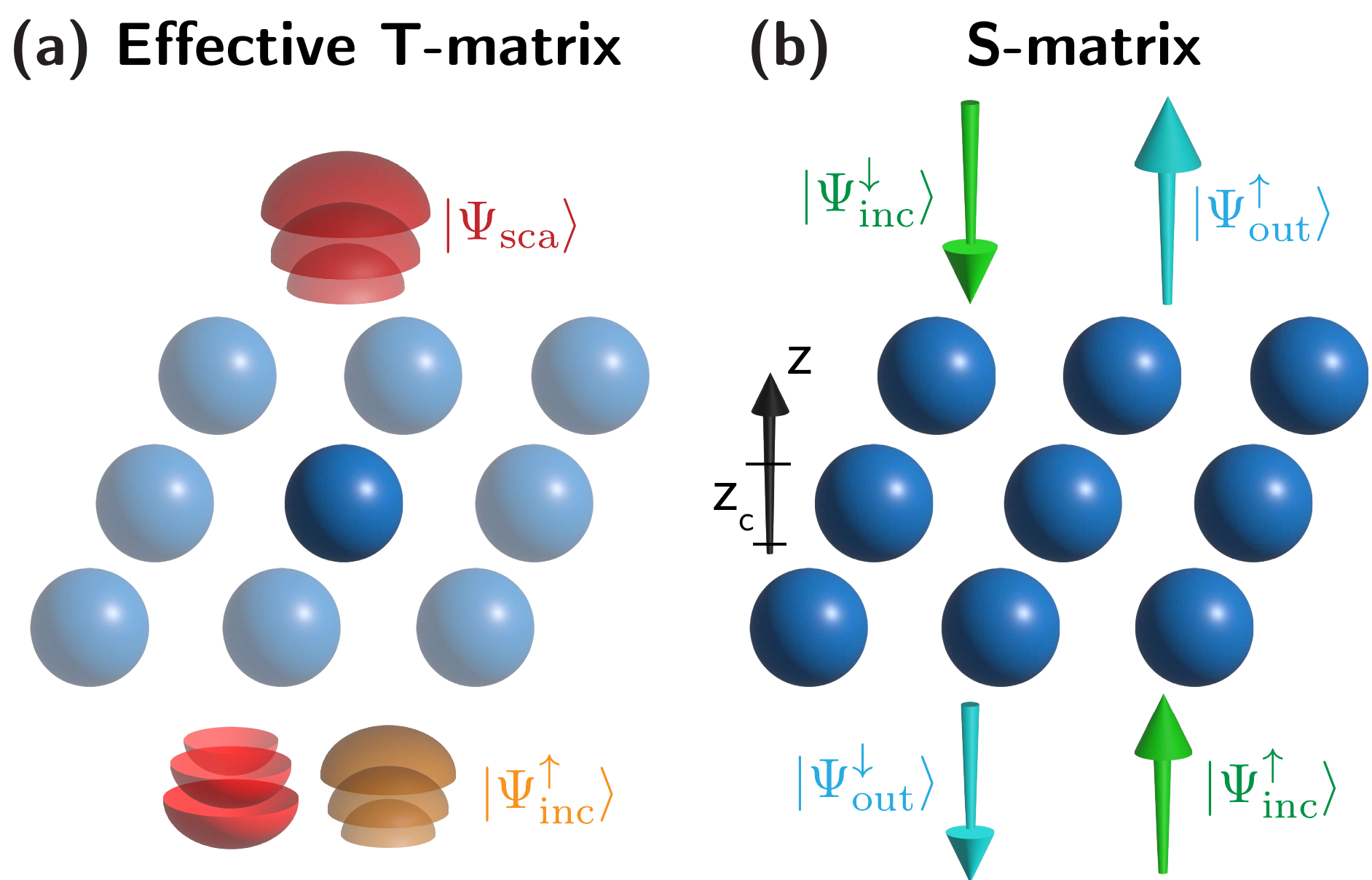}
    \caption{Schematic illustration of the notation and approaches. (a) The effective \textit{T}-matrix describes the acoustic or optical response of a unit cell $\ket{\Psi_{\rm sca}}$ to an incident plane wave $\ket{\Psi_{\rm inc}^d}$ (depicted for $d = \uparrow$), accounting for interactions among unit cells in a spherical or cylindrical multipolar basis. (b) The \textit{S}-matrix links outgoing and incoming plane waves (diffraction orders), while the description is valid for $|z| > z_c/2$, where $z_c$ is the thickness of the metasurface.}
    \label{fig:1}
\end{figure}

As a primary system of interest, we consider a metasurface composed of electromagnetic or acoustic scatterers arranged in a biperiodic lattice lying in the $z = 0$ plane and embedded in a homogeneous isotropic medium (see Fig.~\ref{fig:1}). The medium is characterized by a given speed of light or sound $v$, respectively. The formalisms of the effective \textit{T}-matrix and the \textit{S}-matrix apply to both acoustic and electromagnetic wave scattering. Therefore, this section is formulated in a unified manner that applies to both cases, while explicitly indicating the relevant differences. 

The eigenmodes of a periodic metasurface are classified by the Bloch wave vector $\mathbf{k}_{\parallel}=(k_x, k_y)$ within the first Brillouin zone and the frequency $\omega(\mathbf{k}_{\parallel})$. The set of open scattering (diffraction) channels is then defined by the wave vectors $\left(\mathbf{k}_{\parallel} + \mathbf{G}_{n_x,n_y}\right)$ that satisfy $\left| \mathbf{k}_{\parallel} + \mathbf{G}_{n_x,n_y}\right| < \omega/v$, where $\mathbf{G}_{n_x,n_y}$ denotes a reciprocal lattice vector for a biperiodic lattice, where $n_x,n_y \in \mathbb{Z}$. Electromagnetic and acoustic metasurfaces have been shown to support electromagnetic and acoustic BICs, respectively, at specific Bloch wave vectors where no radiation occurs into any open scattering channels. In the vast majority of cases, BICs occur in systems with only a single open scattering channel corresponding to $\mathbf{G}_{0,0} = \mathbf{0}$, i.e., the zeroth diffraction order. In this case, for the square lattice with a lattice constant $L$, the frequency must satisfy $\left|\mathbf{k}_{\parallel}\right| < \omega/v < \left|\mathbf{k}_{\parallel} + \mathbf{G}_{n_x,n_y}\right|$ for all $|\mathbf{G}_{n_x,n_y}|=2\pi/L$, i.e., $n_x =\pm1$ and $n_y =\pm1$.

\subsection{Notation}
In the following, we assume that the incident field is a plane wave. For electromagnetic waves, the incident field is uniquely characterized by three parameters $(\mathbf{k}_{\parallel}, d, p)$, whereas for acoustic waves, only two parameters $(\mathbf{k}_{\parallel}, d)$ are required. Here, $\mathbf{k}_{\parallel}$ is the projection of the wave vector onto the $xy$ plane, $d = \,\uparrow (+)$ or $\downarrow (-)$ denotes the propagation direction along the $z$ axis, and $p$ specifies the TM ($p = 1$) or TE ($p = 0$) polarization of electromagnetic waves (while acoustic waves are purely longitudinal). The remaining wave-vector component is given by $k_z = d  \sqrt{\omega^2/v^2 - k_{\parallel}^2}$. For $z > z_c/2$, plane waves with $d = \,\downarrow$ and $d = \,\uparrow$ are referred to as incoming and outgoing, respectively, with the convention reversed for $z < -z_c/2$ [cf. Fig.~\ref{fig:1}(b)]. Here, $z_c$ denotes the thickness of the metasurface.

Within the \textit{T}-matrix formalism, we select one of the incoming modes as the incident field and denote the vector of its expansion coefficients in a multipolar basis by $\ket{\Psi_{\rm inc}^d}$. The corresponding scattered field is described by the coefficient vector $\ket{\Psi_{\rm sca}}$ [see Fig.~\ref{fig:1}(a)]. The bra--ket notation is employed to clearly distinguish between column and row vectors and to define the scalar product as $\braket{a}{b} = \sum_i a_i^* b_i$.

In contrast, within the \textit{S}-matrix formalism, we simultaneously consider incoming and outgoing waves in the upper and lower half-spaces. Specifically, we define the vectors of outgoing waves as $\ket{\Psi_{\rm out}} = \left(\ket{\Psi_{\rm out}^\uparrow}, \ket{\Psi_{\rm out}^\downarrow}\right)^{\rm T}$ and incoming waves as $\ket{\Psi_{\rm inc}} = \left(\ket{\Psi_{\rm inc}^\uparrow}, \ket{\Psi_{\rm inc}^\downarrow}\right)^{\rm T}$ [see Fig.~\ref{fig:1}(b)]. These waves are expanded in a plane-wave basis consisting of a countable set of diffraction orders defined by the in-plane wave vectors $\left(\mathbf{k}_{\parallel} + \mathbf{G}_{n_x,n_y}\right)$. If $\left| \mathbf{k}_{\parallel} + \mathbf{G}_{n_x,n_y}\right|$ exceeds $\omega/v$, the corresponding plane wave becomes evanescent in the surrounding medium. 

\subsection{Effective transition matrix}\label{sec:T_eff}
The effective \textit{T}-matrix provides a full-wave solution to the problem of plane-wave scattering by a metasurface~\cite{Beutel2024Apr}. To this end, we expand both the incident and scattered fields of a unit cell into multipolar waves [see Fig.~\ref{fig:1}(a)], which may be spherical or cylindrical depending on the geometry of the problem (see Appendix~\ref{sec:sph_waves}). The effective \textit{T}-matrix (electromagnetic or acoustic) relates the expansion coefficients of the scattered field generated by a unit cell to those of the incident plane wave through the following relation~\cite{Zerulla2023Feb,Rahimzadegan2022May}:
\begin{align}
\label{eq_Teff}
\ket{\Psi_{\rm sca}} = \mathbf{T}_{\rm eff}(\mathbf{k}_{\parallel}, \omega) \ket{\Psi_{\rm inc}^d}\,.
\end{align}
The effective \textit{T}-matrix accounts for mutual interactions among unit cells, which effectively “renormalize” the \textit{T}-matrix of an isolated unit cell through lattice sums of analytically known translation coefficients of vector multipole waves for electromagnetic scatterers~\cite{Beutel2024Apr} or scalar multipole waves for acoustic scatterers~\cite{Li2024Aug}.

Lattice eigenmodes can be sustained by the metasurface in the absence of external excitation, $\ket{\Psi_{\rm inc}^d} \equiv \mathbf{0}$, and are obtained from Eq.~\eqref{eq_Teff} with an additional constraint ensuring the existence of a nontrivial solution~\cite{Necada2021Jan}:
\begin{align}
\label{eq_TeffInv}
\mathbf{T}_{\rm eff}^{-1}(\mathbf{k}_{\parallel}, \omega) \ket{\Psi_{\rm sca}} = \mathbf{0}\,
\Rightarrow \, \mathrm{det} \, \mathbf{T}_{\rm eff}^{-1}(\mathbf{k}_{\parallel}, \omega) = 0\,.
\end{align}
Eigenmodes of a periodic structure are characterized by a dispersion relation $\omega(\mathbf{k}_{\parallel})$, corresponding to those arguments of $\mathbf{T}_{\rm eff}$ for which Eq.~\eqref{eq_TeffInv} has a nontrivial solution. In this framework, a bound state in the continuum (BIC) of a nonabsorptive metasurface is defined as an eigenmode with a purely real eigenfrequency $\omega_{\rm BIC} \in \mathbb{R}$ at a specific Bloch wave vector $\mathbf{k}_{\parallel} = \mathbf{k}_{\parallel}^{\rm BIC}$. 

If the scatterers are dissipative and therefore described by complex-valued material parameters, the eigenfrequency acquires a finite imaginary part $\gamma_{\rm nr} \approx C\tan\delta$ due to absorption, where $C$ is the confinement factor of the mode and $\tan\delta$ is the loss tangent~\cite{Mikhailovskii2024Apr}, while radiative losses remain zero, $\gamma_{\rm r} = 0$. Please note that nonradiative losses $\gamma_{\rm nr}$ can induce radiative losses $\gamma_{\rm r}$ in the case of A-BICs, which are sensitive to breaking time-reversal symmetry~\cite{Hsu2013Jul,Zhen2014Dec}. For example, Fabry--Perot BICs, a type of A-BICs that occur in double-layer metasurfaces (see Sec.~\ref{sec:fabry}), exhibit the following absorption-induced radiative losses $\gamma_{\rm r} \propto \gamma_{\rm nr}^2$~\cite{Semushev2026Jan}. 

To find all possible BICs and their corresponding eigenfrequencies for a given Bloch wave vector $\mathbf{k}_{\parallel}$, we employ the singular value decomposition (SVD) of the effective \textit{T}-matrix~\cite{Horn1985Dec}:
\begin{align}
\label{eq_Teff_svd}
\mathbf{T}_{\rm eff}(\mathbf{k}_{\parallel}, \omega) = \mathbf{U} \mathbf{\Sigma}\mathbf{V}^\dagger\,,
\end{align}
where $\mathbf{\Sigma}$ is a diagonal matrix containing real valued, non-negative singular values $\sigma_1 \geq \sigma_2 \geq ... \geq\sigma_{N_T} \geq 0$ (the dimension $N_T$ is discussed in Appendix~\ref{sec:sph_waves}), and the columns of $\mathbf{U}$ and $\mathbf{V}$ are the left and right singular vectors $\ket{\mathbf{u}_i}$ and $\ket{\mathbf{v}_i}$, respectively. Since $\mathbf{T}_{\rm eff}(\mathbf{k}_{\parallel}, \omega)$ depends on both $\mathbf{k}_{\parallel}$ and $\omega$, its singular values and singular vectors inherit this dependence. It will also be useful to rewrite Eq.~\eqref{eq_Teff_svd} as
\begin{align}
\label{eq_Teff_svdExpansion}
\mathbf{T}_{\rm eff}(\mathbf{k}_{\parallel}, \omega) = \sum_{i=1}^{N_T} \sigma_i \ket{\mathbf{u}_i} \bra{\mathbf{v}_i}\,.
\end{align}
The SVD of the inverse effective \textit{T}-matrix is then given by $\mathbf{T}_{\rm eff}^{-1}(\mathbf{k}_{\parallel}, \omega) = \mathbf{V} \mathbf{\Sigma}^{-1}\mathbf{U}^\dagger$. Consequently, the smallest singular value of $\mathbf{T}_{\rm eff}^{-1}(\mathbf{k}_{\parallel}, \omega)$ equals $\sigma_1^{-1}$. Using the relation between the determinant and the singular values,
$\left| \mathrm{det}\, \mathbf{T}_{\rm eff}^{-1}(\mathbf{k}_{\parallel}, \omega)\right| = \prod\limits_{i=1}^{N_T} \sigma_i^{-1}$~\cite{Horn1985Dec}, Eq.~\eqref{eq_TeffInv} can be reformulated for a BIC as
\begin{align}
\label{eq_sigma1}
\boxed{\sigma^{-1}_1\left[ \mathbf{T}_{\rm eff}(\mathbf{k}_{\parallel}^{\rm BIC},\omega_{\rm BIC})\right] = 0}\,,
\end{align}
where $\sigma_1$ picks the first and largest singular value of $\mathbf{T}_{\rm eff}(\mathbf{k}_{\parallel}, \omega)$ for real arguments. 

When Eq.~\eqref{eq_sigma1} is satisfied, the expansion coefficients of the BIC field in a spherical-wave basis are given by the first left singular vector $\ket{\mathbf{u}_1}$ of $\mathbf{T}_{\rm eff}(\mathbf{k}_{\parallel}, \omega)$. Its decomposition into outgoing (singular) spherical waves reads as follows
\begin{align}
\label{eq_u1_decomp}
\ket{\mathbf{u}_1} =
\sum\limits_{\ell,m,p} u_{\ell,m,p}^{1} \ket{\ell,m,p}\,.
\end{align}
Here, $\ket{\ell,m,p}$ denotes a singular vector spherical wave, where $\ell \in [1, \ell_{\rm max}]$ is the degree, $|m|\leq \ell$ is the order, and $p$ is the TM ($p = 1$) or TE ($p = 0$) polarization for electromagnetic waves only. Moreover, from the coefficients $u_{\ell,m,p}^{1}$ in Eq.~\eqref{eq_u1_decomp}, the Cartesian multipole moments can be obtained using the unitary transformation given in Eq.~(E1) in Ref.~\cite{Rahimzadegan2022May}. For acoustic waves, $\ket{\ell,m,p}$ must be replaced by a scalar spherical wave $\ket{\ell,m}$ (see Appendix~\ref{sec:sph_waves}). An analogous expansion can be written in terms of vector or scalar cylindrical waves (see Appendix~\ref{sec:sph_waves}). 

Because it may be numerically challenging to determine whether $\sigma^{-1}_1$ vanishes exactly (up to machine precision) on a finite parameter mesh, it becomes important to supplement Eq.~\eqref{eq_sigma1} with an additional constraint on the singular vectors $\ket{\mathbf{u}_1}$ and $\ket{\mathbf{v}_1}$. Specifically, a BIC at $\mathbf{k}_{\parallel}^{\rm BIC}$ must be orthogonal to all outgoing plane waves with the same in-plane wave vector, which implies
\begin{align}
\label{eq_rad_losses_u1}
\boxed{\braket{\Psi_{\rm out}^d}{\mathbf{u}_1} = 0\, \quad \forall d \in {\uparrow, \downarrow}}\,.
\end{align}
Here, $\ket{\Psi_{\rm out}^d}$ represents an outgoing plane wave [see Fig.~\ref{fig:1}(b)]. Equation~\eqref{eq_rad_losses_u1} can be satisfied in two distinct scenarios: expansion~\eqref{eq_u1_decomp} either contains only multipoles that do not radiate along the direction $\mathbf{k}_{\parallel}^{\rm BIC}$ (S-BIC), or includes radiating multipoles whose superposition suppresses radiation through destructive interference (A-BIC)~\cite{Sadrieva2019Sep}.

A constraint analogous to Eq.~\eqref{eq_rad_losses_u1} can typically be formulated for the right singular vector $\ket{\mathbf{v}_1}$. Indeed, the set of right singular vectors of $\mathbf{T}_{\rm eff}(\mathbf{k}_{\parallel}, \omega)$ forms an orthonormal basis for incident fields~\cite{Suryadharma2017May}, allowing an external plane wave to be expanded as
\begin{align}
\label{eq_pw_svd}
\ket{\Psi_{\rm inc}^d} = \sum_{i=1}^{N_T}
\underbrace{\braket{\mathbf{v}_i}{\Psi^d_{\rm inc}}}_{W_i^d}
\ket{\mathbf{v}_i}\,,
\end{align}
where $W_i^d$ is referred to as the quasimode coupling strength~\cite{Bulgakov2017Jun}. Notably, the coupling strengths in Eq.~\eqref{eq_pw_svd} quantify the overlap of incident plane waves with the right singular vectors of the effective \textit{T}-matrix $\mathbf{T}_{\rm eff}(\mathbf{k}_{\parallel}, \omega)$, rather than with the left eigenvectors of the interaction matrix $\mathbf{T}_{\rm eff}^{-1}(\mathbf{k}_{\parallel}, \omega)$, as considered in Ref.~\cite{Bulgakov2017Jun}. Substituting Eq.~\eqref{eq_Teff_svdExpansion} into Eq.~\eqref{eq_Teff} and using Eq.~\eqref{eq_pw_svd} yields
\begin{align}
\label{eq_p_svd}
\ket{\Psi_{\rm sca}} = \sum_{i=1}^{N_T} \sigma_i W_i^d \ket{\mathbf{u}_i}\,,
\end{align}
which expresses the scattered field of a unit cell under plane-wave excitation. For a BIC, $\sigma_1$ diverges as a consequence of Eq.~\eqref{eq_sigma1}, and to avoid an unphysical gain of energy upon the scattering, the following condition must be satisfied:
\begin{align}
\label{eq_coupling_v1}
\boxed{W_1^d \equiv \braket{\mathbf{v}_1}{\Psi_{\rm inc}^d} = 0\, \quad \forall d \in {\uparrow, \downarrow}}\,.
\end{align}
Equations~\eqref{eq_rad_losses_u1} and~\eqref{eq_coupling_v1} are equivalent for $\mathcal{PT}$-symmetric systems, reflecting the fact that a nonradiating BIC cannot be excited from the far field by a plane wave and therefore does not contribute to the scattered field in Eq.~\eqref{eq_p_svd}. This decoupling from the radiation continuum is the defining physical property of a BIC. Breaking $\mathcal{PT}$ symmetry, for instance by introducing gain and absorption in the scatterers, generally violates Eq.~\eqref{eq_coupling_v1}, while Eq.~\eqref{eq_rad_losses_u1} remains valid~\cite{Song2020Aug}.

Finally, we consider the case of degenerate BICs. Suppose that $M \leq N_T$ distinct BICs share the same eigenfrequency $\omega_{\rm BIC}$. Equation~\eqref{eq_sigma1} then implies $\sigma_1^{-1} = \sigma_2^{-1} = ... = \sigma_M^{-1} = 0$, while Eqs.~\eqref{eq_rad_losses_u1} and~\eqref{eq_coupling_v1} require $\braket{\Psi_{\rm out}^d}{\mathbf{u}_i} = \braket{\mathbf{v}_i}{\Psi_{\rm inc}^d} = 0$ for all $i\in \{1,\ldots,M\}$ and both propagation directions $d$. Any linear combination of these vectors $\ket{\mathbf{u}_i}$ therefore also represents a BIC, implying that these vectors span an $M$-fold subspace of BICs. In Sec.~\ref{sec:degenerate}, we explicitly consider the case $M = 2$.

\subsection{Scattering matrix}
Next, we consider the \textit{S}-matrix-based description of scattering by a metasurface. In contrast with the effective \textit{T}-matrix, which describes the response of an individual unit cell in the spherical- or cylindrical-wave basis, the \textit{S}-matrix enables the calculation of the fields scattered by the entire metasurface (i.e., all unit cells collectively) in the plane-wave basis, without an additional summation over the unit cells [see Fig.~\ref{fig:1}(b)]. Formally, the scattering of electromagnetic or acoustic waves can be assumed to occur in the plane $z = 0$, allowing the periodic metasurface to be effectively replaced by a planar interface. This interface is then characterized by an \textit{S}-matrix that relates outgoing and incoming plane waves according to
\begin{align}
\label{eq_S}
    \ket{\Psi_{\rm out}} = \mathbf{S}(\mathbf{k}_{\parallel}, \omega) \ket{\Psi_{\rm inc}}\,.
\end{align}

The subsequent analysis of BICs follows from the singular value decomposition (SVD) of the \textit{S}-matrix and closely parallels the approach outlined for the effective \textit{T}-matrix. Lattice eigenmodes can be formally obtained in the absence of incoming waves~\cite{Tikhodeev2002Jul} from
\begin{align}
    \mathbf{S}^{-1}(\mathbf{k}_{\parallel}, \omega) \ket{\Psi_{\rm out}} = \mathbf{0}\,
\Rightarrow \, \mathrm{det} \ \mathbf{S}^{-1}(\mathbf{k}_{\parallel}, \omega) = 0\,.
\end{align}
At first glance, this condition appears to be incompatible with energy conservation in a lossless scattering system, because the determinants must satisfy $\mathrm{det} \ \mathbf{S} = \mathrm{det} \ \mathbf{S}^{-1} = 1$ required by unitarity $\mathbf{S} \mathbf{S}^{\dagger} = \mathbf{S}^{\dagger} \mathbf{S} = \mathbf{I}$, where $\mathbf{I}$ is an identity matrix of the corresponding dimension. However, the \textit{S}-matrix is strictly unitary only when the basis is restricted to propagating diffraction orders. Including evanescent diffraction orders can break unitarity~\cite{Carminati2000Jun}. Because a BIC is intrinsically an evanescent mode, its accurate representation in the plane-wave basis requires the inclusion of evanescent diffraction orders. 

Since BICs typically arise in periodic systems with a single open diffraction channel at $\mathbf{G}_{0,0} = \mathbf{0}$, we additionally include the nearest evanescent diffraction channels with $\mathbf{G}_{n_x,n_y} \neq \mathbf{0}$. For the square lattice, this corresponds to including the zeroth diffraction order $\mathbf{k}_{\parallel}$ and at least four additional orders $\left( \mathbf{k}_{\parallel} + \mathbf{G}_{n_x,n_y} \right)$ with $n_x = \pm1$ or $n_y = \pm1$, so that $\left| \mathbf{G}_{n_x,n_y} \right| = 2\pi / L$. As a result, the dimension of the \textit{S}-matrix (and of its SVD) for the square lattice is at least $N_S = 20$ in the case of electromagnetic waves and $N_S = 10$ in the case of acoustic waves. 

We now perform the SVD of the \textit{S}-matrix
\begin{align}
\label{eq_svd_S}
    \mathbf{S} = \mathbf{H} \mathbf{\Xi}\mathbf{F}^\dagger\,,
\end{align}
where $\mathbf{H}$ and $\mathbf{F}$ contain the left and right singular vectors of $\mathbf{S}$, respectively, and $\mathbf{\Xi}$ is a diagonal matrix of singular values. Two cases can be distinguished:
\begin{itemize}
    \item If $\mathbf{S}$ is unitary, all singular values satisfy $\xi_1  = \xi_2 = ... = \xi_{N_S} = 1$.
    \item If $\mathbf{S}$ is nonunitary (due to the inclusion of evanescent modes), the singular values satisfy $\xi_1 \geq \xi_2 \geq ... \geq\xi_{N_S} \geq 0$.
\end{itemize}
In the latter case, it may be possible to find a real frequency $\omega_{\rm BIC}$ for a nonabsorptive metasurface such that the inverse of the largest singular value vanishes,
\begin{align}
\label{eq_xi1}
    \boxed{\xi^{-1}_1\left[  \mathbf{S}(\mathbf{k}_{\parallel}^{\rm BIC},\omega_{\rm BIC})\right] = 0}\,,
\end{align}
indicating the occurrence of a BIC with Bloch wave vector $\mathbf{k}_{\parallel}^{\rm BIC}$. 

By arguments analogous to those used for the singular vectors of $\mathbf{T}_{\rm eff}(\mathbf{k}_{\parallel}, \omega)$, the corresponding left and right singular vectors of $\mathbf{S}$ must be orthogonal to all propagating plane waves with $\mathbf{k}_{\parallel} = \mathbf{k}_{\parallel}^{\rm BIC}$, arbitrary propagation direction $d$ and polarization $p$. This implies $\braket{\mathbf{h}_1}{\Psi_{\rm inc}} = \braket{\mathbf{f}_1}{\Psi_{\rm inc}} = 0$, which leads to the following BIC criterion:
\begin{align}
\label{eq_h1_f1_braket}
\boxed{h^1_{\mathbf{k}_{\parallel}^{\rm BIC},d,p} = f^1_{\mathbf{k}_{\parallel}^{\rm BIC},d,p} = 0\, \quad \forall {d, p}}\,.
\end{align}
This condition reflects the well-known property that the zeroth Fourier coefficient of a BIC field vanishes~\cite{Koshelev2023May}. Consequently, the singular vectors $\ket{\mathbf{h}_1}$ and $\ket{\mathbf{f}_1}$ have nonzero components only in evanescent diffraction orders.

Using Eq.~\eqref{eq_svd_S}, the \textit{S}-matrix can be expanded as $\mathbf{S} = \sum_i \xi_i \ket{\mathbf{h}_i}\bra{\mathbf{f}_i}$. Substituting this expansion into Eq.~\eqref{eq_S} yields the outgoing field under the BIC conditions in Eqs.~\eqref{eq_xi1} and~\eqref{eq_h1_f1_braket}:
\begin{align}
    \ket{\Psi_{\rm out}} = \underbrace{\xi_1 C_1 \ket{\mathbf{h}_1}}_{=0} + \sum_{i=2}^{N_S} \xi_i C_i \ket{\mathbf{h}_i}\,,
\end{align}
where $C_i = \braket{\mathbf{f}_i}{\Psi_{\rm inc}}$ denotes the quasimode coupling strength. Thus, in the \textit{S}-matrix framework, the contribution associated with the singular vector corresponding to the BIC vanishes ($C_1 = 0$), and only nonresonant singular modes contribute to reflection and transmission. As in the effective \textit{T}-matrix formalism, the \textit{S}-matrix naturally accommodates degenerate BICs: if a periodic system supports $M$-fold degenerate BICs, the first $M$ singular values of $\mathbf{S}$ are equal, and the corresponding singular vectors span an $M$-dimensional BIC subspace (see also Sec.~\ref{sec:degenerate}).

Although the SVD-based description of BICs using the \textit{S}-matrix may appear redundant given its similarities to the effective \textit{T}-matrix formulation, it offers a distinct practical advantage. Specifically, the \textit{S}-matrix formalism is readily applicable to periodic systems on a substrate, as \textit{S}-matrices can be efficiently stacked using the Redheffer star product~\cite{Redheffer1959}. In contrast, the effective \textit{T}-matrix approach is typically restricted to metasurfaces embedded in homogeneous environments, as extending it to substrate-supported systems can be technically cumbersome. It is nevertheless important to emphasize that, for periodic lattices of particles or molecules, the \textit{S}-matrix can be calculated directly from the effective \textit{T}-matrix~\cite{Beutel2024Apr}. Additionally, due to the linear relationship between the spherical-wave and Cartesian-multipole basis~\cite{Rahimzadegan2022May}, the BIC criteria in Sec.~\ref{sec:T_eff}, derived from the SVD of the effective \textit{T}-matrix, can be reformulated in terms of the SVD of the effective multipole polarizabilities~\cite{Allayarov2024Jun}.

\section{Numerical validation}\label{sec:numerics}
Next, we validate the criteria for BIC formation--namely, Eqs.~\eqref{eq_sigma1}, \eqref{eq_rad_losses_u1}, \eqref{eq_coupling_v1}, \eqref{eq_xi1}, and \eqref{eq_h1_f1_braket}--by numerically modeling BICs in electromagnetic and acoustic metasurfaces. To this end, we employ the in-house software packages \textit{treams} (v.~0.4.5)~\cite{Beutel2024Apr} and \textit{acoustotreams} (v.~0.2.13)~\cite{acoustotreams}, which provide full-wave solutions to multiple-scattering problems for electromagnetic and acoustic metasurfaces, respectively. These Python packages enable efficient computation of effective \textit{T}-matrices and \textit{S}-matrices using the Ewald summation technique~\cite{Beutel2023Jan}. The singular value decompositions of these matrices are computed using the \texttt{numpy.linalg.svd} routine in Python~3.

Before presenting the simulation results, we need to remark on the multipole content of BICs in metasurfaces. In practice, however, only a few leading terms are sufficient to accurately describe the eigenmode. For instance, periodic two-dimensional arrays of spheres have been shown to support electromagnetic symmetry-protected BICs (S-BICs) with either TM or TE polarization that can be well described within the dipole approximation. In this case, the metasurface behaves as a lattice of normally oriented electric ($p_z$) or magnetic ($m_z$) dipoles with identical amplitudes and phases~\cite{Evlyukhin2021Dec}. Such modes occur at the $\Gamma$ point, $\mathbf{k}_{\parallel}^{\rm BIC} = \mathbf{0}$. Moreover, by analyzing the poles of the effective dipole polarizability, Evlyukhin \textit{et al.} have identified that this BIC can emerge at the wavelength of a dipole resonance (electric or magnetic) of an isolated sphere $\lambda_{\rm R}$ if the lattice constant is $L/\lambda_{\rm R} = 0.7112...$, offering a practical rule for realizing a BIC at normal incidence for metasurfaces of dipolar scatterers~\cite{Evlyukhin2021Dec}. Please note that if the desired wavelength $\lambda$ differs from $\lambda_{\rm R}$, the ratio $L/\lambda$ required for BIC formation changes accordingly, and also note that taking higher-order multipoles into account may slightly shift the wavelength or lattice constant where the BIC resonance occurs. Nevertheless, in all our computations, a sufficiently large multipolar degree was retained, so that this effect is nearly negligible.

Since the dipole contribution alone is sufficient to describe this S-BIC, we compute the SVD in Secs.~\ref{sec:infinite}–\ref{sec:degenerate} using a basis of vector spherical waves with $\ell_{\rm max} = 1$. In Sec.~\ref{sec:fabry}, we consider Fabry--Perot accidental BICs (A-BICs) in arrays of dielectric rods, using a basis of vector cylindrical waves with $k_z = 0$ and $m_{\rm max} = 2$ (please note the permutation of coordinate axes for cylindrical waves, see Appendix~\ref{sec:sph_waves}). In Sec.~\ref{sec:acoustic}, we study acoustic A-BICs in arrays of spherical particles using a basis of scalar spherical waves with $\ell_{\rm max} = 2$. We emphasize that the results remain essentially unchanged upon increasing $\ell_{\rm max}$ or $m_{\rm max}$, at the expense of increased computational cost (see Appendix~\ref{sec:time}). For the SVD of the \textit{S}-matrix, we include diffraction orders with $|\mathbf{G}_{0,0}| = 0$ and $|\mathbf{G}_{n_x,n_y}| = 2\pi/L$, except in Sec.~\ref{sec:substrate}, where diffraction orders up to $|\mathbf{G}_{n_x,n_y}| = 8\pi/L$ are included to ensure convergence in the presence of a substrate. 

Table~\ref{tab:summary} summarizes the classes of systems that can be modeled using the SVD of the effective \textit{T}-matrix and the \textit{S}-matrix and that are considered as examples in this work.
\begin{table}[h]
    \centering
    \begin{tabular}{l c c}
         \hline
         \hline
         System & Effective \textit{T}-matrix & \textit{S}-matrix \\
         \hline
         Single- and multilayer \\ metasurfaces in free space & Yes & Yes \\
         \hline
         Metasurface on a substrate & No & Yes \\
         \hline
         Finite-size arrays& Yes & No \\
         \hline
         \hline
    \end{tabular}
    \caption{Summary of periodic systems for modeling BICs and quasi-BICs. Columns indicate whether the SVD of the effective \textit{T}-matrix or the \textit{S}-matrix can be applied using the \textit{treams} and \textit{acoustotreams} packages for the calculation of these matrices.}
    \label{tab:summary}
\end{table}

\subsection{Infinite array in free space}\label{sec:infinite}
We begin with a dipole BIC supported by a single array of spherical particles in free space. The particles have a radius of 200 nm and a relative dielectric permittivity of 12.25, corresponding to lossless crystalline silicon at room temperature in the near-infrared spectral range~\cite{Wang2025Aug}. Using Mie theory~\cite{Bohren1998Apr}, one finds that an isolated particle exhibits a magnetic-dipole resonance at wavelength $\lambda_{\rm R} = 1459.16$ nm. According to Ref.~\cite{Evlyukhin2021Dec}, a periodic metasurface with a square lattice supports a TE-polarized S-BIC, which emerges in the dipole approximation when the lattice constant is $L = 1037.779$ nm. Please note that accounting for higher-order multipoles in our case introduces only a minor deviation of $|\delta L| / L \sim 10^{-5}$ from this value (see Appendix~\ref{sec:convergence}). Therefore, the results can be considered converged already for a maximum multipole degree of $\ell_{\rm max} = 1$.

\begin{figure*}
    \centering
    \includegraphics[width=\linewidth]{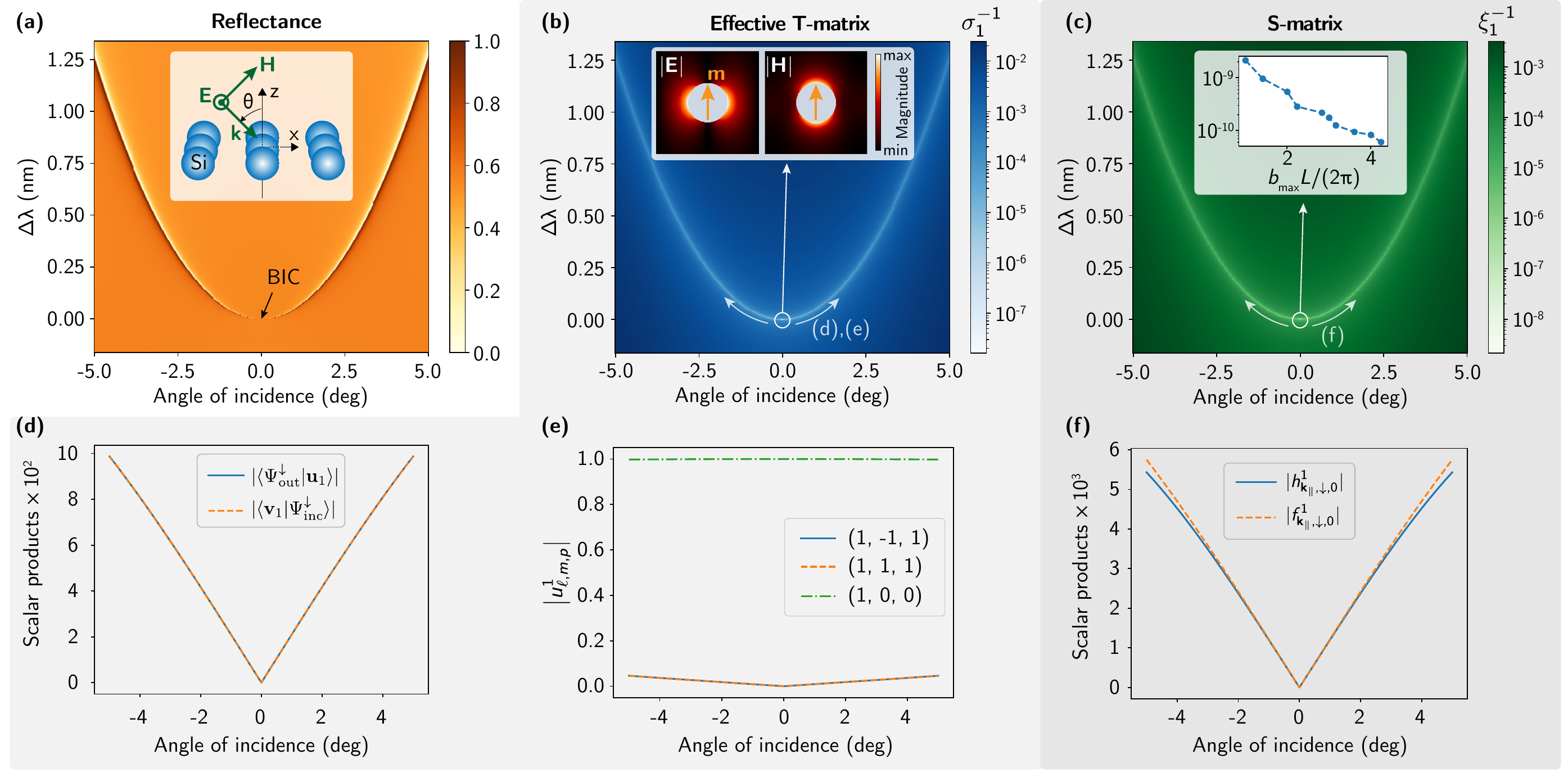}
    \caption{Electromagnetic S-BIC in a single metasurface of spherical particles in free space. (a) Reflectance of the metasurface [for $\ell_{\rm max} = 4$] as a function of the angle of incidence of a TE-polarized plane wave and the wavelength detuning $\Delta \lambda = \lambda - \lambda_{\rm R}$, where $ \lambda_{\rm R} = 1459.16$ nm is the BIC wavelength. The lattice constant is $L = 1037.779$ nm. The inset in panel (a) depicts the metasurface and the illumination conditions. The inverse
    of the largest singular values of (b) the effective \textit{T}-matrix [$\ell_{\rm max} = 1$] and (c) the \textit{S}-matrix [$b_{\rm max}L/ (2\pi) = 1$], where $b_{\rm max}$ sets the upper bound on the magnitude of the reciprocal-lattice vectors $\mathbf{G}_{n_x,n_y}$. The inset in panel (b) shows the magnitudes of the electric and magnetic fields of the BIC. The inset in panel (c) shows $\xi_1^{-1}$ as a function of $b_{\rm max}L/(2\pi)$. (d) Absolute values of scalar products of the singular vectors of $\mathbf{T}_{\rm eff}(\mathbf{k}_{\parallel}, \omega)$ and plane waves: $|\braket{\Psi^\downarrow_{\rm out}}{\mathbf{u}_1}|$ (blue solid) and $|\braket{\mathbf{v}_1}{\Psi^\downarrow_{\rm inc}}|$ (orange dashed), corresponding to minima in panel (b). (e) Absolute values of components of vector $\ket{\mathbf{u}_1}$ in the spherical-wave basis labeled by three indices $(\ell,m,p)$. (f) Absolute values of zeroth-order components of the left and right singular vectors of $\mathbf{S}$, $|h^1_{\mathbf{k}_{\parallel},\downarrow,0}|$ and $|f^1_{\mathbf{k}_{\parallel},\downarrow,0}|$, corresponding to minima in panel (c).}
    \label{fig:2}
\end{figure*}

To confirm this prediction, we plot the reflectance of the metasurface in Fig.~\ref{fig:2}(a) for an excitation by a TE-polarized plane wave $\ket{\Psi^\downarrow_{\rm inc}} = \ket{\mathbf{k}_{\parallel}, \downarrow, 0}$ incident at an angle $\theta$ with respect to the surface normal (the $z$ axis). In this configuration, the in-plane wave vector is $\mathbf{k}_{\parallel} = (k\sin\theta, 0)$, where $k = \omega/c$ [see the inset in Fig.~\ref{fig:2}(a)], and the incident wave carries a magnetic-field component $H_z \propto \sin\theta$, which can excite the out-of-plane magnetic-dipole moment $m_z$. Owing to the mirror symmetry $z \to -z$, the response of the system to $\ket{\Psi^\uparrow_{\rm inc}}$ is identical. As shown in Fig.~\ref{fig:2}(a), the reflectance spectrum exhibits a Fano-type resonance for each nonzero angle of incidence, which disappears at normal incidence when $\lambda = \lambda_{\rm R}$. This behavior signifies the formation of an S-BIC, which is decoupled from a normally incident plane wave. In contrast, the resonances observed at oblique incidence correspond to the quasi-BIC regime.

We now apply the criteria derived in Sec.~\ref{sec:theory} to analyze this BIC. Figures~\ref{fig:2}(b) and~\ref{fig:2}(c) show the inverse largest singular values of the effective \textit{T}-matrix, $\sigma_1^{-1}$, and of the \textit{S}-matrix, $\xi_1^{-1}$, respectively. For each angle of incidence, both quantities exhibit a pronounced minimum at the values of $\theta$ and $\lambda$ corresponding to the quasi-BIC resonance observed in Fig.~\ref{fig:2}(a). The minima do not reach zero, reflecting the fact that quasi-BICs are not genuine eigenmodes of the metasurface at real frequencies [one should consider complex-valued frequencies $\left(\omega-\mathrm{i}\gamma_{\rm r}\right)$]. However, as $(\theta, \lambda) \to (0, \lambda_{\rm R})$, the inverses of both singular values significantly decrease, scaling as $k_{\parallel}^{\alpha}$ with $\alpha \approx 1.8$, thereby validating the BIC criteria in Eqs.~\eqref{eq_sigma1} and~\eqref{eq_xi1}. Since these quantities should vanish in the ideal case, we also examine the convergence of $\xi_1^{-1}$ as the number of diffraction orders in the basis increases. This number is controlled by the parameter $b_{\rm max}$, which defines the radius of a circle in reciprocal space, i.e., the basis only contains diffraction orders satisfying $|\mathbf{G}_{n_x,n_y}| \leq b_{\rm max}$. The inset of Fig.~\ref{fig:2}(c) clearly shows that $\xi_1^{-1}$ decreases with increasing $b_{\rm max}$, demonstrating convergence. Moreover, a slight variation of the lattice constant also reveals the convergence of $\sigma_1^{-1}$ with respect to the maximum multipole degree $\ell_{\rm max}$ (see Appendix~\ref{sec:convergence}).

Additionally, we plot the electric and magnetic field distributions in the $xz$ plane for the considered BIC, corresponding to a magnetic dipole oriented along the $z$ axis [see the inset in Fig.~\ref{fig:2}(b)]. At each point, the electric field contribution from a single unit cell is given by $\mathbf{E}(\mathbf{r})=\braket{\mathbf{r}}{\mathbf{u}_1}$, where $\ket{\mathbf{u}_1}$ is expanded into singular spherical waves as in Eq.~\eqref{eq_u1_decomp}, and their coordinate dependence $\braket{\mathbf{r}}{\ell,m,p}$ is provided in Appendix~\ref{sec:sph_waves}. The total field is obtained by superposing the contributions from all unit cells, which can be efficiently evaluated using the Ewald summation [cf. Eq.~(3.15) in Ref.~\cite{Necada2021Jan}].

Next, we examine the orthogonality of the singular vectors to plane waves when Eqs.~\eqref{eq_sigma1} and~\eqref{eq_xi1} are satisfied. Figure~\ref{fig:2}(d) shows the scalar products with the first left and right singular vectors of the effective \textit{T}-matrix, namely $|\braket{\Psi^\downarrow_{\rm out}}{\mathbf{u}_1}|$ and $|\braket{\mathbf{v}_1}{\Psi^\downarrow_{\rm inc}}|$, evaluated along the quasi-BIC resonance. These quantities are identical because $\ket{\mathbf{v}_1} = \mathrm{e}^{\mathrm{i}\phi_1}\ket{\mathbf{u}_1}$ in this case, and they are equal to zero (numerically to $\sim 10^{-35}$) at $\theta = 0$, thereby confirming the BIC criteria in Eqs.~\eqref{eq_rad_losses_u1} and~\eqref{eq_coupling_v1}. Physically, at normal incidence ($\theta = 0$), the singular vectors $\ket{\mathbf{u}_1}$ and $\ket{\mathbf{v}_1}$ contain only the out-of-plane magnetic dipole component $m_z \propto u^1_{1,0,0}$, which cannot be excited by a normally incident plane wave (therefore, this BIC is referred to as an S-BIC). For $\theta \neq 0$, the scalar products become nonzero because the singular vectors acquire contributions from in-plane electric dipoles $p_y$, which are present in the incident field [specifically, $p_y \propto (u^1_{1,-1,1} + u^1_{1,1,1})$, as shown in Fig.~\ref{fig:2}(e)].

Moreover, the absolute values of the scalar products exhibit a linear dependence on the angle of incidence as $\theta \to 0$. This behavior can be understood by explicitly writing the scalar product while retaining only the dominant dipole contributions: $\braket{\Psi^\downarrow_{\rm out}}{\mathbf{u}_1} = b_{1,-1,1}^* u^{1}_{1,-1,1}  + b_{1,0,0}^* u^{1}_{1,0,0} + b_{1,1,1}^* u^{1}_{1,1,1}$. As shown in Fig.~\ref{fig:2}(e), the coefficients $u^{1}_{1,\pm1,1}$ vary linearly with $\theta$, whereas $u^{1}_{1,0,0}$ is nearly independent of $\theta$. At the same time, the plane-wave expansion coefficients satisfy $b^{1}_{1,\pm1,1} = \mathrm{const}$ and $b^{1}_{1,0,0} \propto \sin\theta \sim \theta$ for $\theta \to 0$~\cite{Rahimzadegan2022May}. Consequently, all terms in $\braket{\Psi^\downarrow_{\rm out}}{\mathbf{u}_1}$ scale linearly with $\theta$, and so does the scalar product itself. Using a classical analog of Fermi’s golden rule, the radiative losses of the quasi-BIC can be expressed as $\gamma_1 \propto \left| \braket{\Psi^\downarrow_{\rm out}}{\mathbf{u}_1} \right|^2$, which yields $\gamma_1 \sim \theta^2 \sim \sin^2\theta \sim k_{\parallel}^2$ in the vicinity of the $\Gamma$ point. Notably, this scaling law coincides with the analytical result reported in Ref.~\cite{Gladyshev2022Jun}.

Finally, Fig.~\ref{fig:2}(f) shows the corresponding scalar products involving the first left and right singular vectors of the \textit{S}-matrix. These quantities also exhibit a linear dependence as $\theta \to 0$ and satisfy the condition in Eq.~\eqref{eq_h1_f1_braket} at normal incidence. Owing to the mirror symmetry $z \to -z$, the scalar products obey $\braket{\Psi^\downarrow_{\rm out}}{\mathbf{u}_1} = \braket{\Psi^\uparrow_{\rm out}}{\mathbf{u}_1}$ and $\braket{\mathbf{v}_1}{\Psi^\downarrow_{\rm inc}} = \braket{\mathbf{v}_1}{\Psi^\uparrow_{\rm inc}}$, and the zeroth-order components of the singular vectors satisfy $h^1_{\mathbf{k}_{\parallel},\downarrow,0} = h^1_{\mathbf{k}_{\parallel},\uparrow,0}$ and $f^1_{\mathbf{k}_{\parallel},\downarrow,0} = f^1_{\mathbf{k}_{\parallel},\uparrow,0}$.

\subsection{Infinite array with absorption}
\begin{figure}
    \centering
    \includegraphics[width=\linewidth]{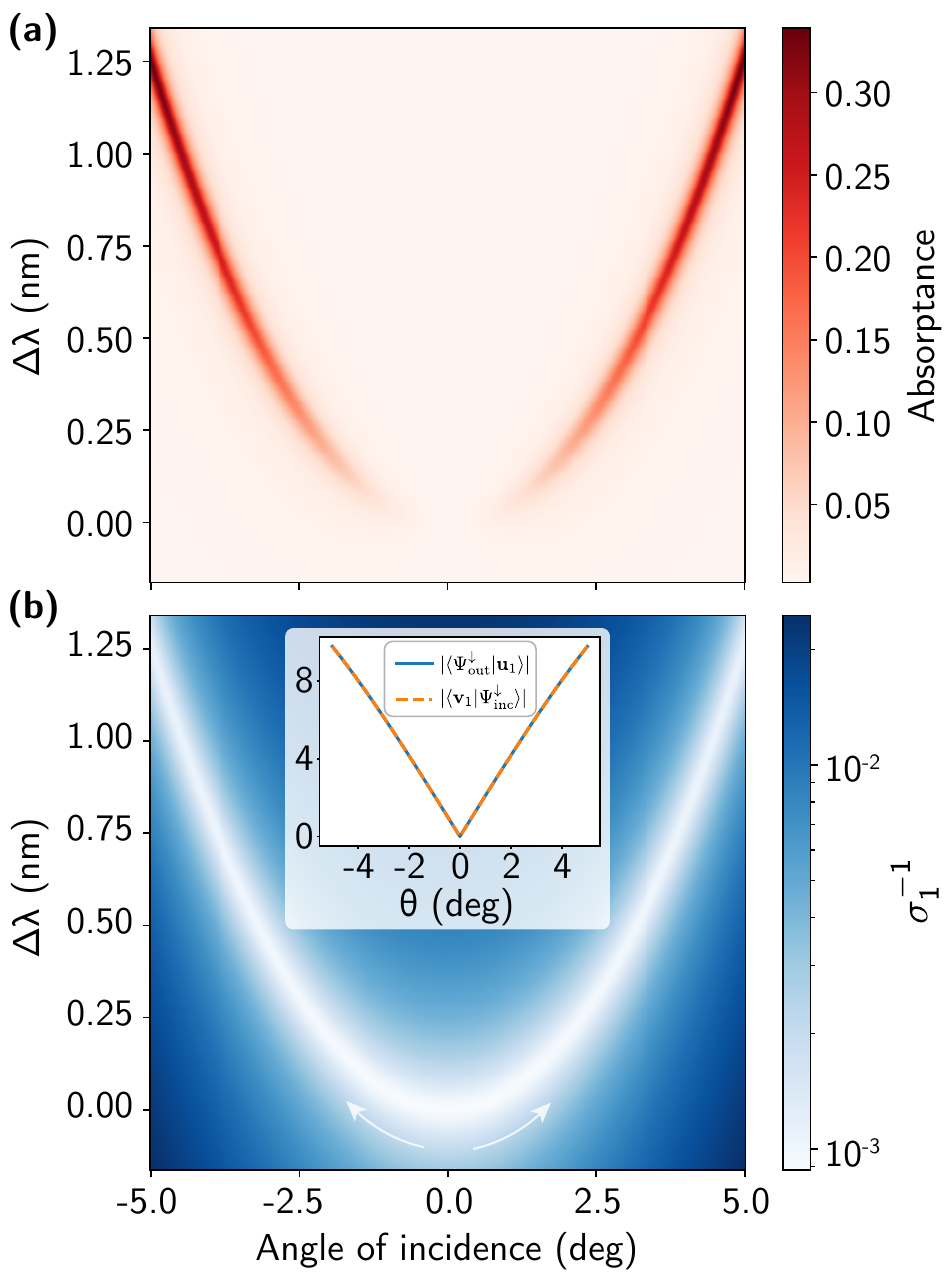}
    \caption{Electromagnetic S-BIC in a single metasurface of spherical particles with absorption [the loss tangent is $\approx4.1 \times 10^{-5}$]. (a) Absorptance of the metasurface [for $\ell_{\rm max} = 4$]. (b) The inverse of the largest singular values of the effective \textit{T}-matrix [$\ell_{\rm max} = 1$]. The inset shows the absolute values of scalar products of the singular vectors of $\mathbf{T}_{\rm eff}(\mathbf{k}_{\parallel}, \omega)$ and plane waves: $|\braket{\Psi^\downarrow_{\rm out}}{\mathbf{u}_1}|$ (blue solid) and $|\braket{\mathbf{v}_1}{\Psi^\downarrow_{\rm inc}}|$ (orange dashed), corresponding to minima in panel (b).}
    \label{fig:abs}
\end{figure}

We now consider the BIC under more realistic conditions by introducing absorption in the particles through a nonzero imaginary part of the dielectric permittivity, $\varepsilon=12.25 + \mathrm{i}10^{-3}$, corresponding to a loss tangent of $\approx 4.1 \times 10^{-5}$. In dissipative systems, a quasi-BIC manifests as a resonance in the absorptance~\cite{Poleva2025May}. Accordingly, we compute the absorptance of the metasurface as $(1-T-R)$, where $T$ and $R$ denote the transmittance and reflectance, respectively, and present the results in Fig.~\ref{fig:abs}(a), which highlights the relatively high absorptance achievable in the quasi-BIC regime. Importantly, the BIC criteria in Eqs.~\eqref{eq_sigma1} and~\eqref{eq_xi1} can be satisfied for real-valued frequencies only in nonabsorptive systems. Nevertheless, our approach remains applicable for identifying the BIC resonance position at real-valued frequencies [see Fig.~\ref{fig:abs}(b)]. Notably, the presence of absorption does not hinder the formation of a BIC with an infinite radiative lifetime and causes only a negligible spectral shift compared with the lossless case. Therefore, to verify the existence of a BIC in an absorptive system, one may set $\Im{\varepsilon}=0$ and check whether the criteria in Eqs.~\eqref{eq_sigma1} and~\eqref{eq_xi1} are fulfilled at a real frequency. Furthermore, the orthogonality conditions in Eqs.~\eqref{eq_rad_losses_u1}, \eqref{eq_coupling_v1}, and \eqref{eq_h1_f1_braket} remain satisfied even in the presence of absorption, as shown in the inset of Fig.~\ref{fig:abs}(b), confirming the nonradiating nature of the BIC. Finally, please note that Fig.~\ref{fig:abs} shows only the SVD of the effective \textit{T}-matrix. Although the SVD of the \textit{S}-matrix is also applicable, it is omitted here for brevity.

\subsection{Finite-size arrays in free space}\label{sec:finite}
\begin{figure}
    \centering
    \includegraphics[width=\linewidth]{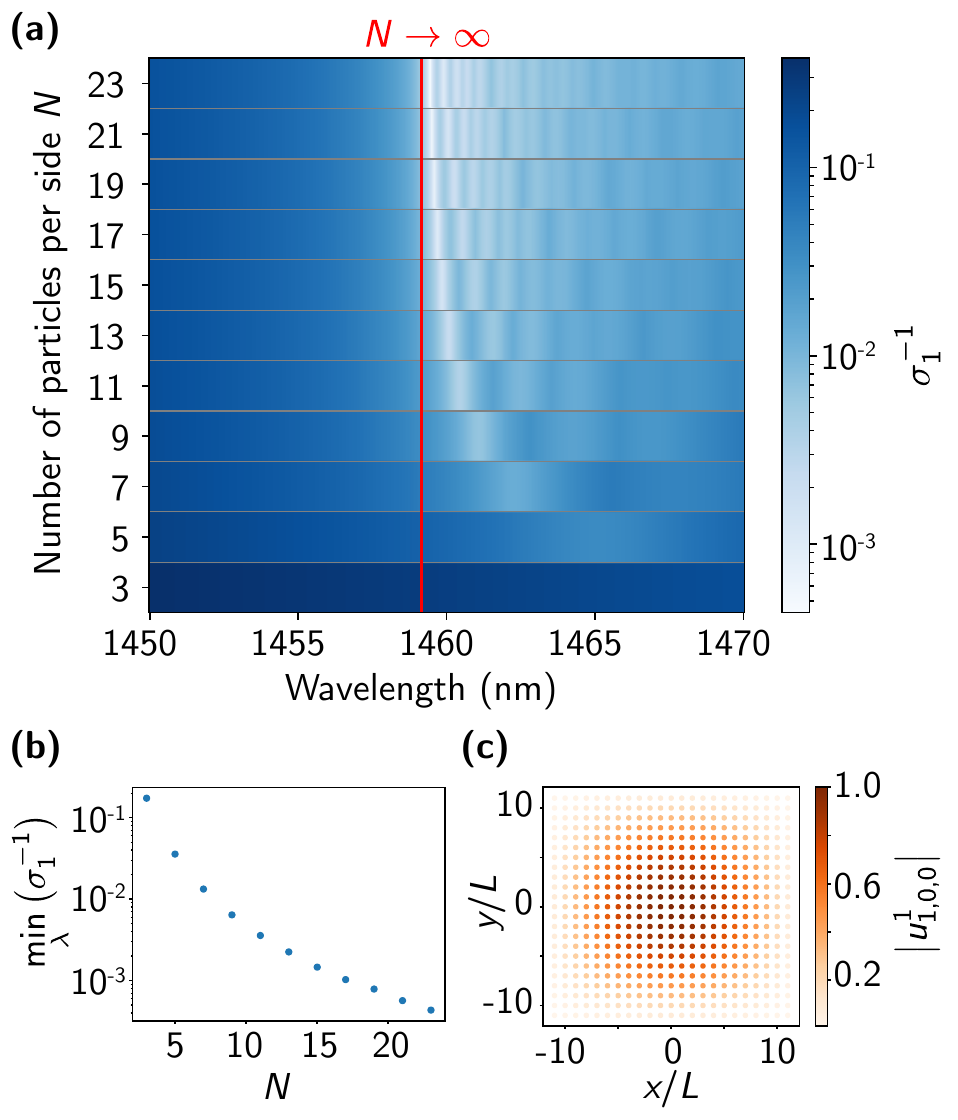}
    \caption{Development of the electromagnetic S-BIC in finite-size arrays of $N \times N$ spheres. (a) Inverse of the largest singular value of the effective \textit{T}-matrix [$\ell_{\rm max} = 1$]. The red line indicates the BIC wavelength in the infinite array ($\lambda_{\rm R}$). (b) Wavelength minimum value of $\sigma_1^{-1}$ in panel (a) as a function of $N$. (c) Normalized absolute values of components $|u^1_{i,1,0,0}|/\max\limits_i|u^1_{i,1,0,0}|$ of the left singular vector $\ket{\mathbf{u}_1}$ for $N = 23$ and $\lambda_{\rm R} = 1459.465$ nm. $u^1_{i,1,0,0}$ is proportional to the $z$ component of the magnetic dipole moment ($m_z$) of the $i$th particle.}
    \label{fig:4}
\end{figure}

In contrast with infinite arrays, finite-size $N \times N$ arrays lack translational symmetry. Such a lack of translational symmetry has two important consequences:
(1) BICs inevitably transform into quasi-BICs due to diffraction from the array edges into the surrounding medium, and
(2) Finite-size arrays cannot be described by an \textit{S}-matrix formulated in the plane-wave basis. By contrast, the SVD of the effective \textit{T}-matrix in the multipolar basis remains applicable to the analysis of quasi-BICs in finite arrays and enables the study of their evolution toward genuine BICs in the limit $N \rightarrow \infty$. Considering the same S-BIC as before, Fig.~\ref{fig:4}(a) shows the inverse of the largest singular value of the effective \textit{T}-matrix for an $N \times N$ array as a function of $N$ and wavelength. For each value of $N$, $\sigma_1^{-1}$ exhibits a minimum that shifts toward the BIC wavelength of the infinite array as $N$ increases. At the same time, the minimum value $\min_{\lambda}(\sigma_1^{-1})$ decreases with $N$ following a power law $N^{-\alpha}$ with $\alpha \approx 3$ [see Fig.~\ref{fig:4}(b)].

It is also well known that quasi-BICs in finite arrays manifest themselves as standing waves~\cite{Bulgakov2019Mar}. Importantly, this behavior is naturally captured by the SVD. In Fig.~\ref{fig:4}(c), we plot the nonzero components of the first left singular vector $\ket{\mathbf{u}_1}$ of the effective \textit{T}-matrix for $N = 23$, corresponding to the minimum of $\sigma_1^{-1}$ in Fig.~\ref{fig:4}(b). The only nonzero components are $u^1_{i,1,0,0}$, which determine the out-of-plane magnetic dipole moments ($m_z$) of the particles labeled by the index $i$. As shown in Fig.~\ref{fig:4}(c), their spatial distribution forms a two-dimensional standing wave with a quantized Bloch wave vector $\mathbf{k}_{\parallel} = \frac{\pi}{N+1}(1,1)$.

\subsection{Infinite array on a substrate}\label{sec:substrate}
\begin{figure}
    \centering
    \includegraphics[width=\linewidth]{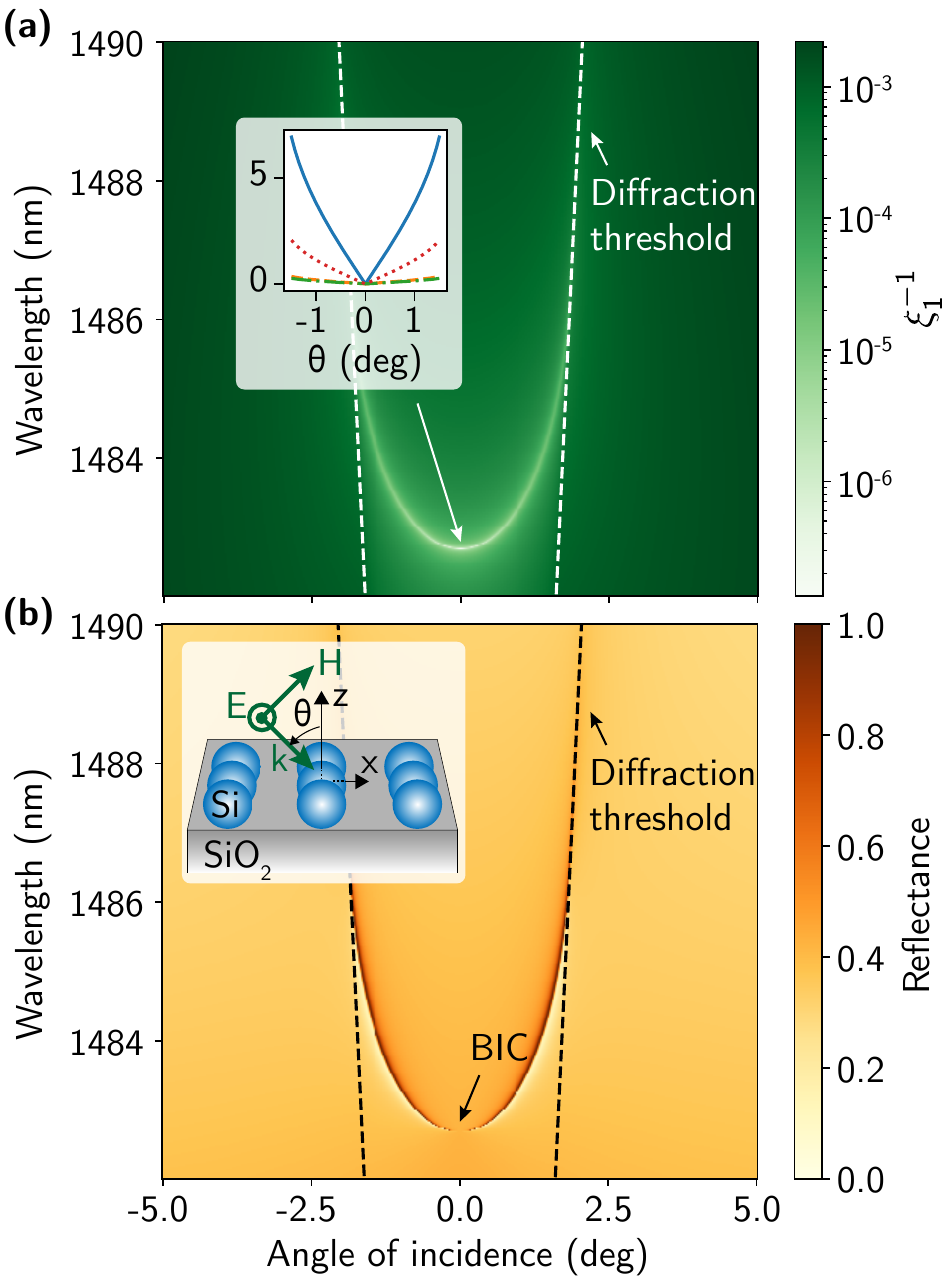}
    \caption{Electromagnetic S-BIC in a single metasurface on a substrate. (a) Inverse of the largest singular value of the total \textit{S}-matrix [$b_{\rm max} L / (2 \pi) = 4$]. Inset shows absolute values of zeroth-order components of the singular vectors $|h^1_{\mathbf{k}_{\parallel},\downarrow,0}|$ (blue solid), $|f^1_{\mathbf{k}_{\parallel},\downarrow,0}|$ (orange dashed), $|h^1_{\mathbf{k}_{\parallel},\uparrow,0}|$ (green dashed-dotted), and $|f^1_{\mathbf{k}_{\parallel},\uparrow,0}|$ (red dotted). (b) Reflectance for a TE-polarized plane wave. Inset: a Si-particle metasurface on a SiO$_2$ semi-infinite substrate and the geometry of incidence. The black and white dashed lines in panels (a) and (b), respectively, indicate the diffraction threshold in the substrate $\lambda(\theta) = L \left( n_{\rm sub} + \left| \sin \theta\right|\right)$, where $n_{\rm sub} = 1.4$.}
    \label{fig:5}
\end{figure}

We now return to infinite arrays and consider an opposite scenario, in which the SVD of the effective \textit{T}-matrix is less practical for predicting the spectral position of a BIC than the SVD of the \textit{S}-matrix. The inset of Fig.~\ref{fig:5}(b) depicts the same metasurface placed on a substrate with refractive index $n_{\rm sub} = 1.4$ (close to SiO$_2$). It is known that S-BICs at the $\Gamma$ point remain robust even when the reflection symmetry $z \to -z$ is broken~\cite{Koshelev2023May}. However, the simple relationship between the lattice constant and the wavelength of a single-particle dipole resonance, $L \approx 0.71\lambda_{\rm R}$, is valid only for metasurfaces in free space.

The presence of a substrate modifies lattice interactions relative to a homogeneous environment, requiring the evaluation of integrals involving Fresnel reflection coefficients~\cite{Czajkowski2020Aug}. In this situation, it is more convenient to apply the \textit{S}-matrix-based approach and use Eq.~\eqref{eq_xi1} to determine the BIC spectral position. In Eq.~\eqref{eq_xi1}, we use the total \textit{S}-matrix of the structure, shown in the inset of Fig.~\ref{fig:5}(b), which is obtained as the Redheffer star product of three \textit{S}-matrices describing the substrate-air interface, propagation in free space along the $z$ axis, and the metasurface. Owing to the substrate, we include a larger number of diffraction orders--specifically those with $|\mathbf{G}_{n_x,n_y}| \leq 8\pi/L$--to ensure convergence.

Numerical simulations of $\xi_1^{-1}$ reveal that its zero shifts from $\lambda_{\rm R} = 1459.16$ nm to $\lambda = 1482.70$ nm in the presence of the substrate [Fig.~\ref{fig:5}(a)]. Moreover, the components of the left and right singular vectors corresponding to the zeroth diffraction order vanish at this wavelength and $\theta = 0$ [see the inset of Fig.~\ref{fig:5}(a)], satisfying both BIC criteria in Eqs.~\eqref{eq_xi1} and \eqref{eq_h1_f1_braket}. As a result, pronounced BIC and quasi-BIC resonances appear in the reflectance spectrum shown in Fig.~\ref{fig:5}(b).

Figures~\ref{fig:5}(a) and~\ref{fig:5}(b) also clearly demonstrate that the effect is destroyed by diffraction into the substrate when the wavelength exceeds the diffraction threshold $\lambda(\theta) = L \left(n_{\rm sub} + |\sin\theta|\right)$. The same phenomenon occurs for the metasurface in the homogeneous environment ($n_{\rm sub} = 1$) if Fig.~\ref{fig:2}(a) is extended to a wider parameter range.

Furthermore, it is also well known that nanostructures on dielectric substrates scatter asymmetrically into the upper ($+z$) and lower ($-z$) half-spaces due to dielectric contrast between the half-spaces~\cite{Novotny2012Sep}. Importantly, the SVD of the \textit{S}-matrix captures this asymmetry: the zero-order components $|h^1_{\mathbf{k}_{\parallel}, \downarrow, 0}|$ and $|f^1_{\mathbf{k}_{\parallel}, \uparrow, 0}|$ of the quasi-BIC are several orders of magnitude larger for the lower-half space than those $|h^1_{\mathbf{k}_{\parallel}, \uparrow, 0}|$ and $|f^1_{\mathbf{k}_{\parallel}, \downarrow, 0}|$ for the upper half-space [see the inset of Fig.~\ref{fig:5}(a)]. Here, $\ket{\mathbf{h}_1}$ and $\ket{\mathbf{f}_1}$ denote the left and right singular vectors corresponding to outgoing and incident modes, respectively.

\subsection{Degenerate BICs}\label{sec:degenerate}
TE- and TM-polarized S-BICs can become degenerate in highly symmetric metasurfaces through various mechanisms~\cite{Doiron2022Dec,Kodigala2017Jan,Evlyukhin2021Dec,Wang2018Dec}. Such degeneracy enables polarization-independent behavior~\cite{Liu2025Mar} and yields pairs of high-$Q$ modes that can be exploited in nonlinear optical processes, including spontaneous parametric down-conversion~\cite{Wang2019Mar} and four-wave mixing~\cite{Colom2019May}. As discussed in Sec.~\ref{sec:theory}, the SVD framework naturally captures degenerate BICs.

To illustrate this, we consider spherical particles with $\varepsilon = \mu = 3.5$, which results in an electromagnetically dual system~\cite{Fernandez-Corbaton2013Aug}. In such systems, electric and magnetic Mie scattering coefficients are identical for all multipole degrees $\ell$~\cite{Kerker1983Jun}. Consequently, electric and magnetic dipole resonances coincide at $\lambda_{\rm R} = 1280.72$ nm, and TE- and TM-polarized dipole BICs emerge at the same wavelength when the lattice constant satisfies $L/\lambda_{\rm R} = 0.7112\ldots$, i.e., $L = 910.9$ nm.

\begin{figure}
    \centering
    \includegraphics[width=\linewidth]{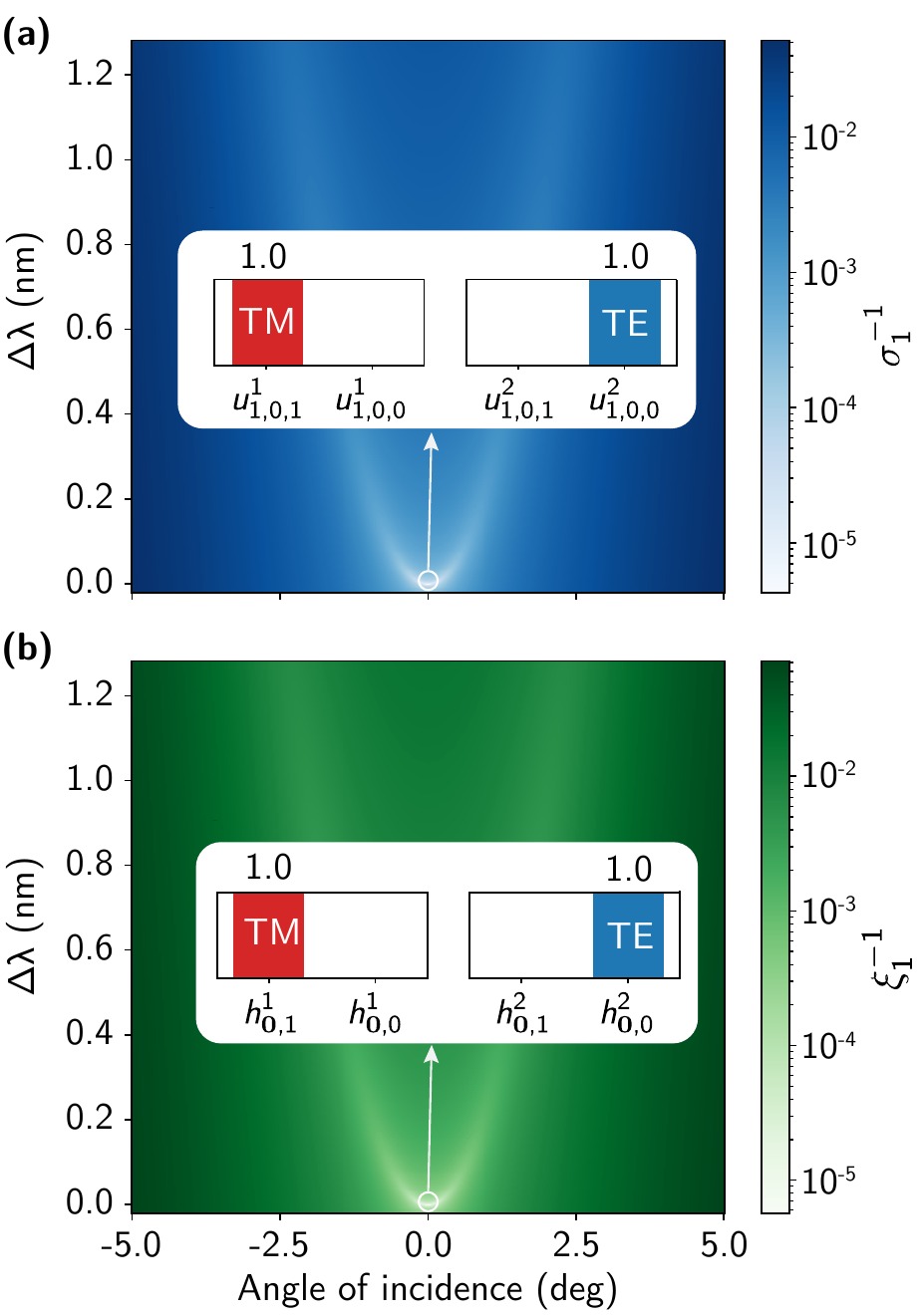}
    \caption{Doubly-degenerate BICs in a single metasurface of particles with $\varepsilon = \mu$. The resonance wavelength is 1280.72 nm, and the lattice constant is 910.90 nm. Panels (a) and (b) show the inverse of the largest singular values of the effective \textit{T}-matrix [$\ell_{\rm max} = 1$] and the \textit{S}-matrix [$b_{\rm max} L / (2 \pi) = 1$], respectively. The insets in panels (a) and (b) show nonzero components of the first and second right singular vectors as degenerate BICs with pure TE and TM polarizations, obtained from the SVD of the effective \textit{T}-matrix and the \textit{S}-matrix, respectively. Please note for the inset in panel (b) that $h^i_{\mathbf{0}, i} = \left( \left| h^i_{\mathbf{0}, \uparrow,i} \right|^2 + \left| h^i_{\mathbf{0}, \downarrow,i} \right|^2 \right)^{1/2}$, where $i=0,1$.}
    \label{fig:6}
\end{figure}

Figures~\ref{fig:6}(a) and~\ref{fig:6}(b) show $\sigma_1^{-1}$ and $\xi_1^{-1}$ for this metasurface, revealing zeros at $\theta = 0$ that correspond to doubly degenerate BICs. Due to the equality of matrix elements in both the effective \textit{T}-matrix and the \textit{S}-matrix, all singular values are doubly degenerate, including $\sigma_1 = \sigma_2$ and $\xi_1 = \xi_2$. Accordingly, the dipole BICs can be represented in the spherical-wave basis by singular vectors $\ket{\mathbf{u}_1} = \ket{1,0,0}$ and $\ket{\mathbf{u}_2} = \ket{1,0,1}$ [see the inset in Fig.~\ref{fig:6}(a)], as well as by any two mutually orthogonal superpositions thereof. An analogous description applies to $\ket{\mathbf{h}_1}$ and $\ket{\mathbf{h}_2}$ in the diffraction-order basis, as illustrated in the inset of Fig.~\ref{fig:6}(b).

\subsection{Fabry--Perot BICs in a double-layer metasurface}\label{sec:fabry}
\begin{figure}
    \centering
    \includegraphics[width=\linewidth]{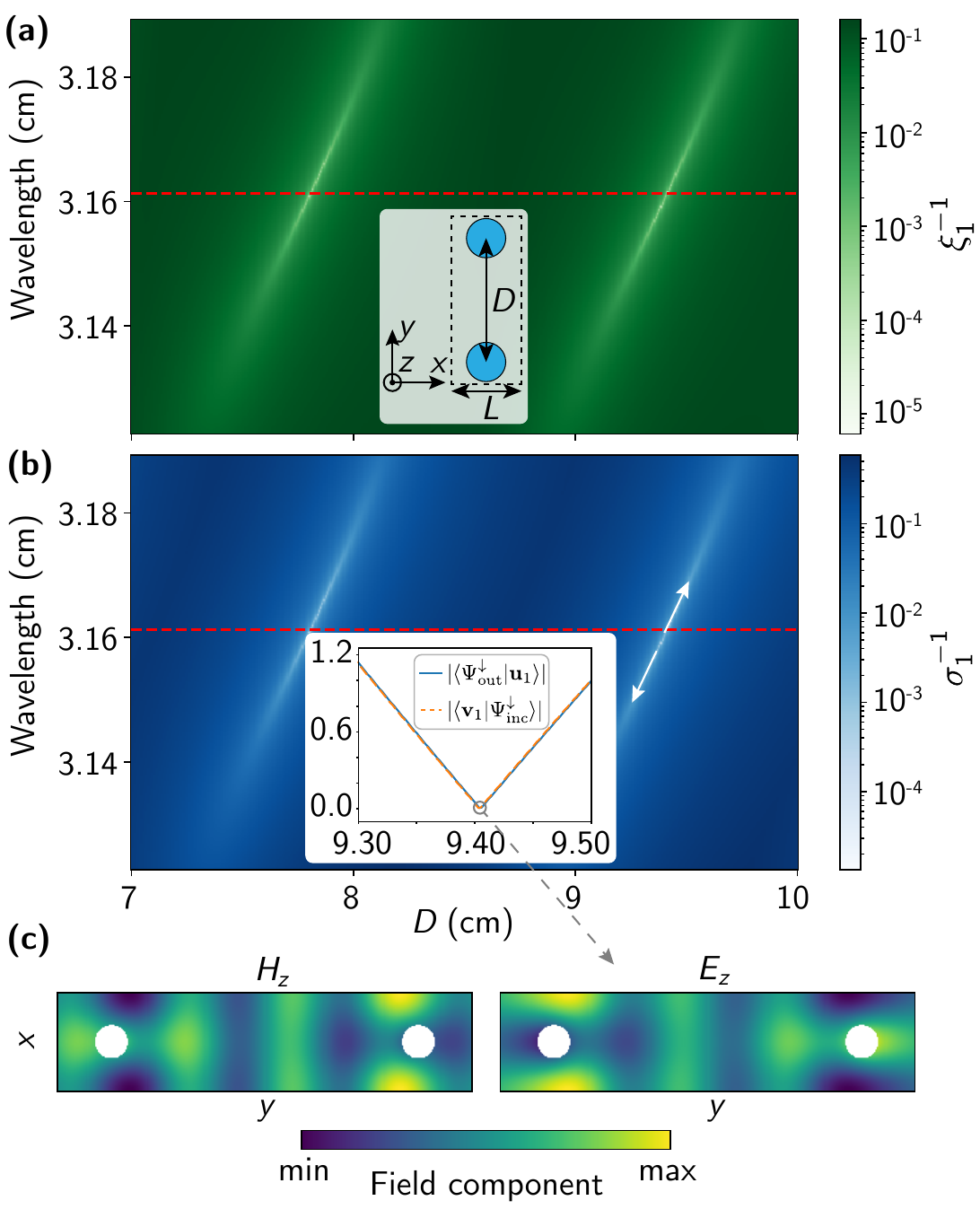}
    \caption{Fabry--Perot BICs in a double-layer metasurface of dielectric rods in free space depicted in the inset in panel (a). The black dashed line outlines a complex unit cell with two rods. Panels (a) and (b) show the inverse of the largest singular values of (a) the \textit{S}-matrix [$b_{\rm max} L / (2 \pi) = 1$] and (b) the effective \textit{T}-matrix in the cylindrical-wave basis [$\left| m_{\rm max}\right| = 2$] as a function of the distance between the layers and the wavelength. Inset in panel (b) shows the absolute values of scalar products $|\braket{\Psi^\downarrow_{\rm out}}{\mathbf{u}_1}|$ (blue solid) and $|\braket{\mathbf{v}_1}{\Psi^\downarrow_{\rm inc}}|$ (orange dashed) for the parameters indicated by the white arrows in panel (b). The red dashed line indicates the wavelength $\lambda_0 = 3.161$ cm of a zero-transmittance resonance of a single lattice. (c) $H_z$ (left) and $E_z$ (right) components of the total electric and magnetic fields of the Fabry--Perot BIC emerging at $D = 9.41$ cm.}
    \label{fig:rods}
\end{figure}

Up to this point, we have focused on symmetry-protected BICs in single-layer metasurfaces. However, the SVD-based approach is equally applicable to accidental BICs. Here, we consider Fabry--Perot BICs~\cite{Ndangali2010Oct}, which arise in double-layer metasurfaces composed of two particle lattices separated by a distance $D$. When each lattice exhibits a reflection resonance at wavelength $\lambda_0$, the system effectively forms a Fabry--Perot cavity with resonant mirrors [see the inset of Fig.~\ref{fig:rods}(a)]. Neglecting near-field coupling between the layers and material losses, the BIC formation condition at $\lambda = \lambda_0$ and $\mathbf{k}_{\parallel}^{\rm BIC} = \mathbf{0}$ is $D = s\lambda_0/2$, where $s \in \mathbb{N}$~\cite{Alagappan2024Nov}.

We consider the structure studied in Ref.~\cite{Semushev2026Jan}, which supports TM-polarized Fabry--Perot BICs in the microwave spectral range. The system consists of two one-dimensional lattices of infinite dielectric rods with radius 0.5~cm, lattice constant 3~cm, and permittivity $\varepsilon = 2.1$ in free space [see the inset of Fig.~\ref{fig:rods}(a)]. Each lattice exhibits a reflection resonance at 9.483~GHz ($\lambda_0 = 3.161$~cm), which is accurately captured by the cylindrical \textit{T}-matrix method with $m_{\rm max} = 2$. For a one-dimensional lattice of infinite rods, we choose the $z$ axis along the rods, the $x$ axis along the lattice vector, and the stacking direction along the $y$ axis.

The SVD of the \textit{S}-matrix for the double-layer metasurface can be computed using the following two approaches: 
\begin{enumerate}
    \item First, one may compute the \textit{S}-matrix of a single lattice from the effective \textit{T}-matrix of a unit cell, and then obtain the total \textit{S}-matrix via the Redheffer star product of three \textit{S}-matrices that describe the lattice, free-space propagation over distance $D$, and the lattice again.
    \item Alternatively, the total \textit{S}-matrix can be computed directly from the effective \textit{T}-matrix of a complex unit cell containing two rods, as shown in the inset of Fig.~\ref{fig:rods}(a), with lattice interactions evaluated using the Ewald summation technique for two sublattices~\cite{Beutel2023Jan}. 
\end{enumerate}

We use the second approach, as it allows us to identify Fabry--Perot BICs via the SVD of both the effective \textit{T}-matrix and the \textit{S}-matrix. Indeed, the formation of two Fabry--Perot BICs is confirmed by the criteria in Eqs.~\eqref{eq_xi1} and~\eqref{eq_sigma1}, as shown in Figs.~\ref{fig:rods}(a) and~\ref{fig:rods}(b), respectively. Furthermore, the inset of Fig.~\ref{fig:rods}(b) demonstrates that one of these BICs becomes orthogonal to plane waves at an interlayer distance $D = 9.41$~cm, with the deviation from the predicted value $D = 3\lambda_0 = 9.483$~cm due to near-field coupling between the layers. Moreover, the inset of Fig.~\ref{fig:rods}(b) also shows that an offset of the interlayer distance $\Delta D$ leads to an increase in radiative losses $\gamma_1 \propto |\braket{\Psi^\downarrow_{\rm out}}{\mathbf{u}_1}|^2 \sim (\Delta D)^2$, indicating a second-order dependence on $\Delta D$, in agreement with Ref.~\cite{Semushev2026Jan}. Similarly to Sec.~\ref{sec:infinite}, we calculate the $z$ components of the electric and magnetic field of this Fabry-Perot BIC in the cylindrical-wave basis and plot them in Fig.~\ref{fig:rods}(c). Please note that in this case, the BIC field is the superposition of the fields radiated by two lattices.

\subsection{Acoustic BICs}\label{sec:acoustic}
Finally, we extend the SVD-based framework to acoustic systems by considering an infinite lattice of lossless high-index acoustic particles ($\rho = 360$~kg/m$^3$, $v = 600$~m/s) embedded in water ($\rho = 998$~kg/m$^3$, $v = 1500$~m/s). Despite being single-layer structures, such lattices can support accidental BICs at $\mathbf{k}_{\parallel}^{\rm BIC} = \mathbf{0}$ due to destructive interference between radiating multipoles of different degrees $\ell$ but identical parity, which were discussed in Ref.~\cite{Ustimenko2026Jan}. Consequently, at least $\ell_{\rm max} = 2$ and $\ell_{\rm max} = 3$ are required to observe even- and odd-parity BICs, respectively. Here, we focus on the even-parity case.

\begin{figure}
    \centering
    \includegraphics[width=\linewidth]{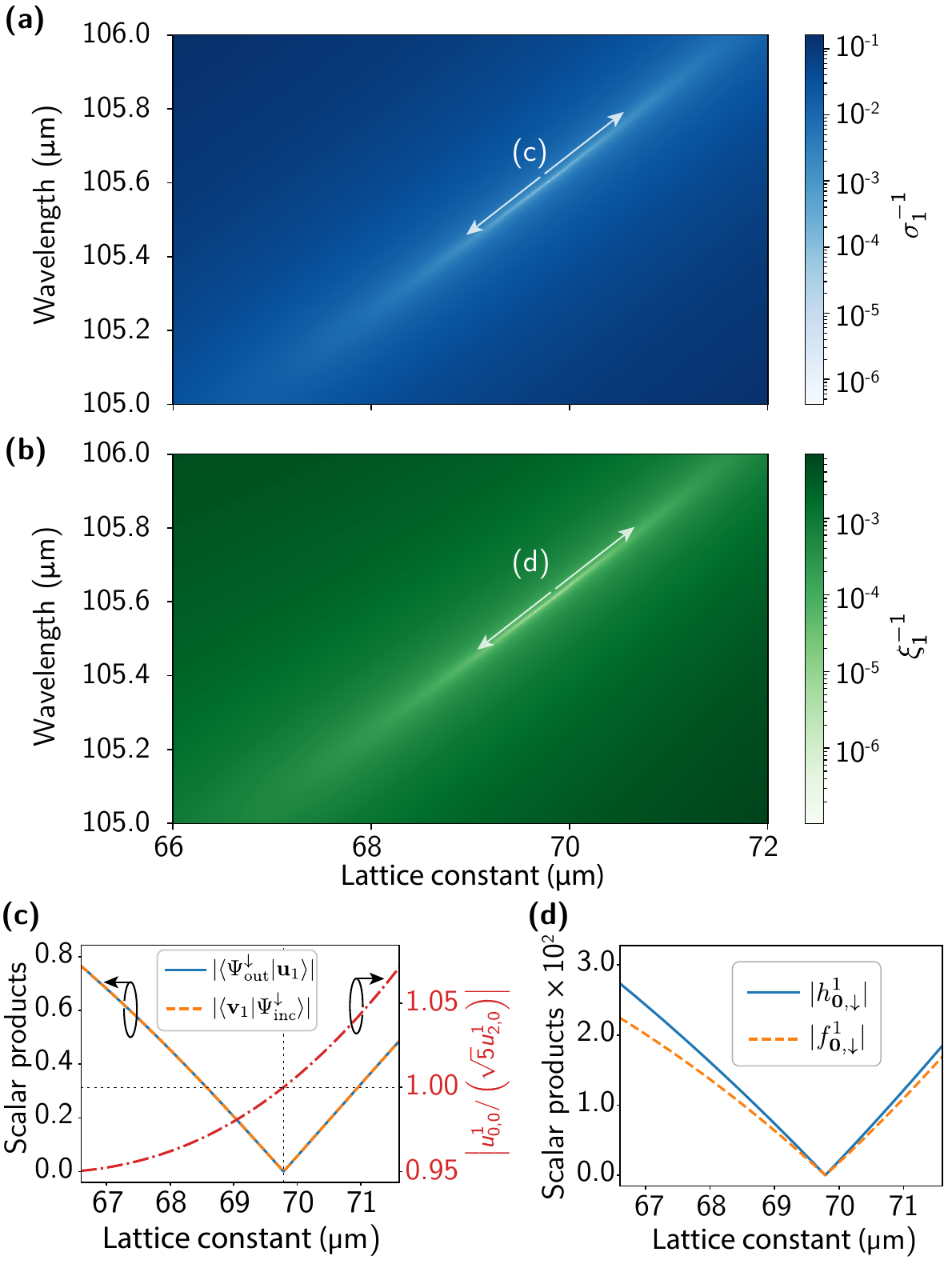}
    \caption{Acoustic A-BIC in a single metasurface of spherical high-index scatterers in water. Panels (a) and (b) show the inverse of the largest singular values of the (a) effective \textit{T}-matrix [$\ell_{\rm max} = 2$] and (b) \textit{S}-matrix [$b_{\rm max} L / (2 \pi) = 1$] as a function of lattice constant and wavelength in water. (c) Absolute values of scalar products  $|\braket{\Psi^\downarrow_{\rm out}}{\mathbf{u}_1}|$ (blue solid) and $|\braket{\mathbf{v}_1}{\Psi^\downarrow_{\rm inc}}|$ (orange dashed), for the parameters indicated with the white arrows in panel (a). The rod dashed-dotted line indicates the absolute value of ratio $u^1_{0,0}/\left(\sqrt{5} u^1_{2,0} \right)$ of the components of vector $\ket{\mathbf{u}_1}$. (d) Absolute values of zeroth-order components of the left and right singular vectors of $\mathbf{S}$, $|h^1_{\mathbf{0},\downarrow}|$ and $|f^1_{\mathbf{0},\downarrow}|$, for the same parameters.}
    \label{fig:8}
\end{figure}

Figures~\ref{fig:8}(a) and~\ref{fig:8}(b) show that the inverses of the largest singular values, $\sigma_1^{-1}$ and $\xi_1^{-1}$, vanish at $L = 69.78~\mu$m and $\lambda = 105.6~\mu$m, confirming BIC formation according to the criteria in Eqs.~\eqref{eq_sigma1} and~\eqref{eq_xi1}. The scalar products in Eqs.~\eqref{eq_rad_losses_u1} and~\eqref{eq_coupling_v1} also vanish at these parameters [see Fig.~\ref{fig:8}(c)] and are equal due to $\ket{\mathbf{v}_1} = \mathrm{e}^{\mathrm{i}\phi_1}\ket{\mathbf{u}_1}$. Although the singular vectors have nonzero overlaps with the incident and outgoing plane waves due to $u^1_{0,m} \neq 0$ and $u^1_{2,m} \neq 0$ for $m=0$, their scalar products vanish owing to the condition $b_{0,0}^* u^1_{0,0} = -b_{2,0}^* u^1_{2,0}$ at the BIC [see Fig.~\ref{fig:8}(c)]. Here, $b_{\ell,0} = (\mathrm{i}d)^\ell \sqrt{2\ell+1}$ are the expansion coefficients of a normally incident acoustic plane wave, where $d = +1\,(\uparrow)$ or $-1 \,(\downarrow)$.

As a function of the offset of the lattice constant $\Delta L$, the absolute value of scalar product $|\braket{\Psi^\downarrow_{\rm out}}{\mathbf{u}_1}|$ varies linearly near the quasi-BIC resonance [see Fig.~\ref{fig:8}(c)], implying radiative losses $\gamma_1 \propto |\braket{\Psi^\downarrow_{\rm out}}{\mathbf{u}_1}|^2 \sim (\Delta L)^2$, in agreement with the simulation results in Ref.~\cite{Ustimenko2026Jan}. Figure~\ref{fig:8}(d) further shows that the zeroth diffraction-order components in Eq.~\eqref{eq_h1_f1_braket} vanish at the BIC, indicating a purely evanescent field.

These results demonstrate that the SVD-based framework applies equally well to acoustic systems. In contrast with Ref.~\cite{Ustimenko2026Jan}, which analyzed the singular values of the acoustic interaction matrix $\mathbf{T}_{\rm eff}^{-1}(\mathbf{k}_{\parallel}, \omega)$ within a two-multipole approximation, we consider the full effective \textit{T}-matrix of size $(\ell_{\rm max}+1)^2 \times (\ell_{\rm max}+1)^2$ (i.e., $9 \times 9$) as well as the SVD of the \textit{S}-matrix. This highlights the generality of our approach, which does not rely on prior knowledge of block-diagonal matrix structures. 

\section{Discussion}\label{sec:discussion}
The results presented demonstrate that the proposed SVD-based approach can capture both genuine BICs and quasi-BICs in periodic metasurfaces. At this stage, an important question arises: How can one reliably distinguish between genuine BICs and quasi-BICs? Although efficient minimization algorithms can be employed to locate minima of $\sigma_1^{-1}$ and $\xi_1^{-1}$ (which correspond to zeros for BICs in non-absorptive systems), our findings indicate that, in practical numerical implementations with a finite spectral mesh, it might be challenging to judge whether $\sigma_1^{-1}$ and $\xi_1^{-1}$ truly vanish (as required for genuine BICs) or merely attain extremely small but finite values (indicative of high-$Q$ quasi-BICs). Indeed, Figs.~\ref{fig:2}(b) and~\ref{fig:2}(c) show that even at parameter values corresponding to genuine BICs, $\sigma_1^{-1}$ and $\xi_1^{-1}$ remain on the order of $\sim10^{-8}$ and $\sim10^{-9}$, respectively, on the chosen mesh. Therefore, the conditions $\sigma_1^{-1} \to 0$ and $\xi_1^{-1} \to 0$ alone are insufficient to unambiguously identify the emergence of a BIC.

In this context, two additional aspects should be taken into account. First, it is necessary to verify the conditions given in Eqs.~\eqref{eq_rad_losses_u1} and~\eqref{eq_h1_f1_braket}, which formally express the orthogonality of BICs to radiative channels in the surrounding medium. In contrast with $\sigma_1^{-1}$ and $\xi_1^{-1}$, the associated scalar products can vanish down to machine precision--approximately $10^{-35}$ in the case considered in Sec.~\ref{sec:infinite}. Second, a genuine BIC is associated with a single frequency (or wavelength) when all other system parameters are fixed, whereas quasi-BIC responses extend over a finite spectral range around the BIC position. As clearly demonstrated by our results, both $\sigma_1^{-1}$ and $\xi_1^{-1}$ decrease as the BIC spectral position is approached.

\section{Conclusion}\label{sec:conclusion}
We have theoretically and numerically analyzed the singular value decomposition of the effective \textit{T}-matrix and the \textit{S}-matrix in both the genuine-BIC and quasi-BIC regimes of periodic metasurfaces. This analysis allowed us to formulate general and practically applicable criteria for the formation of electromagnetic and acoustic BICs, summarized by the boxed equations in Sec.~\ref{sec:theory}. These criteria were validated through numerical simulations using the Python packages \textit{treams} and \textit{acoustotreams} for electromagnetic and acoustic metasurfaces, respectively. In particular, the simulations demonstrate that varying the angle of incidence (for S-BICs) or the system parameters (for A-BICs) induces a transition from a genuine BIC to a quasi-BIC, accompanied by deviations from the ideal BIC conditions. In the quasi-BIC regime, the minimum of the inverse largest singular value accurately identifies the spectral position of the quasi-BIC, while the scalar product between the first left singular vector and outgoing plane waves quantitatively captures the scale of its radiative losses.

The SVD-based descriptions of the effective \textit{T}-matrix and the \textit{S}-matrix are not interchangeable but complementary. They differ both in the classes of systems to which they can be efficiently applied, and in the type of physical information they provide about BICs and quasi-BICs, such as multipole expansions in the \textit{T}-matrix formulation and Fourier (plane-wave) expansions in the \textit{S}-matrix formulation. 

In conclusion, the singular value decomposition of the effective \textit{T}-matrix and the \textit{S}-matrix provides a solid and self-consistent framework for identifying and analyzing genuine BICs in metasurface-based systems, as well as for tracking their evolution in the quasi-BIC regime. We are confident that this approach will facilitate the development of efficient numerical strategies for simulating and optimizing electromagnetic and acoustic metastructures that support high-$Q$ modes enabled by the physics of bound states in the continuum. 
 
\begin{acknowledgments}
The authors thank Puneet Garg and Andrey Bogdanov for fruitful discussions. The work was supported by the Deutsche Forschungsgemeinschaft (DFG, German Research Foundation) under Germany’s Excellence Strategy via the Excellence Cluster 3D Matter Made to Order (EXC-2082/2, Grant No. 390761711) and from the Carl Zeiss Foundation via CZF-Focus@HEiKA.
\end{acknowledgments}

\section*{Data availability}
The data that support the findings of this article are openly available~\cite{svd}.

\appendix
\section{Basis sets}\label{sec:sph_waves}
In this section, we define scalar (respectively vector) spherical, cylindrical, and plane waves that constitute basis sets of solutions to the scalar (respectively vector) Helmholtz equation in spherical $(r,\theta,\varphi)$, cylindrical $(\rho,\varphi,z)$, and Cartesian $(x,y,z)$ coordinate systems~\cite{Morse1953}. 

\subsection{Plane waves}
The coordinate dependence of scalar plane waves with in-plane wave vector $\mathbf{k}_{\parallel} = (k_x, k_y)$ is
\begin{align}
    \braket{\mathbf{r}}{\mathbf{k}_{\parallel}, d}  = \mathrm{e}^{\mathrm{i} \mathbf{k}\cdot\mathbf{r}}\,,
\end{align}
where $\mathbf{k}\cdot\mathbf{r} = k_{x}x + k_{y}y + d \sqrt{\omega^2/v^2-k_{\parallel}^2}z$,
while the coordinate dependence of vector plane waves is
\begin{align}
    \braket{\mathbf{r}}{\mathbf{k}_{\parallel}, d, p}  = 
        \begin{cases}
        \mathbf{N}_{\mathbf{k}}(k\mathbf{r})\,, & p = 1 \,( \text{TM})\, \\
        \mathbf{M}_{\mathbf{k}}(k\mathbf{r})\,, & p = 0 \,(\text{TE})\,, 
    \end{cases}
\end{align}
where $\mathbf{M}_{\mathbf{k}}(k\mathbf{r})$ and $\mathbf{N}_{\mathbf{k}}(k\mathbf{r})$ are defined in Eqs.~(9a) and~(9b) in Ref.~\cite{Beutel2024Apr}, respectively.

\subsection{Spherical waves}
The coordinate dependence of scalar spherical waves, which are used to expand scattered (outgoing) acoustic fields, is
\begin{align}
\label{eq_sph_waves}
    \braket{\mathbf{r}}{\ell,m}  = h_{\ell}^{(1)}(kr) Y_{\ell,m}(\theta,\varphi)\,,
\end{align}
where $\ell \in \mathbb{N}_0$ is the degree, $m \in \{-\ell,-\ell + 1, ..., \ell-1, \ell\}$ is the order, $h_{\ell}^{(1)}(kr)$ is the spherical Hankel function of first kind, and $Y_{\ell,m}(\theta,\varphi)$ is the spherical harmonic [see Eq.~(B.1) in Ref.~\cite{Beutel2024Apr}]. The total number of scalar spherical waves with $\ell \in [0, \ell_{\rm max}]$ is $N_T = (\ell_{\rm max} + 1)^2$. Waves describing incident fields are obtained from Eq.~\eqref{eq_sph_waves} by replacing $h_{\ell}^{(1)}(kr)$ with the spherical Bessel function of the first kind $j_{\ell}(kr)$.

Scalar waves~\eqref{eq_sph_waves} generate vector spherical waves. The transversely polarized vector spherical waves, which expand scattered (outgoing) electromagnetic fields, are defined as
\begin{align}
\label{eq_sph_waves_vec}
    \braket{\mathbf{r}}{\ell,m,p} = 
    \begin{cases}
        \mathbf{N}^{(3)}_{\ell,m}(k\mathbf{r})\,, & p = 1 \,( \text{TM})\,\\
        \mathbf{M}^{(3)}_{\ell,m}(k\mathbf{r})\,, & p = 0 \,(\text{TE})\,, 
    \end{cases}
\end{align}
where the magnetic $\mathbf{M}^{(3)}_{\ell,m}(k\mathbf{r})$ and electric $\mathbf{N}^{(3)}_{\ell,m}(k\mathbf{r})$ spherical waves are given by Eqs.~(11a) and~(11b) in Ref.~\cite{Beutel2024Apr}, respectively. Since the vector fields in Eq.~\eqref{eq_sph_waves_vec} are solenoidal, the degree must satisfy $\ell \geq 1$, i.e., $\ell = 0$ is absent. The total number of vector spherical waves with $\ell \in [1, \ell_{\rm max}]$ and both polarizations is $N_T = 2\ell_{\rm max}(\ell_{\rm max} + 2)$.

\subsection{Cylindrical waves}
To use cylindrical waves in the packages \textit{treams} and \textit{acoustotreams} for scatterers that are infinite along one spatial dimension, we first permute the Cartesian axes from $(x,y,z)$ to $(x,z,y)$. This permutation aligns the elongated scatterers with the $z$ axis, while the $y$ axis becomes the propagation direction. Note that within the \textit{S}-matrix routines of \textit{treams} and \textit{acoustotreams}, the inverse permutation is applied, so that the propagation direction is again aligned with the $z$ axis. 

In the $(x,z,y)$ system, the coordinate dependence of scalar cylindrical waves, which expand scattered acoustic fields, is
\begin{align}
    \braket{\mathbf{r}}{k_z, m}  = H^{(1)}_m (k_{\rho} \rho) \mathrm{e}^{\mathrm{i} \left(m \varphi + k_{z}z\right)}\,,
\end{align}
where $k_{\rho} = \sqrt{k^2 - k_z^2}$ and $H^{(1)}_m (k_{\rho} \rho)$ is the Hankel function of the first kind [replaced by the Bessel function $J_m(k_{\rho} \rho)$ for incident fields]. The total number of scalar cylindrical waves is $N_T = N_{k_z}(2m_{\rm max} + 1)$, where $N_{k_z}$ is the number of $k_z$ components.

The coordinate dependence of outgoing vector cylindrical waves is
\begin{align}
    \braket{\mathbf{r}}{k_z, m, p}  = 
        \begin{cases}
        \mathbf{N}_{k_z,m}^{(3)}(k\mathbf{r})\,, & p = 1 \,( \text{TM})\, \\
        \mathbf{M}_{k_z,m}^{(3)}(k\mathbf{r})\,, & p = 0 \,( \text{TE})\,, 
    \end{cases}
\end{align}
where $\mathbf{M}_{k_z,m}^{(3)}(k\mathbf{r})$ and $\mathbf{N}^{(3)}_{k_z,m}(k\mathbf{r})$ are given by Eqs.~(10a) and~(10b) in Ref.~\cite{Beutel2024Apr}, respectively. The total number of vector cylindrical waves is $N_T = 2N_{k_z}(2m_{\rm max} + 1)$.

 \begin{figure}
     \centering
     \includegraphics[width=\linewidth]{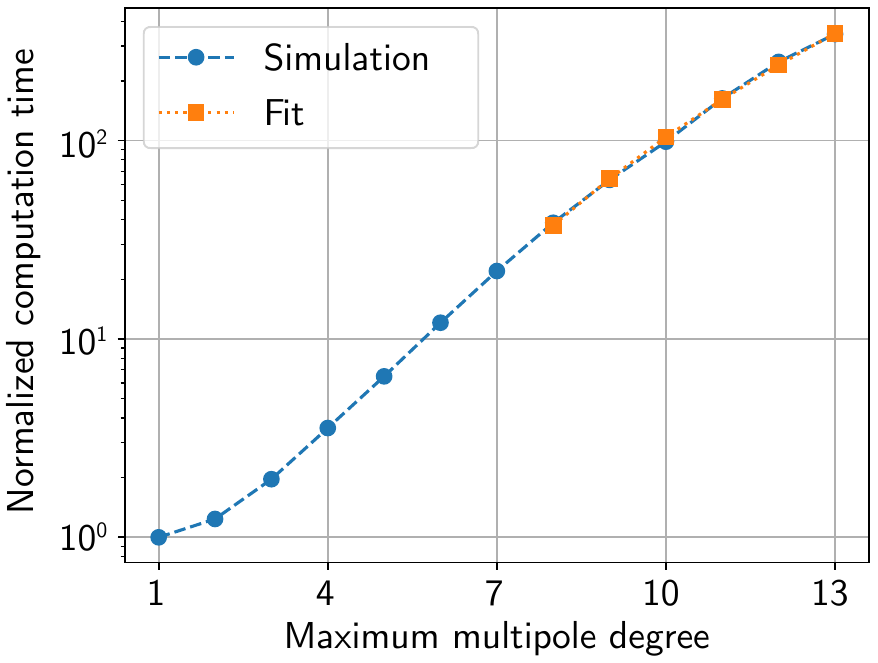}
     \caption{Normalized time $\tau_{\ell_{\rm max}}/\tau_1$ to compute the effective \textit{T}-matrix and its SVD in \textit{treams} (blue) and the polynomial fit of the time (orange) as a function of maximum multipole degree $\ell_{\rm max}$.}
     \label{fig:time}
 \end{figure}
 
\section{Analysis of the computation time}\label{sec:time}

To estimate the computational complexity of our approach, we use the \texttt{time} package to measure the elapsed time $\tau_{\ell}$ required to construct the electromagnetic \textit{T}-matrix of a sphere up to a maximum multipole degree $\ell_{\rm max}$ in \textit{treams}, compute the effective \textit{T}-matrix, and perform its SVD for a single wavelength, lattice constant, and angle of incidence. Figure~\ref{fig:time} shows $\tau_{\ell_{\rm max}}$ normalized by $\tau_{\ell_{\rm max}=1}$ as a function of $\ell_{\rm max}$. For our setup (Intel(R) Xeon(R) CPU E5-2650 v4, 2.20 GHz, 48 cores; 504 GB RAM), $\tau_{\ell_{\rm max}=1} = 8.54$ ms. We also fit the data in Fig.~\ref{fig:time} using the scaling relation $\tau_{\ell_{\rm max}}/\tau_{1} \sim \ell_{\rm max}^\alpha$, obtaining $\alpha \approx 4.59$ for $\ell_{\rm max} \in [7,13]$.

\section{Analysis of the convergence}\label{sec:convergence}

\begin{figure}
    \centering
    \includegraphics[width=\linewidth]{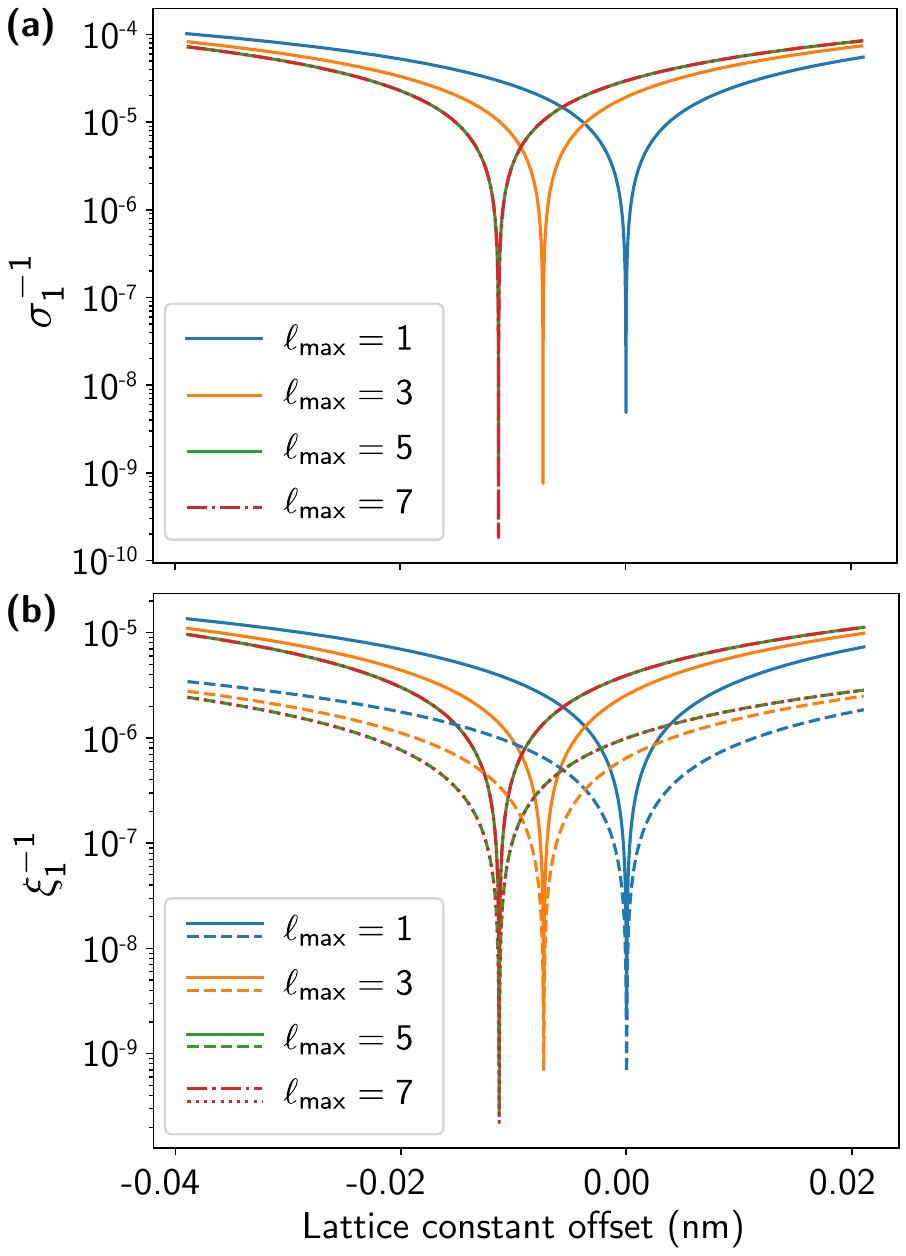}
    \caption{Inverse of the largest singular value of (a) the effective \textit{T}-matrix and (b) the \textit{S}-matrix of the metasurface considered in Sec.~\ref{sec:infinite}, for different values of the maximum multipole degree $\ell_{\rm max}$. The lattice constant offset is defined relative to the value $1037.779$ nm obtained for $\ell_{\rm max} = 1$. In panel (b), solid and dashed lines for $\ell_{\rm max}=1,3,$ and 5 correspond to the plane-wave basis with $b_{\rm max} L/(2\pi) = 1$ and $b_{\rm max} L/(2\pi) = 2$, respectively. For $\ell_{\rm max} = 7$, these lines become dashed-dotted and dotted, respectively.}
    \label{fig:10}
\end{figure}

To analyze the convergence of the results presented in Fig.~\ref{fig:2} with respect to $\ell_{\rm max}$, we plot $\sigma_1^{-1}$ at the wavelength $\lambda_{\rm R} = 1459.16$ nm as a function of the lattice constant offset $\delta L$ in Fig.~\ref{fig:10}(a) for different values of $\ell_{\rm max}$ (with $\delta L = 0$ for $\ell_{\rm max} = 1$). A BIC resonance is observed for each value of $\ell_{\rm max}$; its position shifts to $\delta L < 0$ for $\ell_{\rm max} > 1$ and converges to $\delta L \approx -0.01$ nm at $\ell_{\rm max} = 5$. Further increasing $\ell_{\rm max}$ only reduces $\sigma_1^{-1}$, thereby confirming convergence.

We recall that the \textit{S}-matrix of the metasurface is computed from the effective \textit{T}-matrix. Therefore, for a given $\ell_{\rm max}$, $\xi_1^{-1}$ vanishes at the same lattice constant as $\sigma_1^{-1}$, independently of the number of diffraction orders, which is controlled by $b_{\rm max}$ [see Fig.~\ref{fig:10}(b)]. However, increasing $b_{\rm max}$ reduces $\xi_1^{-1}$ for each value of $\ell_{\rm max}$ [cf. the inset of Fig.~\ref{fig:2}(c)].

\bibliography{main_revised}

\begin{thebibliography}{93}%
\makeatletter
\providecommand \@ifxundefined [1]{%
 \@ifx{#1\undefined}
}%
\providecommand \@ifnum [1]{%
 \ifnum #1\expandafter \@firstoftwo
 \else \expandafter \@secondoftwo
 \fi
}%
\providecommand \@ifx [1]{%
 \ifx #1\expandafter \@firstoftwo
 \else \expandafter \@secondoftwo
 \fi
}%
\providecommand \natexlab [1]{#1}%
\providecommand \enquote  [1]{``#1''}%
\providecommand \bibnamefont  [1]{#1}%
\providecommand \bibfnamefont [1]{#1}%
\providecommand \citenamefont [1]{#1}%
\providecommand \href@noop [0]{\@secondoftwo}%
\providecommand \href [0]{\begingroup \@sanitize@url \@href}%
\providecommand \@href[1]{\@@startlink{#1}\@@href}%
\providecommand \@@href[1]{\endgroup#1\@@endlink}%
\providecommand \@sanitize@url [0]{\catcode `\\12\catcode `\$12\catcode `\&12\catcode `\#12\catcode `\^12\catcode `\_12\catcode `\%12\relax}%
\providecommand \@@startlink[1]{}%
\providecommand \@@endlink[0]{}%
\providecommand \url  [0]{\begingroup\@sanitize@url \@url }%
\providecommand \@url [1]{\endgroup\@href {#1}{\urlprefix }}%
\providecommand \urlprefix  [0]{URL }%
\providecommand \Eprint [0]{\href }%
\providecommand \doibase [0]{https://doi.org/}%
\providecommand \selectlanguage [0]{\@gobble}%
\providecommand \bibinfo  [0]{\@secondoftwo}%
\providecommand \bibfield  [0]{\@secondoftwo}%
\providecommand \translation [1]{[#1]}%
\providecommand \BibitemOpen [0]{}%
\providecommand \bibitemStop [0]{}%
\providecommand \bibitemNoStop [0]{.\EOS\space}%
\providecommand \EOS [0]{\spacefactor3000\relax}%
\providecommand \BibitemShut  [1]{\csname bibitem#1\endcsname}%
\let\auto@bib@innerbib\@empty
\bibitem [{\citenamefont {Koshelev}\ \emph {et~al.}(2023)\citenamefont {Koshelev}, \citenamefont {Sadrieva}, \citenamefont {Shcherbakov}, \citenamefont {Kivshar},\ and\ \citenamefont {Bogdanov}}]{Koshelev2023May}%
  \BibitemOpen
  \bibfield  {author} {\bibinfo {author} {\bibfnamefont {K.~L.}\ \bibnamefont {Koshelev}}, \bibinfo {author} {\bibfnamefont {Z.~F.}\ \bibnamefont {Sadrieva}}, \bibinfo {author} {\bibfnamefont {A.~A.}\ \bibnamefont {Shcherbakov}}, \bibinfo {author} {\bibfnamefont {{\relax Yu}.~S.}\ \bibnamefont {Kivshar}},\ and\ \bibinfo {author} {\bibfnamefont {A.~A.}\ \bibnamefont {Bogdanov}},\ }\bibfield  {title} {\bibinfo {title} {{Bound states in the continuum in photonic structures}},\ }\href {https://ufn.ru/en/articles/2023/5/c} {\bibfield  {journal} {\bibinfo  {journal} {Phys. Usp.}\ }\textbf {\bibinfo {volume} {66}},\ \bibinfo {pages} {494} (\bibinfo {year} {2023})}\BibitemShut {NoStop}%
\bibitem [{\citenamefont {Evans}\ \emph {et~al.}(1994)\citenamefont {Evans}, \citenamefont {Levitin},\ and\ \citenamefont {Vassiliev}}]{Evans1994Feb}%
  \BibitemOpen
  \bibfield  {author} {\bibinfo {author} {\bibfnamefont {D.~V.}\ \bibnamefont {Evans}}, \bibinfo {author} {\bibfnamefont {M.}~\bibnamefont {Levitin}},\ and\ \bibinfo {author} {\bibfnamefont {D.}~\bibnamefont {Vassiliev}},\ }\bibfield  {title} {\bibinfo {title} {{Existence theorems for trapped modes}},\ }\href {https://doi.org/10.1017/S0022112094000236} {\bibfield  {journal} {\bibinfo  {journal} {J. Fluid Mech.}\ }\textbf {\bibinfo {volume} {261}},\ \bibinfo {pages} {21} (\bibinfo {year} {1994})}\BibitemShut {NoStop}%
\bibitem [{\citenamefont {Prokhorov}\ \emph {et~al.}(2022)\citenamefont {Prokhorov}, \citenamefont {Terekhov}, \citenamefont {Gubin}, \citenamefont {Shesterikov}, \citenamefont {Ni}, \citenamefont {Tuz},\ and\ \citenamefont {Evlyukhin}}]{Prokhorov2022Dec}%
  \BibitemOpen
  \bibfield  {author} {\bibinfo {author} {\bibfnamefont {A.~V.}\ \bibnamefont {Prokhorov}}, \bibinfo {author} {\bibfnamefont {P.~D.}\ \bibnamefont {Terekhov}}, \bibinfo {author} {\bibfnamefont {M.~{\relax Yu}.}\ \bibnamefont {Gubin}}, \bibinfo {author} {\bibfnamefont {A.~V.}\ \bibnamefont {Shesterikov}}, \bibinfo {author} {\bibfnamefont {X.}~\bibnamefont {Ni}}, \bibinfo {author} {\bibfnamefont {V.~R.}\ \bibnamefont {Tuz}},\ and\ \bibinfo {author} {\bibfnamefont {A.~B.}\ \bibnamefont {Evlyukhin}},\ }\bibfield  {title} {\bibinfo {title} {{Resonant Light Trapping via Lattice-Induced Multipole Coupling in Symmetrical Metasurfaces}},\ }\href {https://doi.org/10.1021/acsphotonics.2c01066} {\bibfield  {journal} {\bibinfo  {journal} {ACS Photonics}\ }\textbf {\bibinfo {volume} {9}},\ \bibinfo {pages} {3869} (\bibinfo {year} {2022})}\BibitemShut {NoStop}%
\bibitem [{\citenamefont {Hsu}\ \emph {et~al.}(2013)\citenamefont {Hsu}, \citenamefont {Zhen}, \citenamefont {Lee}, \citenamefont {Chua}, \citenamefont {Johnson}, \citenamefont {Joannopoulos},\ and\ \citenamefont {Solja{\ifmmode\check{c}\else\v{c}\fi}i{\ifmmode\acute{c}\else\'{c}\fi}}}]{Hsu2013Jul}%
  \BibitemOpen
  \bibfield  {author} {\bibinfo {author} {\bibfnamefont {C.~W.}\ \bibnamefont {Hsu}}, \bibinfo {author} {\bibfnamefont {B.}~\bibnamefont {Zhen}}, \bibinfo {author} {\bibfnamefont {J.}~\bibnamefont {Lee}}, \bibinfo {author} {\bibfnamefont {S.-L.}\ \bibnamefont {Chua}}, \bibinfo {author} {\bibfnamefont {S.~G.}\ \bibnamefont {Johnson}}, \bibinfo {author} {\bibfnamefont {J.~D.}\ \bibnamefont {Joannopoulos}},\ and\ \bibinfo {author} {\bibfnamefont {M.}~\bibnamefont {Solja{\ifmmode\check{c}\else\v{c}\fi}i{\ifmmode\acute{c}\else\'{c}\fi}}},\ }\bibfield  {title} {\bibinfo {title} {{Observation of trapped light within the radiation continuum}},\ }\href {https://doi.org/10.1038/nature12289} {\bibfield  {journal} {\bibinfo  {journal} {Nature}\ }\textbf {\bibinfo {volume} {499}},\ \bibinfo {pages} {188} (\bibinfo {year} {2013})}\BibitemShut {NoStop}%
\bibitem [{\citenamefont {Hsu}\ \emph {et~al.}(2016)\citenamefont {Hsu}, \citenamefont {Zhen}, \citenamefont {Stone}, \citenamefont {Joannopoulos},\ and\ \citenamefont {Solja{\ifmmode\check{c}\else\v{c}\fi}i{\ifmmode\acute{c}\else\'{c}\fi}}}]{Hsu2016Jul}%
  \BibitemOpen
  \bibfield  {author} {\bibinfo {author} {\bibfnamefont {C.~W.}\ \bibnamefont {Hsu}}, \bibinfo {author} {\bibfnamefont {B.}~\bibnamefont {Zhen}}, \bibinfo {author} {\bibfnamefont {A.~D.}\ \bibnamefont {Stone}}, \bibinfo {author} {\bibfnamefont {J.~D.}\ \bibnamefont {Joannopoulos}},\ and\ \bibinfo {author} {\bibfnamefont {M.}~\bibnamefont {Solja{\ifmmode\check{c}\else\v{c}\fi}i{\ifmmode\acute{c}\else\'{c}\fi}}},\ }\bibfield  {title} {\bibinfo {title} {{Bound states in the continuum}},\ }\href {https://doi.org/10.1038/natrevmats.2016.48} {\bibfield  {journal} {\bibinfo  {journal} {Nat. Rev. Mater.}\ }\textbf {\bibinfo {volume} {1}},\ \bibinfo {pages} {16048} (\bibinfo {year} {2016})}\BibitemShut {NoStop}%
\bibitem [{\citenamefont {He}\ \emph {et~al.}(2025)\citenamefont {He}, \citenamefont {Sun}, \citenamefont {Abraham-Ekeroth}, \citenamefont {Jin}, \citenamefont {Xiang},\ and\ \citenamefont {Torrent}}]{He2025Jul}%
  \BibitemOpen
  \bibfield  {author} {\bibinfo {author} {\bibfnamefont {L.}~\bibnamefont {He}}, \bibinfo {author} {\bibfnamefont {S.}~\bibnamefont {Sun}}, \bibinfo {author} {\bibfnamefont {R.~M.}\ \bibnamefont {Abraham-Ekeroth}}, \bibinfo {author} {\bibfnamefont {Y.}~\bibnamefont {Jin}}, \bibinfo {author} {\bibfnamefont {Y.}~\bibnamefont {Xiang}},\ and\ \bibinfo {author} {\bibfnamefont {D.}~\bibnamefont {Torrent}},\ }\bibfield  {title} {\bibinfo {title} {{Theoretical developments and experimental insights of acoustic and elastic bound states in the continuum}},\ }\href {https://doi.org/10.1038/s44384-025-00013-w} {\bibfield  {journal} {\bibinfo  {journal} {npj Acoust.}\ }\textbf {\bibinfo {volume} {1}},\ \bibinfo {pages} {11} (\bibinfo {year} {2025})}\BibitemShut {NoStop}%
\bibitem [{\citenamefont {Bulgakov}\ and\ \citenamefont {Sadreev}(2011)}]{Bulgakov2011Jun}%
  \BibitemOpen
  \bibfield  {author} {\bibinfo {author} {\bibfnamefont {E.}~\bibnamefont {Bulgakov}}\ and\ \bibinfo {author} {\bibfnamefont {A.}~\bibnamefont {Sadreev}},\ }\bibfield  {title} {\bibinfo {title} {{Formation of bound states in the continuum for a quantum dot with variable width}},\ }\href {https://doi.org/10.1103/PhysRevB.83.235321} {\bibfield  {journal} {\bibinfo  {journal} {Phys. Rev. B}\ }\textbf {\bibinfo {volume} {83}},\ \bibinfo {pages} {235321} (\bibinfo {year} {2011})}\BibitemShut {NoStop}%
\bibitem [{\citenamefont {Taghizadeh}\ and\ \citenamefont {Chung}(2017)}]{Taghizadeh2017Jul}%
  \BibitemOpen
  \bibfield  {author} {\bibinfo {author} {\bibfnamefont {A.}~\bibnamefont {Taghizadeh}}\ and\ \bibinfo {author} {\bibfnamefont {I.-S.}\ \bibnamefont {Chung}},\ }\bibfield  {title} {\bibinfo {title} {{Quasi bound states in the continuum with few unit cells of photonic crystal slab}},\ }\href {https://doi.org/10.1063/1.4990753} {\bibfield  {journal} {\bibinfo  {journal} {Appl. Phys. Lett.}\ }\textbf {\bibinfo {volume} {111}},\ \bibinfo {pages} {031114} (\bibinfo {year} {2017})}\BibitemShut {NoStop}%
\bibitem [{\citenamefont {Sadrieva}\ \emph {et~al.}(2019{\natexlab{a}})\citenamefont {Sadrieva}, \citenamefont {Belyakov}, \citenamefont {Balezin}, \citenamefont {Kapitanova}, \citenamefont {Nenasheva}, \citenamefont {Sadreev},\ and\ \citenamefont {Bogdanov}}]{Sadrieva2019May}%
  \BibitemOpen
  \bibfield  {author} {\bibinfo {author} {\bibfnamefont {Z.~F.}\ \bibnamefont {Sadrieva}}, \bibinfo {author} {\bibfnamefont {M.~A.}\ \bibnamefont {Belyakov}}, \bibinfo {author} {\bibfnamefont {M.~A.}\ \bibnamefont {Balezin}}, \bibinfo {author} {\bibfnamefont {P.~V.}\ \bibnamefont {Kapitanova}}, \bibinfo {author} {\bibfnamefont {E.~A.}\ \bibnamefont {Nenasheva}}, \bibinfo {author} {\bibfnamefont {A.~F.}\ \bibnamefont {Sadreev}},\ and\ \bibinfo {author} {\bibfnamefont {A.~A.}\ \bibnamefont {Bogdanov}},\ }\bibfield  {title} {\bibinfo {title} {{Experimental observation of a symmetry-protected bound state in the continuum in a chain of dielectric disks}},\ }\href {https://doi.org/10.1103/PhysRevA.99.053804} {\bibfield  {journal} {\bibinfo  {journal} {Phys. Rev. A}\ }\textbf {\bibinfo {volume} {99}},\ \bibinfo {pages} {053804} (\bibinfo {year} {2019}{\natexlab{a}})}\BibitemShut {NoStop}%
\bibitem [{\citenamefont {Maksimov}\ \emph {et~al.}(2020{\natexlab{a}})\citenamefont {Maksimov}, \citenamefont {Romano},\ and\ \citenamefont {Polyutov}}]{Maksimov2020Dec}%
  \BibitemOpen
  \bibfield  {author} {\bibinfo {author} {\bibfnamefont {V.~S.}\ \bibnamefont {Maksimov}, \bibfnamefont {Dmitrii N.~Gerasimov}}, \bibinfo {author} {\bibfnamefont {S.}~\bibnamefont {Romano}},\ and\ \bibinfo {author} {\bibfnamefont {S.~P.}\ \bibnamefont {Polyutov}},\ }\bibfield  {title} {\bibinfo {title} {{Refractive index sensing with optical bound states in the continuum}},\ }\href {https://doi.org/10.1364/OE.411749} {\bibfield  {journal} {\bibinfo  {journal} {Opt. Express}\ }\textbf {\bibinfo {volume} {28}},\ \bibinfo {pages} {38907} (\bibinfo {year} {2020}{\natexlab{a}})}\BibitemShut {NoStop}%
\bibitem [{\citenamefont {Joseph}\ \emph {et~al.}(2021{\natexlab{a}})\citenamefont {Joseph}, \citenamefont {Sarkar}, \citenamefont {Khan},\ and\ \citenamefont {Joseph}}]{Joseph2021Apr}%
  \BibitemOpen
  \bibfield  {author} {\bibinfo {author} {\bibfnamefont {S.}~\bibnamefont {Joseph}}, \bibinfo {author} {\bibfnamefont {S.}~\bibnamefont {Sarkar}}, \bibinfo {author} {\bibfnamefont {S.}~\bibnamefont {Khan}},\ and\ \bibinfo {author} {\bibfnamefont {J.}~\bibnamefont {Joseph}},\ }\bibfield  {title} {\bibinfo {title} {{Exploring the Optical Bound State in the Continuum in a Dielectric Grating Coupled Plasmonic Hybrid System}},\ }\href {https://doi.org/10.1002/adom.202001895} {\bibfield  {journal} {\bibinfo  {journal} {Adv. Opt. Mater.}\ }\textbf {\bibinfo {volume} {9}},\ \bibinfo {pages} {2001895} (\bibinfo {year} {2021}{\natexlab{a}})}\BibitemShut {NoStop}%
\bibitem [{\citenamefont {Fedotov}\ \emph {et~al.}(2007)\citenamefont {Fedotov}, \citenamefont {Rose}, \citenamefont {Prosvirnin}, \citenamefont {Papasimakis},\ and\ \citenamefont {Zheludev}}]{Fedotov2007Oct}%
  \BibitemOpen
  \bibfield  {author} {\bibinfo {author} {\bibfnamefont {V.~A.}\ \bibnamefont {Fedotov}}, \bibinfo {author} {\bibfnamefont {M.}~\bibnamefont {Rose}}, \bibinfo {author} {\bibfnamefont {S.~L.}\ \bibnamefont {Prosvirnin}}, \bibinfo {author} {\bibfnamefont {N.}~\bibnamefont {Papasimakis}},\ and\ \bibinfo {author} {\bibfnamefont {N.~I.}\ \bibnamefont {Zheludev}},\ }\bibfield  {title} {\bibinfo {title} {{Sharp Trapped-Mode Resonances in Planar Metamaterials with a Broken Structural Symmetry}},\ }\href {https://doi.org/10.1103/PhysRevLett.99.147401} {\bibfield  {journal} {\bibinfo  {journal} {Phys. Rev. Lett.}\ }\textbf {\bibinfo {volume} {99}},\ \bibinfo {pages} {147401} (\bibinfo {year} {2007})}\BibitemShut {NoStop}%
\bibitem [{\citenamefont {Koshelev}\ \emph {et~al.}(2018)\citenamefont {Koshelev}, \citenamefont {Lepeshov}, \citenamefont {Liu}, \citenamefont {Bogdanov},\ and\ \citenamefont {Kivshar}}]{Koshelev2018Nov}%
  \BibitemOpen
  \bibfield  {author} {\bibinfo {author} {\bibfnamefont {K.}~\bibnamefont {Koshelev}}, \bibinfo {author} {\bibfnamefont {S.}~\bibnamefont {Lepeshov}}, \bibinfo {author} {\bibfnamefont {M.}~\bibnamefont {Liu}}, \bibinfo {author} {\bibfnamefont {A.}~\bibnamefont {Bogdanov}},\ and\ \bibinfo {author} {\bibfnamefont {Y.}~\bibnamefont {Kivshar}},\ }\bibfield  {title} {\bibinfo {title} {{Asymmetric Metasurfaces with High-$Q$ Resonances Governed by Bound States in the Continuum}},\ }\href {https://doi.org/10.1103/PhysRevLett.121.193903} {\bibfield  {journal} {\bibinfo  {journal} {Phys. Rev. Lett.}\ }\textbf {\bibinfo {volume} {121}},\ \bibinfo {pages} {193903} (\bibinfo {year} {2018})}\BibitemShut {NoStop}%
\bibitem [{\citenamefont {Allayarov}\ \emph {et~al.}(2024)\citenamefont {Allayarov}, \citenamefont {Lesina},\ and\ \citenamefont {Evlyukhin}}]{Allayarov2024Jun}%
  \BibitemOpen
  \bibfield  {author} {\bibinfo {author} {\bibfnamefont {I.}~\bibnamefont {Allayarov}}, \bibinfo {author} {\bibfnamefont {A.~C.}\ \bibnamefont {Lesina}},\ and\ \bibinfo {author} {\bibfnamefont {A.~B.}\ \bibnamefont {Evlyukhin}},\ }\bibfield  {title} {\bibinfo {title} {{Anapole mechanism of bound states in the continuum in symmetric dielectric metasurfaces}},\ }\href {https://doi.org/10.1103/PhysRevB.109.L241405} {\bibfield  {journal} {\bibinfo  {journal} {Phys. Rev. B}\ }\textbf {\bibinfo {volume} {109}},\ \bibinfo {pages} {L241405} (\bibinfo {year} {2024})}\BibitemShut {NoStop}%
\bibitem [{\citenamefont {Joseph}\ \emph {et~al.}(2021{\natexlab{b}})\citenamefont {Joseph}, \citenamefont {Pandey}, \citenamefont {Sarkar},\ and\ \citenamefont {Joseph}}]{Joseph2021Dec}%
  \BibitemOpen
  \bibfield  {author} {\bibinfo {author} {\bibfnamefont {S.}~\bibnamefont {Joseph}}, \bibinfo {author} {\bibfnamefont {S.}~\bibnamefont {Pandey}}, \bibinfo {author} {\bibfnamefont {S.}~\bibnamefont {Sarkar}},\ and\ \bibinfo {author} {\bibfnamefont {J.}~\bibnamefont {Joseph}},\ }\bibfield  {title} {\bibinfo {title} {{Bound states in the continuum in resonant nanostructures: an overview of engineered materials for tailored applications}},\ }\href {https://doi.org/10.1515/nanoph-2021-0387} {\bibfield  {journal} {\bibinfo  {journal} {Nanophotonics}\ }\textbf {\bibinfo {volume} {10}},\ \bibinfo {pages} {4175} (\bibinfo {year} {2021}{\natexlab{b}})}\BibitemShut {NoStop}%
\bibitem [{\citenamefont {Bulgakov}\ and\ \citenamefont {Maksimov}(2017)}]{Bulgakov2017Jun}%
  \BibitemOpen
  \bibfield  {author} {\bibinfo {author} {\bibfnamefont {E.~N.}\ \bibnamefont {Bulgakov}}\ and\ \bibinfo {author} {\bibfnamefont {D.~N.}\ \bibnamefont {Maksimov}},\ }\bibfield  {title} {\bibinfo {title} {{Light enhancement by quasi-bound states in the continuum in dielectric arrays}},\ }\href {https://doi.org/10.1364/OE.25.014134} {\bibfield  {journal} {\bibinfo  {journal} {Opt. Express}\ }\textbf {\bibinfo {volume} {25}},\ \bibinfo {pages} {14134} (\bibinfo {year} {2017})}\BibitemShut {NoStop}%
\bibitem [{\citenamefont {Zograf}\ \emph {et~al.}(2022)\citenamefont {Zograf}, \citenamefont {Koshelev}, \citenamefont {Zalogina}, \citenamefont {Korolev}, \citenamefont {Hollinger}, \citenamefont {Choi}, \citenamefont {Zuerch}, \citenamefont {Spielmann}, \citenamefont {Luther-Davies}, \citenamefont {Kartashov}, \citenamefont {Makarov}, \citenamefont {Kruk},\ and\ \citenamefont {Kivshar}}]{Zograf2022Feb}%
  \BibitemOpen
  \bibfield  {author} {\bibinfo {author} {\bibfnamefont {G.}~\bibnamefont {Zograf}}, \bibinfo {author} {\bibfnamefont {K.}~\bibnamefont {Koshelev}}, \bibinfo {author} {\bibfnamefont {A.}~\bibnamefont {Zalogina}}, \bibinfo {author} {\bibfnamefont {V.}~\bibnamefont {Korolev}}, \bibinfo {author} {\bibfnamefont {R.}~\bibnamefont {Hollinger}}, \bibinfo {author} {\bibfnamefont {D.-Y.}\ \bibnamefont {Choi}}, \bibinfo {author} {\bibfnamefont {M.}~\bibnamefont {Zuerch}}, \bibinfo {author} {\bibfnamefont {C.}~\bibnamefont {Spielmann}}, \bibinfo {author} {\bibfnamefont {B.}~\bibnamefont {Luther-Davies}}, \bibinfo {author} {\bibfnamefont {D.}~\bibnamefont {Kartashov}}, \bibinfo {author} {\bibfnamefont {S.~V.}\ \bibnamefont {Makarov}}, \bibinfo {author} {\bibfnamefont {S.~S.}\ \bibnamefont {Kruk}},\ and\ \bibinfo {author} {\bibfnamefont {Y.}~\bibnamefont {Kivshar}},\ }\bibfield  {title} {\bibinfo {title} {{High-Harmonic Generation from Resonant Dielectric Metasurfaces Empowered by Bound States in the Continuum}},\ }\href
  {https://doi.org/10.1021/acsphotonics.1c01511} {\bibfield  {journal} {\bibinfo  {journal} {ACS Photonics}\ }\textbf {\bibinfo {volume} {9}},\ \bibinfo {pages} {567} (\bibinfo {year} {2022})}\BibitemShut {NoStop}%
\bibitem [{\citenamefont {Kodigala}\ \emph {et~al.}(2017)\citenamefont {Kodigala}, \citenamefont {Lepetit}, \citenamefont {Gu}, \citenamefont {Bahari}, \citenamefont {Fainman},\ and\ \citenamefont {Kant{\ifmmode\acute{e}\else\'{e}\fi}}}]{Kodigala2017Jan}%
  \BibitemOpen
  \bibfield  {author} {\bibinfo {author} {\bibfnamefont {A.}~\bibnamefont {Kodigala}}, \bibinfo {author} {\bibfnamefont {T.}~\bibnamefont {Lepetit}}, \bibinfo {author} {\bibfnamefont {Q.}~\bibnamefont {Gu}}, \bibinfo {author} {\bibfnamefont {B.}~\bibnamefont {Bahari}}, \bibinfo {author} {\bibfnamefont {Y.}~\bibnamefont {Fainman}},\ and\ \bibinfo {author} {\bibfnamefont {B.}~\bibnamefont {Kant{\ifmmode\acute{e}\else\'{e}\fi}}},\ }\bibfield  {title} {\bibinfo {title} {{Lasing action from photonic bound states in continuum}},\ }\href {https://doi.org/10.1038/nature20799} {\bibfield  {journal} {\bibinfo  {journal} {Nature}\ }\textbf {\bibinfo {volume} {541}},\ \bibinfo {pages} {196} (\bibinfo {year} {2017})}\BibitemShut {NoStop}%
\bibitem [{\citenamefont {Lalanne}\ \emph {et~al.}(2018)\citenamefont {Lalanne}, \citenamefont {Yan}, \citenamefont {Vynck}, \citenamefont {Sauvan},\ and\ \citenamefont {Hugonin}}]{Lalanne2018May}%
  \BibitemOpen
  \bibfield  {author} {\bibinfo {author} {\bibfnamefont {P.}~\bibnamefont {Lalanne}}, \bibinfo {author} {\bibfnamefont {W.}~\bibnamefont {Yan}}, \bibinfo {author} {\bibfnamefont {K.}~\bibnamefont {Vynck}}, \bibinfo {author} {\bibfnamefont {C.}~\bibnamefont {Sauvan}},\ and\ \bibinfo {author} {\bibfnamefont {J.-P.}\ \bibnamefont {Hugonin}},\ }\bibfield  {title} {\bibinfo {title} {{Light Interaction with Photonic and Plasmonic Resonances}},\ }\href {https://doi.org/10.1002/lpor.201700113} {\bibfield  {journal} {\bibinfo  {journal} {Laser Photonics Rev.}\ }\textbf {\bibinfo {volume} {12}},\ \bibinfo {pages} {1700113} (\bibinfo {year} {2018})}\BibitemShut {NoStop}%
\bibitem [{\citenamefont {Sauvan}\ \emph {et~al.}(2013)\citenamefont {Sauvan}, \citenamefont {Hugonin}, \citenamefont {Maksymov},\ and\ \citenamefont {Lalanne}}]{Sauvan2013Jun}%
  \BibitemOpen
  \bibfield  {author} {\bibinfo {author} {\bibfnamefont {C.}~\bibnamefont {Sauvan}}, \bibinfo {author} {\bibfnamefont {J.~P.}\ \bibnamefont {Hugonin}}, \bibinfo {author} {\bibfnamefont {I.~S.}\ \bibnamefont {Maksymov}},\ and\ \bibinfo {author} {\bibfnamefont {P.}~\bibnamefont {Lalanne}},\ }\bibfield  {title} {\bibinfo {title} {{Theory of the Spontaneous Optical Emission of Nanosize Photonic and Plasmon Resonators}},\ }\href {https://doi.org/10.1103/PhysRevLett.110.237401} {\bibfield  {journal} {\bibinfo  {journal} {Phys. Rev. Lett.}\ }\textbf {\bibinfo {volume} {110}},\ \bibinfo {pages} {237401} (\bibinfo {year} {2013})}\BibitemShut {NoStop}%
\bibitem [{\citenamefont {Doost}\ \emph {et~al.}(2014)\citenamefont {Doost}, \citenamefont {Langbein},\ and\ \citenamefont {Muljarov}}]{Doost2014Jul}%
  \BibitemOpen
  \bibfield  {author} {\bibinfo {author} {\bibfnamefont {M.~B.}\ \bibnamefont {Doost}}, \bibinfo {author} {\bibfnamefont {W.}~\bibnamefont {Langbein}},\ and\ \bibinfo {author} {\bibfnamefont {E.~A.}\ \bibnamefont {Muljarov}},\ }\bibfield  {title} {\bibinfo {title} {{Resonant-state expansion applied to three-dimensional open optical systems}},\ }\href {https://doi.org/10.1103/PhysRevA.90.013834} {\bibfield  {journal} {\bibinfo  {journal} {Phys. Rev. A}\ }\textbf {\bibinfo {volume} {90}},\ \bibinfo {pages} {013834} (\bibinfo {year} {2014})}\BibitemShut {NoStop}%
\bibitem [{\citenamefont {Muljarov}\ and\ \citenamefont {Langbein}(2016)}]{Muljarov2016Feb}%
  \BibitemOpen
  \bibfield  {author} {\bibinfo {author} {\bibfnamefont {E.~A.}\ \bibnamefont {Muljarov}}\ and\ \bibinfo {author} {\bibfnamefont {W.}~\bibnamefont {Langbein}},\ }\bibfield  {title} {\bibinfo {title} {{Resonant-state expansion of dispersive open optical systems: Creating gold from sand}},\ }\href {https://doi.org/10.1103/PhysRevB.93.075417} {\bibfield  {journal} {\bibinfo  {journal} {Phys. Rev. B}\ }\textbf {\bibinfo {volume} {93}},\ \bibinfo {pages} {075417} (\bibinfo {year} {2016})}\BibitemShut {NoStop}%
\bibitem [{\citenamefont {Both}\ and\ \citenamefont {Weiss}(2021)}]{Both2021Dec}%
  \BibitemOpen
  \bibfield  {author} {\bibinfo {author} {\bibfnamefont {S.}~\bibnamefont {Both}}\ and\ \bibinfo {author} {\bibfnamefont {T.}~\bibnamefont {Weiss}},\ }\bibfield  {title} {\bibinfo {title} {{Resonant states and their role in nanophotonics}},\ }\href {https://doi.org/10.1088/1361-6641/ac3290} {\bibfield  {journal} {\bibinfo  {journal} {Semicond. Sci. Technol.}\ }\textbf {\bibinfo {volume} {37}},\ \bibinfo {pages} {013002} (\bibinfo {year} {2021})}\BibitemShut {NoStop}%
\bibitem [{\citenamefont {Laude}\ and\ \citenamefont {Wang}(2023)}]{Laude2023Apr}%
  \BibitemOpen
  \bibfield  {author} {\bibinfo {author} {\bibfnamefont {V.}~\bibnamefont {Laude}}\ and\ \bibinfo {author} {\bibfnamefont {Y.-F.}\ \bibnamefont {Wang}},\ }\bibfield  {title} {\bibinfo {title} {{Quasinormal mode representation of radiating resonators in open phononic systems}},\ }\href {https://doi.org/10.1103/PhysRevB.107.144301} {\bibfield  {journal} {\bibinfo  {journal} {Phys. Rev. B}\ }\textbf {\bibinfo {volume} {107}},\ \bibinfo {pages} {144301} (\bibinfo {year} {2023})}\BibitemShut {NoStop}%
\bibitem [{\citenamefont {Neale}\ and\ \citenamefont {Muljarov}(2021)}]{Neale2021Apr}%
  \BibitemOpen
  \bibfield  {author} {\bibinfo {author} {\bibfnamefont {S.}~\bibnamefont {Neale}}\ and\ \bibinfo {author} {\bibfnamefont {E.~A.}\ \bibnamefont {Muljarov}},\ }\bibfield  {title} {\bibinfo {title} {{Accidental and symmetry-protected bound states in the continuum in a photonic-crystal slab: A resonant-state expansion study}},\ }\href {https://doi.org/10.1103/PhysRevB.103.155112} {\bibfield  {journal} {\bibinfo  {journal} {Phys. Rev. B}\ }\textbf {\bibinfo {volume} {103}},\ \bibinfo {pages} {155112} (\bibinfo {year} {2021})}\BibitemShut {NoStop}%
\bibitem [{\citenamefont {Sadreev}\ \emph {et~al.}(2006)\citenamefont {Sadreev}, \citenamefont {Bulgakov},\ and\ \citenamefont {Rotter}}]{Sadreev2006Jun}%
  \BibitemOpen
  \bibfield  {author} {\bibinfo {author} {\bibfnamefont {A.~F.}\ \bibnamefont {Sadreev}}, \bibinfo {author} {\bibfnamefont {E.~N.}\ \bibnamefont {Bulgakov}},\ and\ \bibinfo {author} {\bibfnamefont {I.}~\bibnamefont {Rotter}},\ }\bibfield  {title} {\bibinfo {title} {{Bound states in the continuum in open quantum billiards with a variable shape}},\ }\href {https://doi.org/10.1103/PhysRevB.73.235342} {\bibfield  {journal} {\bibinfo  {journal} {Phys. Rev. B}\ }\textbf {\bibinfo {volume} {73}},\ \bibinfo {pages} {235342} (\bibinfo {year} {2006})}\BibitemShut {NoStop}%
\bibitem [{\citenamefont {Bulgakov}\ and\ \citenamefont {Sadreev}(2008)}]{Bulgakov2008Aug}%
  \BibitemOpen
  \bibfield  {author} {\bibinfo {author} {\bibfnamefont {E.~N.}\ \bibnamefont {Bulgakov}}\ and\ \bibinfo {author} {\bibfnamefont {A.~F.}\ \bibnamefont {Sadreev}},\ }\bibfield  {title} {\bibinfo {title} {{Bound states in the continuum in photonic waveguides inspired by defects}},\ }\href {https://doi.org/10.1103/PhysRevB.78.075105} {\bibfield  {journal} {\bibinfo  {journal} {Phys. Rev. B}\ }\textbf {\bibinfo {volume} {78}},\ \bibinfo {pages} {075105} (\bibinfo {year} {2008})}\BibitemShut {NoStop}%
\bibitem [{\citenamefont {Can{\ifmmode\acute{o}\else\'{o}\fi}s~Valero}\ \emph {et~al.}(2025)\citenamefont {Can{\ifmmode\acute{o}\else\'{o}\fi}s~Valero}, \citenamefont {Sztranyovszky}, \citenamefont {Muljarov}, \citenamefont {Bogdanov},\ and\ \citenamefont {Weiss}}]{CanosValero2025Mar}%
  \BibitemOpen
  \bibfield  {author} {\bibinfo {author} {\bibfnamefont {A.}~\bibnamefont {Can{\ifmmode\acute{o}\else\'{o}\fi}s~Valero}}, \bibinfo {author} {\bibfnamefont {Z.}~\bibnamefont {Sztranyovszky}}, \bibinfo {author} {\bibfnamefont {E.~A.}\ \bibnamefont {Muljarov}}, \bibinfo {author} {\bibfnamefont {A.}~\bibnamefont {Bogdanov}},\ and\ \bibinfo {author} {\bibfnamefont {T.}~\bibnamefont {Weiss}},\ }\bibfield  {title} {\bibinfo {title} {{Exceptional Bound States in the Continuum}},\ }\href {https://doi.org/10.1103/PhysRevLett.134.103802} {\bibfield  {journal} {\bibinfo  {journal} {Phys. Rev. Lett.}\ }\textbf {\bibinfo {volume} {134}},\ \bibinfo {pages} {103802} (\bibinfo {year} {2025})}\BibitemShut {NoStop}%
\bibitem [{\citenamefont {Maksimov}\ \emph {et~al.}(2020{\natexlab{b}})\citenamefont {Maksimov}, \citenamefont {Bogdanov},\ and\ \citenamefont {Bulgakov}}]{Maksimov2020Sep}%
  \BibitemOpen
  \bibfield  {author} {\bibinfo {author} {\bibfnamefont {D.~N.}\ \bibnamefont {Maksimov}}, \bibinfo {author} {\bibfnamefont {A.~A.}\ \bibnamefont {Bogdanov}},\ and\ \bibinfo {author} {\bibfnamefont {E.~N.}\ \bibnamefont {Bulgakov}},\ }\bibfield  {title} {\bibinfo {title} {{Optical bistability with bound states in the continuum in dielectric gratings}},\ }\href {https://doi.org/10.1103/PhysRevA.102.033511} {\bibfield  {journal} {\bibinfo  {journal} {Phys. Rev. A}\ }\textbf {\bibinfo {volume} {102}},\ \bibinfo {pages} {033511} (\bibinfo {year} {2020}{\natexlab{b}})}\BibitemShut {NoStop}%
\bibitem [{\citenamefont {Maksimov}\ \emph {et~al.}(2025)\citenamefont {Maksimov}, \citenamefont {Pankin}, \citenamefont {Kim}, \citenamefont {Song}, \citenamefont {Peng},\ and\ \citenamefont {Bogdanov}}]{Maksimov2025Dec}%
  \BibitemOpen
  \bibfield  {author} {\bibinfo {author} {\bibfnamefont {D.~N.}\ \bibnamefont {Maksimov}}, \bibinfo {author} {\bibfnamefont {P.~S.}\ \bibnamefont {Pankin}}, \bibinfo {author} {\bibfnamefont {D.-W.}\ \bibnamefont {Kim}}, \bibinfo {author} {\bibfnamefont {M.}~\bibnamefont {Song}}, \bibinfo {author} {\bibfnamefont {C.}~\bibnamefont {Peng}},\ and\ \bibinfo {author} {\bibfnamefont {A.~A.}\ \bibnamefont {Bogdanov}},\ }\bibfield  {title} {\bibinfo {title} {{Temporal coupled mode theory: From bound states in the continuum to uniguided resonances}},\ }\href {https://doi.org/10.1103/rq46-g378} {\bibfield  {journal} {\bibinfo  {journal} {Phys. Rev. B}\ }\textbf {\bibinfo {volume} {112}},\ \bibinfo {pages} {235303} (\bibinfo {year} {2025})}\BibitemShut {NoStop}%
\bibitem [{\citenamefont {M{\ifmmode\acute{a}\else\'{a}\fi}{\ifmmode\tilde{n}\else\~{n}\fi}ez-Espina}\ \emph {et~al.}(2025)\citenamefont {M{\ifmmode\acute{a}\else\'{a}\fi}{\ifmmode\tilde{n}\else\~{n}\fi}ez-Espina}, \citenamefont {Amrahi}, \citenamefont {Asadchy},\ and\ \citenamefont {D{\ifmmode\acute{\imath}\else\'{\i}\fi}az-Rubio}}]{Manez-Espina2025Dec}%
  \BibitemOpen
  \bibfield  {author} {\bibinfo {author} {\bibfnamefont {L.~M.}\ \bibnamefont {M{\ifmmode\acute{a}\else\'{a}\fi}{\ifmmode\tilde{n}\else\~{n}\fi}ez-Espina}}, \bibinfo {author} {\bibfnamefont {B.}~\bibnamefont {Amrahi}}, \bibinfo {author} {\bibfnamefont {V.}~\bibnamefont {Asadchy}},\ and\ \bibinfo {author} {\bibfnamefont {A.}~\bibnamefont {D{\ifmmode\acute{\imath}\else\'{\i}\fi}az-Rubio}},\ }\bibfield  {title} {\bibinfo {title} {{Strong Mode Coupling via Quasi-Bound States in the Continuum in Bianisotropic Metasurfaces}},\ }\bibfield  {journal} {\bibinfo  {journal} {arXiv}\ }\href {https://doi.org/10.48550/arXiv.2512.08927} {10.48550/arXiv.2512.08927} (\bibinfo {year} {2025}),\ \Eprint {https://arxiv.org/abs/2512.08927} {2512.08927} \BibitemShut {NoStop}%
\bibitem [{\citenamefont {Sadreev}(2021)}]{Sadreev2021Apr}%
  \BibitemOpen
  \bibfield  {author} {\bibinfo {author} {\bibfnamefont {A.~F.}\ \bibnamefont {Sadreev}},\ }\bibfield  {title} {\bibinfo {title} {{Interference traps waves in an open system: bound states in the continuum}},\ }\href {https://doi.org/10.1088/1361-6633/abefb9} {\bibfield  {journal} {\bibinfo  {journal} {Rep. Prog. Phys.}\ }\textbf {\bibinfo {volume} {84}},\ \bibinfo {pages} {055901} (\bibinfo {year} {2021})}\BibitemShut {NoStop}%
\bibitem [{\citenamefont {Maksimov}\ and\ \citenamefont {Bogdanov}(2025)}]{Maksimov2025Jul}%
  \BibitemOpen
  \bibfield  {author} {\bibinfo {author} {\bibfnamefont {D.~N.}\ \bibnamefont {Maksimov}}\ and\ \bibinfo {author} {\bibfnamefont {A.~A.}\ \bibnamefont {Bogdanov}},\ }\bibfield  {title} {\bibinfo {title} {{Quantum noise suppression in non-Hermitian resonators at an exceptional point}},\ }\href {https://doi.org/10.1103/kcm2-2mz4} {\bibfield  {journal} {\bibinfo  {journal} {Phys. Rev. B}\ }\textbf {\bibinfo {volume} {112}},\ \bibinfo {pages} {L041302} (\bibinfo {year} {2025})}\BibitemShut {NoStop}%
\bibitem [{\citenamefont {Wu}\ and\ \citenamefont {Lalanne}(2024)}]{Wu2024Jun}%
  \BibitemOpen
  \bibfield  {author} {\bibinfo {author} {\bibfnamefont {T.}~\bibnamefont {Wu}}\ and\ \bibinfo {author} {\bibfnamefont {P.}~\bibnamefont {Lalanne}},\ }\bibfield  {title} {\bibinfo {title} {{Exact Maxwell evolution equation of resonator dynamics: temporal coupled-mode theory revisited}},\ }\href {https://doi.org/10.1364/OE.517237} {\bibfield  {journal} {\bibinfo  {journal} {Opt. Express}\ }\textbf {\bibinfo {volume} {32}},\ \bibinfo {pages} {20904} (\bibinfo {year} {2024})}\BibitemShut {NoStop}%
\bibitem [{\citenamefont {Nevi{\ifmmode\grave{e}\else\`{e}\fi}re}\ \emph {et~al.}(1995)\citenamefont {Nevi{\ifmmode\grave{e}\else\`{e}\fi}re}, \citenamefont {Popov},\ and\ \citenamefont {Reinisch}}]{Neviere1995Mar}%
  \BibitemOpen
  \bibfield  {author} {\bibinfo {author} {\bibfnamefont {M.}~\bibnamefont {Nevi{\ifmmode\grave{e}\else\`{e}\fi}re}}, \bibinfo {author} {\bibfnamefont {E.}~\bibnamefont {Popov}},\ and\ \bibinfo {author} {\bibfnamefont {R.}~\bibnamefont {Reinisch}},\ }\bibfield  {title} {\bibinfo {title} {{Electromagnetic resonances in linear and nonlinear optics: phenomenological study of grating behavior through the poles and zeros of the scattering operator}},\ }\href {https://doi.org/10.1364/JOSAA.12.000513} {\bibfield  {journal} {\bibinfo  {journal} {J. Opt. Soc. Am. A}\ }\textbf {\bibinfo {volume} {12}},\ \bibinfo {pages} {513} (\bibinfo {year} {1995})}\BibitemShut {NoStop}%
\bibitem [{\citenamefont {Alpeggiani}\ \emph {et~al.}(2017)\citenamefont {Alpeggiani}, \citenamefont {Parappurath}, \citenamefont {Verhagen},\ and\ \citenamefont {Kuipers}}]{Alpeggiani2017Jun}%
  \BibitemOpen
  \bibfield  {author} {\bibinfo {author} {\bibfnamefont {F.}~\bibnamefont {Alpeggiani}}, \bibinfo {author} {\bibfnamefont {N.}~\bibnamefont {Parappurath}}, \bibinfo {author} {\bibfnamefont {E.}~\bibnamefont {Verhagen}},\ and\ \bibinfo {author} {\bibfnamefont {L.}~\bibnamefont {Kuipers}},\ }\bibfield  {title} {\bibinfo {title} {{Quasinormal-Mode Expansion of the Scattering Matrix}},\ }\href {https://doi.org/10.1103/PhysRevX.7.021035} {\bibfield  {journal} {\bibinfo  {journal} {Phys. Rev. X}\ }\textbf {\bibinfo {volume} {7}},\ \bibinfo {pages} {021035} (\bibinfo {year} {2017})}\BibitemShut {NoStop}%
\bibitem [{\citenamefont {Colom}\ \emph {et~al.}(2023)\citenamefont {Colom}, \citenamefont {Mikheeva}, \citenamefont {Achouri}, \citenamefont {Zuniga-Perez}, \citenamefont {Bonod}, \citenamefont {Martin}, \citenamefont {Burger},\ and\ \citenamefont {Genevet}}]{Colom2023Jun}%
  \BibitemOpen
  \bibfield  {author} {\bibinfo {author} {\bibfnamefont {R.}~\bibnamefont {Colom}}, \bibinfo {author} {\bibfnamefont {E.}~\bibnamefont {Mikheeva}}, \bibinfo {author} {\bibfnamefont {K.}~\bibnamefont {Achouri}}, \bibinfo {author} {\bibfnamefont {J.}~\bibnamefont {Zuniga-Perez}}, \bibinfo {author} {\bibfnamefont {N.}~\bibnamefont {Bonod}}, \bibinfo {author} {\bibfnamefont {O.~J.~F.}\ \bibnamefont {Martin}}, \bibinfo {author} {\bibfnamefont {S.}~\bibnamefont {Burger}},\ and\ \bibinfo {author} {\bibfnamefont {P.}~\bibnamefont {Genevet}},\ }\bibfield  {title} {\bibinfo {title} {{Crossing of the Branch Cut: The Topological Origin of a Universal 2{$\pi$}-Phase Retardation in Non-Hermitian Metasurfaces}},\ }\href {https://doi.org/10.1002/lpor.202200976} {\bibfield  {journal} {\bibinfo  {journal} {Laser Photonics Rev.}\ }\textbf {\bibinfo {volume} {17}},\ \bibinfo {pages} {2200976} (\bibinfo {year} {2023})}\BibitemShut {NoStop}%
\bibitem [{\citenamefont {Blanchard}\ \emph {et~al.}(2016)\citenamefont {Blanchard}, \citenamefont {Hugonin},\ and\ \citenamefont {Sauvan}}]{Blanchard2016Oct}%
  \BibitemOpen
  \bibfield  {author} {\bibinfo {author} {\bibfnamefont {C.}~\bibnamefont {Blanchard}}, \bibinfo {author} {\bibfnamefont {J.-P.}\ \bibnamefont {Hugonin}},\ and\ \bibinfo {author} {\bibfnamefont {C.}~\bibnamefont {Sauvan}},\ }\bibfield  {title} {\bibinfo {title} {{Fano resonances in photonic crystal slabs near optical bound states in the continuum}},\ }\href {https://doi.org/10.1103/PhysRevB.94.155303} {\bibfield  {journal} {\bibinfo  {journal} {Phys. Rev. B}\ }\textbf {\bibinfo {volume} {94}},\ \bibinfo {pages} {155303} (\bibinfo {year} {2016})}\BibitemShut {NoStop}%
\bibitem [{\citenamefont {Liu}\ \emph {et~al.}(2025{\natexlab{a}})\citenamefont {Liu}, \citenamefont {Zeng}, \citenamefont {Zhao}, \citenamefont {Yang}, \citenamefont {Zhang}, \citenamefont {Cui}, \citenamefont {Wu},\ and\ \citenamefont {Chan}}]{Liu2025Dec}%
  \BibitemOpen
  \bibfield  {author} {\bibinfo {author} {\bibfnamefont {W.}~\bibnamefont {Liu}}, \bibinfo {author} {\bibfnamefont {Y.-S.}\ \bibnamefont {Zeng}}, \bibinfo {author} {\bibfnamefont {J.}~\bibnamefont {Zhao}}, \bibinfo {author} {\bibfnamefont {C.}~\bibnamefont {Yang}}, \bibinfo {author} {\bibfnamefont {R.-Y.}\ \bibnamefont {Zhang}}, \bibinfo {author} {\bibfnamefont {X.}~\bibnamefont {Cui}}, \bibinfo {author} {\bibfnamefont {G.-B.}\ \bibnamefont {Wu}},\ and\ \bibinfo {author} {\bibfnamefont {C.~T.}\ \bibnamefont {Chan}},\ }\bibfield  {title} {\bibinfo {title} {{Bound States in the Continuum as Nodal Chain Points of Scattering Matrices}},\ }\href {https://doi.org/10.1103/rpb2-ryyk} {\bibfield  {journal} {\bibinfo  {journal} {Phys. Rev. Lett.}\ }\textbf {\bibinfo {volume} {135}},\ \bibinfo {pages} {243804} (\bibinfo {year} {2025}{\natexlab{a}})}\BibitemShut {NoStop}%
\bibitem [{\citenamefont {Horn}\ and\ \citenamefont {Johnson}(1985)}]{Horn1985Dec}%
  \BibitemOpen
  \bibfield  {author} {\bibinfo {author} {\bibfnamefont {R.~A.}\ \bibnamefont {Horn}}\ and\ \bibinfo {author} {\bibfnamefont {C.~R.}\ \bibnamefont {Johnson}},\ }\href {https://doi.org/10.1017/CBO9780511810817} {\emph {\bibinfo {title} {{Matrix Analysis}}}}\ (\bibinfo  {publisher} {Cambridge University Press},\ \bibinfo {address} {Cambridge, England, UK},\ \bibinfo {year} {1985})\BibitemShut {NoStop}%
\bibitem [{\citenamefont {Guo}\ and\ \citenamefont {Fan}(2022)}]{Guo2022Jun}%
  \BibitemOpen
  \bibfield  {author} {\bibinfo {author} {\bibfnamefont {C.}~\bibnamefont {Guo}}\ and\ \bibinfo {author} {\bibfnamefont {S.}~\bibnamefont {Fan}},\ }\bibfield  {title} {\bibinfo {title} {{Reciprocity Constraints on Reflection}},\ }\href {https://doi.org/10.1103/PhysRevLett.128.256101} {\bibfield  {journal} {\bibinfo  {journal} {Phys. Rev. Lett.}\ }\textbf {\bibinfo {volume} {128}},\ \bibinfo {pages} {256101} (\bibinfo {year} {2022})}\BibitemShut {NoStop}%
\bibitem [{\citenamefont {Zhang}\ \emph {et~al.}(2025)\citenamefont {Zhang}, \citenamefont {Delaporte},\ and\ \citenamefont {Ma}}]{Zhang2025Sep}%
  \BibitemOpen
  \bibfield  {author} {\bibinfo {author} {\bibfnamefont {H.}~\bibnamefont {Zhang}}, \bibinfo {author} {\bibfnamefont {T.}~\bibnamefont {Delaporte}},\ and\ \bibinfo {author} {\bibfnamefont {G.}~\bibnamefont {Ma}},\ }\bibfield  {title} {\bibinfo {title} {{Scattering matrices of two-dimensional complex acoustic media}},\ }\href {https://doi.org/10.1038/s44384-025-00020-x} {\bibfield  {journal} {\bibinfo  {journal} {npj Acoust.}\ }\textbf {\bibinfo {volume} {1}},\ \bibinfo {pages} {17} (\bibinfo {year} {2025})}\BibitemShut {NoStop}%
\bibitem [{\citenamefont {Miller}\ \emph {et~al.}(2017)\citenamefont {Miller}, \citenamefont {Zhu},\ and\ \citenamefont {Fan}}]{Miller2017Apr}%
  \BibitemOpen
  \bibfield  {author} {\bibinfo {author} {\bibfnamefont {D.~A.~B.}\ \bibnamefont {Miller}}, \bibinfo {author} {\bibfnamefont {L.}~\bibnamefont {Zhu}},\ and\ \bibinfo {author} {\bibfnamefont {S.}~\bibnamefont {Fan}},\ }\bibfield  {title} {\bibinfo {title} {{Universal modal radiation laws for all thermal emitters}},\ }\href {https://doi.org/10.1073/pnas.1701606114} {\bibfield  {journal} {\bibinfo  {journal} {Proc. Natl. Acad. Sci. U. S. A.}\ }\textbf {\bibinfo {volume} {114}},\ \bibinfo {pages} {4336} (\bibinfo {year} {2017})}\BibitemShut {NoStop}%
\bibitem [{\citenamefont {Guo}\ and\ \citenamefont {Fan}(2023)}]{Guo2023Apr}%
  \BibitemOpen
  \bibfield  {author} {\bibinfo {author} {\bibfnamefont {C.}~\bibnamefont {Guo}}\ and\ \bibinfo {author} {\bibfnamefont {S.}~\bibnamefont {Fan}},\ }\bibfield  {title} {\bibinfo {title} {{Majorization Theory for Unitary Control of Optical Absorption and Emission}},\ }\href {https://doi.org/10.1103/PhysRevLett.130.146202} {\bibfield  {journal} {\bibinfo  {journal} {Phys. Rev. Lett.}\ }\textbf {\bibinfo {volume} {130}},\ \bibinfo {pages} {146202} (\bibinfo {year} {2023})}\BibitemShut {NoStop}%
\bibitem [{\citenamefont {Miller}(2019)}]{Miller2019Sep}%
  \BibitemOpen
  \bibfield  {author} {\bibinfo {author} {\bibfnamefont {D.~A.~B.}\ \bibnamefont {Miller}},\ }\bibfield  {title} {\bibinfo {title} {{Waves, modes, communications, and optics: a tutorial}},\ }\href {https://doi.org/10.1364/AOP.11.000679} {\bibfield  {journal} {\bibinfo  {journal} {Adv. Opt. Photonics}\ }\textbf {\bibinfo {volume} {11}},\ \bibinfo {pages} {679} (\bibinfo {year} {2019})}\BibitemShut {NoStop}%
\bibitem [{\citenamefont {Baranov}\ \emph {et~al.}(2017)\citenamefont {Baranov}, \citenamefont {Krasnok}, \citenamefont {Shegai}, \citenamefont {Al{\ifmmode\grave{u}\else\`{u}\fi}},\ and\ \citenamefont {Chong}}]{Baranov2017Oct}%
  \BibitemOpen
  \bibfield  {author} {\bibinfo {author} {\bibfnamefont {D.~G.}\ \bibnamefont {Baranov}}, \bibinfo {author} {\bibfnamefont {A.}~\bibnamefont {Krasnok}}, \bibinfo {author} {\bibfnamefont {T.}~\bibnamefont {Shegai}}, \bibinfo {author} {\bibfnamefont {A.}~\bibnamefont {Al{\ifmmode\grave{u}\else\`{u}\fi}}},\ and\ \bibinfo {author} {\bibfnamefont {Y.}~\bibnamefont {Chong}},\ }\bibfield  {title} {\bibinfo {title} {{Coherent perfect absorbers: linear control of light with light}},\ }\href {https://doi.org/10.1038/natrevmats.2017.64} {\bibfield  {journal} {\bibinfo  {journal} {Nat. Rev. Mater.}\ }\textbf {\bibinfo {volume} {2}},\ \bibinfo {pages} {17064} (\bibinfo {year} {2017})}\BibitemShut {NoStop}%
\bibitem [{\citenamefont {Guo}\ \emph {et~al.}(2023)\citenamefont {Guo}, \citenamefont {Li}, \citenamefont {Xiao},\ and\ \citenamefont {Fan}}]{Guo2023Oct}%
  \BibitemOpen
  \bibfield  {author} {\bibinfo {author} {\bibfnamefont {C.}~\bibnamefont {Guo}}, \bibinfo {author} {\bibfnamefont {J.}~\bibnamefont {Li}}, \bibinfo {author} {\bibfnamefont {M.}~\bibnamefont {Xiao}},\ and\ \bibinfo {author} {\bibfnamefont {S.}~\bibnamefont {Fan}},\ }\bibfield  {title} {\bibinfo {title} {{Singular topology of scattering matrices}},\ }\href {https://doi.org/10.1103/PhysRevB.108.155418} {\bibfield  {journal} {\bibinfo  {journal} {Phys. Rev. B}\ }\textbf {\bibinfo {volume} {108}},\ \bibinfo {pages} {155418} (\bibinfo {year} {2023})}\BibitemShut {NoStop}%
\bibitem [{\citenamefont {Sadrieva}\ \emph {et~al.}(2019{\natexlab{b}})\citenamefont {Sadrieva}, \citenamefont {Frizyuk}, \citenamefont {Petrov}, \citenamefont {Kivshar},\ and\ \citenamefont {Bogdanov}}]{Sadrieva2019Sep}%
  \BibitemOpen
  \bibfield  {author} {\bibinfo {author} {\bibfnamefont {Z.}~\bibnamefont {Sadrieva}}, \bibinfo {author} {\bibfnamefont {K.}~\bibnamefont {Frizyuk}}, \bibinfo {author} {\bibfnamefont {M.}~\bibnamefont {Petrov}}, \bibinfo {author} {\bibfnamefont {Y.}~\bibnamefont {Kivshar}},\ and\ \bibinfo {author} {\bibfnamefont {A.}~\bibnamefont {Bogdanov}},\ }\bibfield  {title} {\bibinfo {title} {{Multipolar origin of bound states in the continuum}},\ }\href {https://doi.org/10.1103/PhysRevB.100.115303} {\bibfield  {journal} {\bibinfo  {journal} {Phys. Rev. B}\ }\textbf {\bibinfo {volume} {100}},\ \bibinfo {pages} {115303} (\bibinfo {year} {2019}{\natexlab{b}})}\BibitemShut {NoStop}%
\bibitem [{\citenamefont {Waterman}(1965)}]{Waterman1965Aug}%
  \BibitemOpen
  \bibfield  {author} {\bibinfo {author} {\bibfnamefont {P.~C.}\ \bibnamefont {Waterman}},\ }\bibfield  {title} {\bibinfo {title} {{Matrix formulation of electromagnetic scattering}},\ }\href {https://doi.org/10.1109/PROC.1965.4058} {\bibfield  {journal} {\bibinfo  {journal} {Proc. IEEE}\ }\textbf {\bibinfo {volume} {53}},\ \bibinfo {pages} {805} (\bibinfo {year} {1965})}\BibitemShut {NoStop}%
\bibitem [{\citenamefont {Waterman}(1969)}]{Waterman1969Jun}%
  \BibitemOpen
  \bibfield  {author} {\bibinfo {author} {\bibfnamefont {P.~C.}\ \bibnamefont {Waterman}},\ }\bibfield  {title} {\bibinfo {title} {{New Formulation of Acoustic Scattering}},\ }\href {https://doi.org/10.1121/1.1911619} {\bibfield  {journal} {\bibinfo  {journal} {J. Acoust. Soc. Am.}\ }\textbf {\bibinfo {volume} {45}},\ \bibinfo {pages} {1417} (\bibinfo {year} {1969})}\BibitemShut {NoStop}%
\bibitem [{\citenamefont {Asadova~\textit{et al.}}(2025)}]{Asadova2025Mar}%
  \BibitemOpen
  \bibfield  {author} {\bibinfo {author} {\bibfnamefont {N.}~\bibnamefont {Asadova~\textit{et al.}}},\ }\bibfield  {title} {\bibinfo {title} {{T-matrix representation of optical scattering response: Suggestion for a data format}},\ }\href {https://doi.org/10.1016/j.jqsrt.2024.109310} {\bibfield  {journal} {\bibinfo  {journal} {J. Quant. Spectrosc. Radiat. Transfer}\ }\textbf {\bibinfo {volume} {333}},\ \bibinfo {pages} {109310} (\bibinfo {year} {2025})}\BibitemShut {NoStop}%
\bibitem [{\citenamefont {Xu}(2013)}]{Xu2013Jun}%
  \BibitemOpen
  \bibfield  {author} {\bibinfo {author} {\bibfnamefont {Y.-L.}\ \bibnamefont {Xu}},\ }\bibfield  {title} {\bibinfo {title} {{Scattering of electromagnetic waves by periodic particle arrays}},\ }\href {https://doi.org/10.1364/JOSAA.30.001053} {\bibfield  {journal} {\bibinfo  {journal} {J. Opt. Soc. Am. A}\ }\textbf {\bibinfo {volume} {30}},\ \bibinfo {pages} {1053} (\bibinfo {year} {2013})}\BibitemShut {NoStop}%
\bibitem [{\citenamefont {Li}\ \emph {et~al.}(2024)\citenamefont {Li}, \citenamefont {Han},\ and\ \citenamefont {Hu}}]{Li2024Aug}%
  \BibitemOpen
  \bibfield  {author} {\bibinfo {author} {\bibfnamefont {Z.}~\bibnamefont {Li}}, \bibinfo {author} {\bibfnamefont {P.}~\bibnamefont {Han}},\ and\ \bibinfo {author} {\bibfnamefont {G.}~\bibnamefont {Hu}},\ }\bibfield  {title} {\bibinfo {title} {{Willis dynamic homogenization method for acoustic metamaterials based on multiple scattering theory}},\ }\href {https://doi.org/10.1016/j.jmps.2024.105692} {\bibfield  {journal} {\bibinfo  {journal} {J. Mech. Phys. Solids}\ }\textbf {\bibinfo {volume} {189}},\ \bibinfo {pages} {105692} (\bibinfo {year} {2024})}\BibitemShut {NoStop}%
\bibitem [{\citenamefont {Zerulla}\ \emph {et~al.}(2023)\citenamefont {Zerulla}, \citenamefont {Venkitakrishnan}, \citenamefont {Beutel}, \citenamefont {Krsti{\ifmmode\acute{c}\else\'{c}\fi}}, \citenamefont {Holzer}, \citenamefont {Rockstuhl},\ and\ \citenamefont {Fernandez-Corbaton}}]{Zerulla2023Feb}%
  \BibitemOpen
  \bibfield  {author} {\bibinfo {author} {\bibfnamefont {B.}~\bibnamefont {Zerulla}}, \bibinfo {author} {\bibfnamefont {R.}~\bibnamefont {Venkitakrishnan}}, \bibinfo {author} {\bibfnamefont {D.}~\bibnamefont {Beutel}}, \bibinfo {author} {\bibfnamefont {M.}~\bibnamefont {Krsti{\ifmmode\acute{c}\else\'{c}\fi}}}, \bibinfo {author} {\bibfnamefont {C.}~\bibnamefont {Holzer}}, \bibinfo {author} {\bibfnamefont {C.}~\bibnamefont {Rockstuhl}},\ and\ \bibinfo {author} {\bibfnamefont {I.}~\bibnamefont {Fernandez-Corbaton}},\ }\bibfield  {title} {\bibinfo {title} {{A T-Matrix Based Approach to Homogenize Artificial Materials}},\ }\href {https://doi.org/10.1002/adom.202201564} {\bibfield  {journal} {\bibinfo  {journal} {Adv. Opt. Mater.}\ }\textbf {\bibinfo {volume} {11}},\ \bibinfo {pages} {2201564} (\bibinfo {year} {2023})}\BibitemShut {NoStop}%
\bibitem [{\citenamefont {Beutel}\ \emph {et~al.}(2021)\citenamefont {Beutel}, \citenamefont {Groner}, \citenamefont {Rockstuhl},\ and\ \citenamefont {Fernandez-Corbaton}}]{Beutel2021Jun}%
  \BibitemOpen
  \bibfield  {author} {\bibinfo {author} {\bibfnamefont {D.}~\bibnamefont {Beutel}}, \bibinfo {author} {\bibfnamefont {A.}~\bibnamefont {Groner}}, \bibinfo {author} {\bibfnamefont {C.}~\bibnamefont {Rockstuhl}},\ and\ \bibinfo {author} {\bibfnamefont {I.}~\bibnamefont {Fernandez-Corbaton}},\ }\bibfield  {title} {\bibinfo {title} {{Efficient simulation of biperiodic, layered structures based on the T-matrix method}},\ }\href {https://doi.org/10.1364/JOSAB.419645} {\bibfield  {journal} {\bibinfo  {journal} {J. Opt. Soc. Am. B}\ }\textbf {\bibinfo {volume} {38}},\ \bibinfo {pages} {1782} (\bibinfo {year} {2021})}\BibitemShut {NoStop}%
\bibitem [{\citenamefont {Rahimzadegan}\ \emph {et~al.}(2022)\citenamefont {Rahimzadegan}, \citenamefont {Karamanos}, \citenamefont {Alaee}, \citenamefont {Lamprianidis}, \citenamefont {Beutel}, \citenamefont {Boyd},\ and\ \citenamefont {Rockstuhl}}]{Rahimzadegan2022May}%
  \BibitemOpen
  \bibfield  {author} {\bibinfo {author} {\bibfnamefont {A.}~\bibnamefont {Rahimzadegan}}, \bibinfo {author} {\bibfnamefont {T.~D.}\ \bibnamefont {Karamanos}}, \bibinfo {author} {\bibfnamefont {R.}~\bibnamefont {Alaee}}, \bibinfo {author} {\bibfnamefont {A.~G.}\ \bibnamefont {Lamprianidis}}, \bibinfo {author} {\bibfnamefont {D.}~\bibnamefont {Beutel}}, \bibinfo {author} {\bibfnamefont {R.~W.}\ \bibnamefont {Boyd}},\ and\ \bibinfo {author} {\bibfnamefont {C.}~\bibnamefont {Rockstuhl}},\ }\bibfield  {title} {\bibinfo {title} {{A Comprehensive Multipolar Theory for Periodic Metasurfaces}},\ }\href {https://doi.org/10.1002/adom.202102059} {\bibfield  {journal} {\bibinfo  {journal} {Adv. Opt. Mater.}\ }\textbf {\bibinfo {volume} {10}},\ \bibinfo {pages} {2102059} (\bibinfo {year} {2022})}\BibitemShut {NoStop}%
\bibitem [{\citenamefont {Bulgakov}\ and\ \citenamefont {Sadreev}(2014)}]{Bulgakov2014Nov}%
  \BibitemOpen
  \bibfield  {author} {\bibinfo {author} {\bibfnamefont {E.~N.}\ \bibnamefont {Bulgakov}}\ and\ \bibinfo {author} {\bibfnamefont {A.~F.}\ \bibnamefont {Sadreev}},\ }\bibfield  {title} {\bibinfo {title} {{Bloch bound states in the radiation continuum in a periodic array of dielectric rods}},\ }\href {https://doi.org/10.1103/PhysRevA.90.053801} {\bibfield  {journal} {\bibinfo  {journal} {Phys. Rev. A}\ }\textbf {\bibinfo {volume} {90}},\ \bibinfo {pages} {053801} (\bibinfo {year} {2014})}\BibitemShut {NoStop}%
\bibitem [{\citenamefont {Bulgakov}\ and\ \citenamefont {Maksimov}(2018)}]{Bulgakov2018Oct}%
  \BibitemOpen
  \bibfield  {author} {\bibinfo {author} {\bibfnamefont {E.~N.}\ \bibnamefont {Bulgakov}}\ and\ \bibinfo {author} {\bibfnamefont {D.~N.}\ \bibnamefont {Maksimov}},\ }\bibfield  {title} {\bibinfo {title} {{Optical response induced by bound states in the continuum in arrays of dielectric spheres}},\ }\href {https://doi.org/10.1364/JOSAB.35.002443} {\bibfield  {journal} {\bibinfo  {journal} {J. Opt. Soc. Am. B}\ }\textbf {\bibinfo {volume} {35}},\ \bibinfo {pages} {2443} (\bibinfo {year} {2018})}\BibitemShut {NoStop}%
\bibitem [{\citenamefont {Suryadharma}\ \emph {et~al.}(2017)\citenamefont {Suryadharma}, \citenamefont {Fruhnert}, \citenamefont {Rockstuhl},\ and\ \citenamefont {Fernandez-Corbaton}}]{Suryadharma2017May}%
  \BibitemOpen
  \bibfield  {author} {\bibinfo {author} {\bibfnamefont {R.~N.~S.}\ \bibnamefont {Suryadharma}}, \bibinfo {author} {\bibfnamefont {M.}~\bibnamefont {Fruhnert}}, \bibinfo {author} {\bibfnamefont {C.}~\bibnamefont {Rockstuhl}},\ and\ \bibinfo {author} {\bibfnamefont {I.}~\bibnamefont {Fernandez-Corbaton}},\ }\bibfield  {title} {\bibinfo {title} {{Singular-value decomposition for electromagnetic-scattering analysis}},\ }\href {https://doi.org/10.1103/PhysRevA.95.053834} {\bibfield  {journal} {\bibinfo  {journal} {Phys. Rev. A}\ }\textbf {\bibinfo {volume} {95}},\ \bibinfo {pages} {053834} (\bibinfo {year} {2017})}\BibitemShut {NoStop}%
\bibitem [{\citenamefont {Wang}\ \emph {et~al.}(2025{\natexlab{a}})\citenamefont {Wang}, \citenamefont {Garg}, \citenamefont {Mirmoosa}, \citenamefont {Lamprianidis}, \citenamefont {Rockstuhl},\ and\ \citenamefont {Asadchy}}]{Wang2025Feb}%
  \BibitemOpen
  \bibfield  {author} {\bibinfo {author} {\bibfnamefont {X.}~\bibnamefont {Wang}}, \bibinfo {author} {\bibfnamefont {P.}~\bibnamefont {Garg}}, \bibinfo {author} {\bibfnamefont {M.~S.}\ \bibnamefont {Mirmoosa}}, \bibinfo {author} {\bibfnamefont {A.~G.}\ \bibnamefont {Lamprianidis}}, \bibinfo {author} {\bibfnamefont {C.}~\bibnamefont {Rockstuhl}},\ and\ \bibinfo {author} {\bibfnamefont {V.~S.}\ \bibnamefont {Asadchy}},\ }\bibfield  {title} {\bibinfo {title} {{Expanding momentum bandgaps in photonic time crystals through resonances}},\ }\href {https://doi.org/10.1038/s41566-024-01563-3} {\bibfield  {journal} {\bibinfo  {journal} {Nat. Photonics}\ }\textbf {\bibinfo {volume} {19}},\ \bibinfo {pages} {149} (\bibinfo {year} {2025}{\natexlab{a}})}\BibitemShut {NoStop}%
\bibitem [{\citenamefont {Cao}\ and\ \citenamefont {Assouar}(2025)}]{Cao2025Dec}%
  \BibitemOpen
  \bibfield  {author} {\bibinfo {author} {\bibfnamefont {L.}~\bibnamefont {Cao}}\ and\ \bibinfo {author} {\bibfnamefont {B.}~\bibnamefont {Assouar}},\ }\bibfield  {title} {\bibinfo {title} {{Asymmetric Elastic Bound State in the Continuum by an Exceptional Point}},\ }\href {https://doi.org/10.1103/1v48-2t7x} {\bibfield  {journal} {\bibinfo  {journal} {Phys. Rev. Lett.}\ }\textbf {\bibinfo {volume} {135}},\ \bibinfo {pages} {246301} (\bibinfo {year} {2025})}\BibitemShut {NoStop}%
\bibitem [{\citenamefont {Liu}\ \emph {et~al.}(2025{\natexlab{b}})\citenamefont {Liu}, \citenamefont {Cao}, \citenamefont {Guo}, \citenamefont {Wang},\ and\ \citenamefont {Wang}}]{Liu2025Oct}%
  \BibitemOpen
  \bibfield  {author} {\bibinfo {author} {\bibfnamefont {F.}~\bibnamefont {Liu}}, \bibinfo {author} {\bibfnamefont {L.}~\bibnamefont {Cao}}, \bibinfo {author} {\bibfnamefont {W.-H.}\ \bibnamefont {Guo}}, \bibinfo {author} {\bibfnamefont {Y.-F.}\ \bibnamefont {Wang}},\ and\ \bibinfo {author} {\bibfnamefont {Y.-S.}\ \bibnamefont {Wang}},\ }\bibfield  {title} {\bibinfo {title} {{Elastic bound states in the continuum with multi-polarization hybridization}},\ }\href {https://doi.org/10.1038/s42005-025-02301-z} {\bibfield  {journal} {\bibinfo  {journal} {Commun. Phys.}\ }\textbf {\bibinfo {volume} {8}},\ \bibinfo {pages} {392} (\bibinfo {year} {2025}{\natexlab{b}})}\BibitemShut {NoStop}%
\bibitem [{svd(2026)}]{svd}%
  \BibitemOpen
  \href@noop {} {\bibinfo {title} {See the following link to find a part of the codes that were used for examples considered in this work}},\ \bibinfo {howpublished} {\url{https://github.com/NikUstimenko/svd.git}} (\bibinfo {year} {2026})\BibitemShut {NoStop}%
\bibitem [{\citenamefont {Beutel}\ \emph {et~al.}(2024)\citenamefont {Beutel}, \citenamefont {Fernandez-Corbaton},\ and\ \citenamefont {Rockstuhl}}]{Beutel2024Apr}%
  \BibitemOpen
  \bibfield  {author} {\bibinfo {author} {\bibfnamefont {D.}~\bibnamefont {Beutel}}, \bibinfo {author} {\bibfnamefont {I.}~\bibnamefont {Fernandez-Corbaton}},\ and\ \bibinfo {author} {\bibfnamefont {C.}~\bibnamefont {Rockstuhl}},\ }\bibfield  {title} {\bibinfo {title} {{treams {\textendash} a T-matrix-based scattering code for nanophotonics}},\ }\href {https://doi.org/10.1016/j.cpc.2023.109076} {\bibfield  {journal} {\bibinfo  {journal} {Comput. Phys. Commun.}\ }\textbf {\bibinfo {volume} {297}},\ \bibinfo {pages} {109076} (\bibinfo {year} {2024})}\BibitemShut {NoStop}%
\bibitem [{\citenamefont {Ne{\ifmmode\check{c}\else\v{c}\fi}ada}\ and\ \citenamefont {T{\ifmmode\ddot{o}\else\"{o}\fi}rm{\ifmmode\ddot{a}\else\"{a}\fi}}(2021)}]{Necada2021Jan}%
  \BibitemOpen
  \bibfield  {author} {\bibinfo {author} {\bibfnamefont {M.}~\bibnamefont {Ne{\ifmmode\check{c}\else\v{c}\fi}ada}}\ and\ \bibinfo {author} {\bibfnamefont {P.}~\bibnamefont {T{\ifmmode\ddot{o}\else\"{o}\fi}rm{\ifmmode\ddot{a}\else\"{a}\fi}}},\ }\bibfield  {title} {\bibinfo {title} {{Multiple-Scattering $T$-Matrix Simulations for Nanophotonics: Symmetries and Periodic Lattices}},\ }\href {https://doi.org/10.4208/cicp.OA-2020-0136} {\bibfield  {journal} {\bibinfo  {journal} {Comm. Comput. Phys.}\ }\textbf {\bibinfo {volume} {30}},\ \bibinfo {pages} {357} (\bibinfo {year} {2021})}\BibitemShut {NoStop}%
\bibitem [{\citenamefont {Mikhailovskii}\ \emph {et~al.}(2024)\citenamefont {Mikhailovskii}, \citenamefont {Poleva}, \citenamefont {Solodovchenko}, \citenamefont {Sidorenko}, \citenamefont {Sadrieva}, \citenamefont {Petrov}, \citenamefont {Bogdanov},\ and\ \citenamefont {Savelev}}]{Mikhailovskii2024Apr}%
  \BibitemOpen
  \bibfield  {author} {\bibinfo {author} {\bibfnamefont {M.~S.}\ \bibnamefont {Mikhailovskii}}, \bibinfo {author} {\bibfnamefont {M.~A.}\ \bibnamefont {Poleva}}, \bibinfo {author} {\bibfnamefont {N.~S.}\ \bibnamefont {Solodovchenko}}, \bibinfo {author} {\bibfnamefont {M.~S.}\ \bibnamefont {Sidorenko}}, \bibinfo {author} {\bibfnamefont {Z.~F.}\ \bibnamefont {Sadrieva}}, \bibinfo {author} {\bibfnamefont {M.~I.}\ \bibnamefont {Petrov}}, \bibinfo {author} {\bibfnamefont {A.~A.}\ \bibnamefont {Bogdanov}},\ and\ \bibinfo {author} {\bibfnamefont {R.~S.}\ \bibnamefont {Savelev}},\ }\bibfield  {title} {\bibinfo {title} {{Engineering of High-Q States via Collective Mode Coupling in Chains of Mie Resonators}},\ }\href {https://doi.org/10.1021/acsphotonics.3c01874} {\bibfield  {journal} {\bibinfo  {journal} {ACS Photonics}\ }\textbf {\bibinfo {volume} {11}},\ \bibinfo {pages} {1657} (\bibinfo {year} {2024})}\BibitemShut {NoStop}%
\bibitem [{\citenamefont {Zhen}\ \emph {et~al.}(2014)\citenamefont {Zhen}, \citenamefont {Hsu}, \citenamefont {Lu}, \citenamefont {Stone},\ and\ \citenamefont {Solja{\ifmmode\check{c}\else\v{c}\fi}i{\ifmmode\acute{c}\else\'{c}\fi}}}]{Zhen2014Dec}%
  \BibitemOpen
  \bibfield  {author} {\bibinfo {author} {\bibfnamefont {B.}~\bibnamefont {Zhen}}, \bibinfo {author} {\bibfnamefont {C.~W.}\ \bibnamefont {Hsu}}, \bibinfo {author} {\bibfnamefont {L.}~\bibnamefont {Lu}}, \bibinfo {author} {\bibfnamefont {A.~D.}\ \bibnamefont {Stone}},\ and\ \bibinfo {author} {\bibfnamefont {M.}~\bibnamefont {Solja{\ifmmode\check{c}\else\v{c}\fi}i{\ifmmode\acute{c}\else\'{c}\fi}}},\ }\bibfield  {title} {\bibinfo {title} {{Topological Nature of Optical Bound States in the Continuum}},\ }\href {https://doi.org/10.1103/PhysRevLett.113.257401} {\bibfield  {journal} {\bibinfo  {journal} {Phys. Rev. Lett.}\ }\textbf {\bibinfo {volume} {113}},\ \bibinfo {pages} {257401} (\bibinfo {year} {2014})}\BibitemShut {NoStop}%
\bibitem [{\citenamefont {Semushev}\ \emph {et~al.}(2026)\citenamefont {Semushev}, \citenamefont {Zhao}, \citenamefont {Proskurin}, \citenamefont {Song}, \citenamefont {Liu}, \citenamefont {Rybin}, \citenamefont {Maslova},\ and\ \citenamefont {Bogdanov}}]{Semushev2026Jan}%
  \BibitemOpen
  \bibfield  {author} {\bibinfo {author} {\bibfnamefont {K.~V.}\ \bibnamefont {Semushev}}, \bibinfo {author} {\bibfnamefont {Z.}~\bibnamefont {Zhao}}, \bibinfo {author} {\bibfnamefont {A.}~\bibnamefont {Proskurin}}, \bibinfo {author} {\bibfnamefont {M.}~\bibnamefont {Song}}, \bibinfo {author} {\bibfnamefont {X.}~\bibnamefont {Liu}}, \bibinfo {author} {\bibfnamefont {M.~V.}\ \bibnamefont {Rybin}}, \bibinfo {author} {\bibfnamefont {E.~E.}\ \bibnamefont {Maslova}},\ and\ \bibinfo {author} {\bibfnamefont {A.~A.}\ \bibnamefont {Bogdanov}},\ }\bibfield  {title} {\bibinfo {title} {{Robustness of bound states in the continuum in bilayer structures against symmetry breaking}},\ }\href {https://doi.org/10.1103/qj87-5xz9} {\bibfield  {journal} {\bibinfo  {journal} {Phys. Rev. Appl.}\ }\textbf {\bibinfo {volume} {25}},\ \bibinfo {pages} {014038} (\bibinfo {year} {2026})}\BibitemShut {NoStop}%
\bibitem [{\citenamefont {Song}\ \emph {et~al.}(2020)\citenamefont {Song}, \citenamefont {Hu}, \citenamefont {Dai}, \citenamefont {Zheng}, \citenamefont {Han}, \citenamefont {Zi}, \citenamefont {Zhang},\ and\ \citenamefont {Chan}}]{Song2020Aug}%
  \BibitemOpen
  \bibfield  {author} {\bibinfo {author} {\bibfnamefont {Q.}~\bibnamefont {Song}}, \bibinfo {author} {\bibfnamefont {J.}~\bibnamefont {Hu}}, \bibinfo {author} {\bibfnamefont {S.}~\bibnamefont {Dai}}, \bibinfo {author} {\bibfnamefont {C.}~\bibnamefont {Zheng}}, \bibinfo {author} {\bibfnamefont {D.}~\bibnamefont {Han}}, \bibinfo {author} {\bibfnamefont {J.}~\bibnamefont {Zi}}, \bibinfo {author} {\bibfnamefont {Z.~Q.}\ \bibnamefont {Zhang}},\ and\ \bibinfo {author} {\bibfnamefont {C.~T.}\ \bibnamefont {Chan}},\ }\bibfield  {title} {\bibinfo {title} {{Coexistence of a new type of bound state in the continuum and a lasing threshold mode induced by PT symmetry}},\ }\href {https://www.science.org/doi/10.1126/sciadv.abc1160} {\bibfield  {journal} {\bibinfo  {journal} {Sci. Adv.}\ }\textbf {\bibinfo {volume} {6}},\ \bibinfo {pages} {eabc1160} (\bibinfo {year} {2020})}\BibitemShut {NoStop}%
\bibitem [{\citenamefont {Tikhodeev}\ \emph {et~al.}(2002)\citenamefont {Tikhodeev}, \citenamefont {Yablonskii}, \citenamefont {Muljarov}, \citenamefont {Gippius},\ and\ \citenamefont {Ishihara}}]{Tikhodeev2002Jul}%
  \BibitemOpen
  \bibfield  {author} {\bibinfo {author} {\bibfnamefont {S.~G.}\ \bibnamefont {Tikhodeev}}, \bibinfo {author} {\bibfnamefont {A.~L.}\ \bibnamefont {Yablonskii}}, \bibinfo {author} {\bibfnamefont {E.~A.}\ \bibnamefont {Muljarov}}, \bibinfo {author} {\bibfnamefont {N.~A.}\ \bibnamefont {Gippius}},\ and\ \bibinfo {author} {\bibfnamefont {T.}~\bibnamefont {Ishihara}},\ }\bibfield  {title} {\bibinfo {title} {{Quasiguided modes and optical properties of photonic crystal slabs}},\ }\href {https://doi.org/10.1103/PhysRevB.66.045102} {\bibfield  {journal} {\bibinfo  {journal} {Phys. Rev. B}\ }\textbf {\bibinfo {volume} {66}},\ \bibinfo {pages} {045102} (\bibinfo {year} {2002})}\BibitemShut {NoStop}%
\bibitem [{\citenamefont {Carminati}\ \emph {et~al.}(2000)\citenamefont {Carminati}, \citenamefont {S{\ifmmode\acute{a}\else\'{a}\fi}enz}, \citenamefont {Greffet},\ and\ \citenamefont {Nieto-Vesperinas}}]{Carminati2000Jun}%
  \BibitemOpen
  \bibfield  {author} {\bibinfo {author} {\bibfnamefont {R.}~\bibnamefont {Carminati}}, \bibinfo {author} {\bibfnamefont {J.~J.}\ \bibnamefont {S{\ifmmode\acute{a}\else\'{a}\fi}enz}}, \bibinfo {author} {\bibfnamefont {J.-J.}\ \bibnamefont {Greffet}},\ and\ \bibinfo {author} {\bibfnamefont {M.}~\bibnamefont {Nieto-Vesperinas}},\ }\bibfield  {title} {\bibinfo {title} {{Reciprocity, unitarity, and time-reversal symmetry of the S matrix of fields containing evanescent components}},\ }\href {https://doi.org/10.1103/PhysRevA.62.012712} {\bibfield  {journal} {\bibinfo  {journal} {Phys. Rev. A}\ }\textbf {\bibinfo {volume} {62}},\ \bibinfo {pages} {012712} (\bibinfo {year} {2000})}\BibitemShut {NoStop}%
\bibitem [{\citenamefont {Redheffer}(1959)}]{Redheffer1959}%
  \BibitemOpen
  \bibfield  {author} {\bibinfo {author} {\bibfnamefont {R.}~\bibnamefont {Redheffer}},\ }\bibfield  {title} {\bibinfo {title} {{Inequalities for a Matrix Riccati Equation}},\ }\href {https://www.jstor.org/stable/24900576} {\bibfield  {journal} {\bibinfo  {journal} {J. Math. Mech.}\ }\textbf {\bibinfo {volume} {8}},\ \bibinfo {pages} {349} (\bibinfo {year} {1959})}\BibitemShut {NoStop}%
\bibitem [{\citenamefont {Ustimenko}(2025)}]{acoustotreams}%
  \BibitemOpen
  \bibfield  {author} {\bibinfo {author} {\bibfnamefont {N.}~\bibnamefont {Ustimenko}},\ }\href@noop {} {\bibinfo {title} {acoustotreams (v. 0.2.13)}},\ \bibinfo {howpublished} {\url{https://github.com/NikUstimenko/acoustotreams.git}} (\bibinfo {year} {2025})\BibitemShut {NoStop}%
\bibitem [{\citenamefont {Beutel}\ \emph {et~al.}(2023)\citenamefont {Beutel}, \citenamefont {Fernandez-Corbaton},\ and\ \citenamefont {Rockstuhl}}]{Beutel2023Jan}%
  \BibitemOpen
  \bibfield  {author} {\bibinfo {author} {\bibfnamefont {D.}~\bibnamefont {Beutel}}, \bibinfo {author} {\bibfnamefont {I.}~\bibnamefont {Fernandez-Corbaton}},\ and\ \bibinfo {author} {\bibfnamefont {C.}~\bibnamefont {Rockstuhl}},\ }\bibfield  {title} {\bibinfo {title} {{Unified lattice sums accommodating multiple sublattices for solutions of the Helmholtz equation in two and three dimensions}},\ }\href {https://doi.org/10.1103/PhysRevA.107.013508} {\bibfield  {journal} {\bibinfo  {journal} {Phys. Rev. A}\ }\textbf {\bibinfo {volume} {107}},\ \bibinfo {pages} {013508} (\bibinfo {year} {2023})}\BibitemShut {NoStop}%
\bibitem [{\citenamefont {Evlyukhin}\ \emph {et~al.}(2021)\citenamefont {Evlyukhin}, \citenamefont {Poleva}, \citenamefont {Prokhorov}, \citenamefont {Baryshnikova}, \citenamefont {Miroshnichenko},\ and\ \citenamefont {Chichkov}}]{Evlyukhin2021Dec}%
  \BibitemOpen
  \bibfield  {author} {\bibinfo {author} {\bibfnamefont {A.~B.}\ \bibnamefont {Evlyukhin}}, \bibinfo {author} {\bibfnamefont {M.~A.}\ \bibnamefont {Poleva}}, \bibinfo {author} {\bibfnamefont {A.~V.}\ \bibnamefont {Prokhorov}}, \bibinfo {author} {\bibfnamefont {K.~V.}\ \bibnamefont {Baryshnikova}}, \bibinfo {author} {\bibfnamefont {A.~E.}\ \bibnamefont {Miroshnichenko}},\ and\ \bibinfo {author} {\bibfnamefont {B.~N.}\ \bibnamefont {Chichkov}},\ }\bibfield  {title} {\bibinfo {title} {{Polarization Switching Between Electric and Magnetic Quasi-Trapped Modes in Bianisotropic All-Dielectric Metasurfaces}},\ }\href {https://doi.org/10.1002/lpor.202100206} {\bibfield  {journal} {\bibinfo  {journal} {Laser Photonics Rev.}\ }\textbf {\bibinfo {volume} {15}},\ \bibinfo {pages} {2100206} (\bibinfo {year} {2021})}\BibitemShut {NoStop}%
\bibitem [{\citenamefont {Wang}\ \emph {et~al.}(2025{\natexlab{b}})\citenamefont {Wang}, \citenamefont {Shi}, \citenamefont {Zheng},\ and\ \citenamefont {Thompson}}]{Wang2025Aug}%
  \BibitemOpen
  \bibfield  {author} {\bibinfo {author} {\bibfnamefont {X.}~\bibnamefont {Wang}}, \bibinfo {author} {\bibfnamefont {C.}~\bibnamefont {Shi}}, \bibinfo {author} {\bibfnamefont {Q.}~\bibnamefont {Zheng}},\ and\ \bibinfo {author} {\bibfnamefont {D.}~\bibnamefont {Thompson}},\ }\bibfield  {title} {\bibinfo {title} {{Resonant enhancement of the thermoreflectance response of silicon nanodisks}},\ }\href {https://doi.org/10.1063/5.0276240} {\bibfield  {journal} {\bibinfo  {journal} {Appl. Phys. Lett.}\ }\textbf {\bibinfo {volume} {127}},\ \bibinfo {pages} {061103} (\bibinfo {year} {2025}{\natexlab{b}})}\BibitemShut {NoStop}%
\bibitem [{\citenamefont {Bohren}\ and\ \citenamefont {Huffman}(1998)}]{Bohren1998Apr}%
  \BibitemOpen
  \bibfield  {author} {\bibinfo {author} {\bibfnamefont {C.~F.}\ \bibnamefont {Bohren}}\ and\ \bibinfo {author} {\bibfnamefont {D.~R.}\ \bibnamefont {Huffman}},\ }\href {https://doi.org/10.1002/9783527618156} {\emph {\bibinfo {title} {{Absorption and Scattering of Light by Small Particles}}}}\ (\bibinfo  {publisher} {WILEY-VCH Verlag GmbH \& Co. KGaA, Weinheim, Germany},\ \bibinfo {year} {1998})\BibitemShut {NoStop}%
\bibitem [{\citenamefont {Gladyshev}\ \emph {et~al.}(2022)\citenamefont {Gladyshev}, \citenamefont {Shalev}, \citenamefont {Frizyuk}, \citenamefont {Ladutenko},\ and\ \citenamefont {Bogdanov}}]{Gladyshev2022Jun}%
  \BibitemOpen
  \bibfield  {author} {\bibinfo {author} {\bibfnamefont {S.}~\bibnamefont {Gladyshev}}, \bibinfo {author} {\bibfnamefont {A.}~\bibnamefont {Shalev}}, \bibinfo {author} {\bibfnamefont {K.}~\bibnamefont {Frizyuk}}, \bibinfo {author} {\bibfnamefont {K.}~\bibnamefont {Ladutenko}},\ and\ \bibinfo {author} {\bibfnamefont {A.}~\bibnamefont {Bogdanov}},\ }\bibfield  {title} {\bibinfo {title} {{Bound states in the continuum in multipolar lattices}},\ }\href {https://doi.org/10.1103/PhysRevB.105.L241301} {\bibfield  {journal} {\bibinfo  {journal} {Phys. Rev. B}\ }\textbf {\bibinfo {volume} {105}},\ \bibinfo {pages} {L241301} (\bibinfo {year} {2022})}\BibitemShut {NoStop}%
\bibitem [{\citenamefont {Poleva}\ \emph {et~al.}(2025)\citenamefont {Poleva}, \citenamefont {Rockstuhl},\ and\ \citenamefont {Evlyukhin}}]{Poleva2025May}%
  \BibitemOpen
  \bibfield  {author} {\bibinfo {author} {\bibfnamefont {M.~A.}\ \bibnamefont {Poleva}}, \bibinfo {author} {\bibfnamefont {C.}~\bibnamefont {Rockstuhl}},\ and\ \bibinfo {author} {\bibfnamefont {A.~B.}\ \bibnamefont {Evlyukhin}},\ }\bibfield  {title} {\bibinfo {title} {{Material induced bianisotropy of hybrid nanostructures: From a single meta-atom resonance to metasurfaces with trapped modes}},\ }\href {https://doi.org/10.1103/PhysRevB.111.195417} {\bibfield  {journal} {\bibinfo  {journal} {Phys. Rev. B}\ }\textbf {\bibinfo {volume} {111}},\ \bibinfo {pages} {195417} (\bibinfo {year} {2025})}\BibitemShut {NoStop}%
\bibitem [{\citenamefont {Bulgakov}\ and\ \citenamefont {Sadreev}(2019)}]{Bulgakov2019Mar}%
  \BibitemOpen
  \bibfield  {author} {\bibinfo {author} {\bibfnamefont {E.~N.}\ \bibnamefont {Bulgakov}}\ and\ \bibinfo {author} {\bibfnamefont {A.~F.}\ \bibnamefont {Sadreev}},\ }\bibfield  {title} {\bibinfo {title} {{High-$Q$ resonant modes in a finite array of dielectric particles}},\ }\href {https://doi.org/10.1103/PhysRevA.99.033851} {\bibfield  {journal} {\bibinfo  {journal} {Phys. Rev. A}\ }\textbf {\bibinfo {volume} {99}},\ \bibinfo {pages} {033851} (\bibinfo {year} {2019})}\BibitemShut {NoStop}%
\bibitem [{\citenamefont {Czajkowski}\ \emph {et~al.}(2020)\citenamefont {Czajkowski}, \citenamefont {Bancerek},\ and\ \citenamefont {Antosiewicz}}]{Czajkowski2020Aug}%
  \BibitemOpen
  \bibfield  {author} {\bibinfo {author} {\bibfnamefont {K.~M.}\ \bibnamefont {Czajkowski}}, \bibinfo {author} {\bibfnamefont {M.}~\bibnamefont {Bancerek}},\ and\ \bibinfo {author} {\bibfnamefont {T.~J.}\ \bibnamefont {Antosiewicz}},\ }\bibfield  {title} {\bibinfo {title} {{Multipole analysis of substrate-supported dielectric nanoresonator metasurfaces via the $T$-matrix method}},\ }\href {https://doi.org/10.1103/PhysRevB.102.085431} {\bibfield  {journal} {\bibinfo  {journal} {Phys. Rev. B}\ }\textbf {\bibinfo {volume} {102}},\ \bibinfo {pages} {085431} (\bibinfo {year} {2020})}\BibitemShut {NoStop}%
\bibitem [{\citenamefont {Novotny}\ and\ \citenamefont {Hecht}(2012)}]{Novotny2012Sep}%
  \BibitemOpen
  \bibfield  {author} {\bibinfo {author} {\bibfnamefont {L.}~\bibnamefont {Novotny}}\ and\ \bibinfo {author} {\bibfnamefont {B.}~\bibnamefont {Hecht}},\ }\href {https://doi.org/10.1017/CBO9780511794193} {\emph {\bibinfo {title} {{Principles of Nano-Optics}}}}\ (\bibinfo  {publisher} {Cambridge University Press},\ \bibinfo {address} {Cambridge, England, UK},\ \bibinfo {year} {2012})\BibitemShut {NoStop}%
\bibitem [{\citenamefont {Doiron}\ \emph {et~al.}(2022)\citenamefont {Doiron}, \citenamefont {Brener},\ and\ \citenamefont {Cerjan}}]{Doiron2022Dec}%
  \BibitemOpen
  \bibfield  {author} {\bibinfo {author} {\bibfnamefont {C.~F.}\ \bibnamefont {Doiron}}, \bibinfo {author} {\bibfnamefont {I.}~\bibnamefont {Brener}},\ and\ \bibinfo {author} {\bibfnamefont {A.}~\bibnamefont {Cerjan}},\ }\bibfield  {title} {\bibinfo {title} {{Realizing symmetry-guaranteed pairs of bound states in the continuum in metasurfaces}},\ }\href {https://doi.org/10.1038/s41467-022-35246-w} {\bibfield  {journal} {\bibinfo  {journal} {Nat. Commun.}\ }\textbf {\bibinfo {volume} {13}},\ \bibinfo {pages} {7534} (\bibinfo {year} {2022})}\BibitemShut {NoStop}%
\bibitem [{\citenamefont {Wang}\ \emph {et~al.}(2018)\citenamefont {Wang}, \citenamefont {Gupta}, \citenamefont {Zhu}, \citenamefont {Lu}, \citenamefont {Liu},\ and\ \citenamefont {Chen}}]{Wang2018Dec}%
  \BibitemOpen
  \bibfield  {author} {\bibinfo {author} {\bibfnamefont {H.-F.}\ \bibnamefont {Wang}}, \bibinfo {author} {\bibfnamefont {S.~K.}\ \bibnamefont {Gupta}}, \bibinfo {author} {\bibfnamefont {X.-Y.}\ \bibnamefont {Zhu}}, \bibinfo {author} {\bibfnamefont {M.-H.}\ \bibnamefont {Lu}}, \bibinfo {author} {\bibfnamefont {X.-P.}\ \bibnamefont {Liu}},\ and\ \bibinfo {author} {\bibfnamefont {Y.-F.}\ \bibnamefont {Chen}},\ }\bibfield  {title} {\bibinfo {title} {{Bound states in the continuum in a bilayer photonic crystal with TE-TM cross coupling}},\ }\href {https://doi.org/10.1103/PhysRevB.98.214101} {\bibfield  {journal} {\bibinfo  {journal} {Phys. Rev. B}\ }\textbf {\bibinfo {volume} {98}},\ \bibinfo {pages} {214101} (\bibinfo {year} {2018})}\BibitemShut {NoStop}%
\bibitem [{\citenamefont {Liu}\ \emph {et~al.}(2025{\natexlab{c}})\citenamefont {Liu}, \citenamefont {Qin}, \citenamefont {Qiu}, \citenamefont {Tu}, \citenamefont {Qiu}, \citenamefont {Wu}, \citenamefont {Yu}, \citenamefont {Liu},\ and\ \citenamefont {Xiao}}]{Liu2025Mar}%
  \BibitemOpen
  \bibfield  {author} {\bibinfo {author} {\bibfnamefont {T.}~\bibnamefont {Liu}}, \bibinfo {author} {\bibfnamefont {M.}~\bibnamefont {Qin}}, \bibinfo {author} {\bibfnamefont {J.}~\bibnamefont {Qiu}}, \bibinfo {author} {\bibfnamefont {X.}~\bibnamefont {Tu}}, \bibinfo {author} {\bibfnamefont {H.}~\bibnamefont {Qiu}}, \bibinfo {author} {\bibfnamefont {F.}~\bibnamefont {Wu}}, \bibinfo {author} {\bibfnamefont {T.}~\bibnamefont {Yu}}, \bibinfo {author} {\bibfnamefont {Q.}~\bibnamefont {Liu}},\ and\ \bibinfo {author} {\bibfnamefont {S.}~\bibnamefont {Xiao}},\ }\bibfield  {title} {\bibinfo {title} {{Polarization-Independent Enhancement of Third-Harmonic Generation Empowered by Doubly Degenerate Quasi-Bound States in the Continuum}},\ }\href {https://doi.org/10.1021/acs.nanolett.5c00146} {\bibfield  {journal} {\bibinfo  {journal} {Nano Lett.}\ }\textbf {\bibinfo {volume} {25}},\ \bibinfo {pages} {3646} (\bibinfo {year} {2025}{\natexlab{c}})}\BibitemShut {NoStop}%
\bibitem [{\citenamefont {Wang}\ \emph {et~al.}(2019)\citenamefont {Wang}, \citenamefont {Li},\ and\ \citenamefont {Zhang}}]{Wang2019Mar}%
  \BibitemOpen
  \bibfield  {author} {\bibinfo {author} {\bibfnamefont {T.}~\bibnamefont {Wang}}, \bibinfo {author} {\bibfnamefont {Z.}~\bibnamefont {Li}},\ and\ \bibinfo {author} {\bibfnamefont {X.}~\bibnamefont {Zhang}},\ }\bibfield  {title} {\bibinfo {title} {{Improved generation of correlated photon pairs from monolayer WS2 based on bound states in the continuum}},\ }\href {https://doi.org/10.1364/PRJ.7.000341} {\bibfield  {journal} {\bibinfo  {journal} {Photonics Res.}\ }\textbf {\bibinfo {volume} {7}},\ \bibinfo {pages} {341} (\bibinfo {year} {2019})}\BibitemShut {NoStop}%
\bibitem [{\citenamefont {Colom}\ \emph {et~al.}(2019)\citenamefont {Colom}, \citenamefont {Xu}, \citenamefont {Marini}, \citenamefont {Bedu}, \citenamefont {Ozerov}, \citenamefont {Begou}, \citenamefont {Lumeau}, \citenamefont {Miroshnishenko}, \citenamefont {Neshev}, \citenamefont {Kuhlmey}, \citenamefont {Palomba},\ and\ \citenamefont {Bonod}}]{Colom2019May}%
  \BibitemOpen
  \bibfield  {author} {\bibinfo {author} {\bibfnamefont {R.}~\bibnamefont {Colom}}, \bibinfo {author} {\bibfnamefont {L.}~\bibnamefont {Xu}}, \bibinfo {author} {\bibfnamefont {L.}~\bibnamefont {Marini}}, \bibinfo {author} {\bibfnamefont {F.}~\bibnamefont {Bedu}}, \bibinfo {author} {\bibfnamefont {I.}~\bibnamefont {Ozerov}}, \bibinfo {author} {\bibfnamefont {T.}~\bibnamefont {Begou}}, \bibinfo {author} {\bibfnamefont {J.}~\bibnamefont {Lumeau}}, \bibinfo {author} {\bibfnamefont {A.~E.}\ \bibnamefont {Miroshnishenko}}, \bibinfo {author} {\bibfnamefont {D.}~\bibnamefont {Neshev}}, \bibinfo {author} {\bibfnamefont {B.~T.}\ \bibnamefont {Kuhlmey}}, \bibinfo {author} {\bibfnamefont {S.}~\bibnamefont {Palomba}},\ and\ \bibinfo {author} {\bibfnamefont {N.}~\bibnamefont {Bonod}},\ }\bibfield  {title} {\bibinfo {title} {{Enhanced Four-Wave Mixing in Doubly Resonant Si Nanoresonators}},\ }\href {https://doi.org/10.1021/acsphotonics.9b00442} {\bibfield  {journal} {\bibinfo  {journal} {ACS Photonics}\ }\textbf {\bibinfo
  {volume} {6}},\ \bibinfo {pages} {1295} (\bibinfo {year} {2019})}\BibitemShut {NoStop}%
\bibitem [{\citenamefont {Fernandez-Corbaton}\ \emph {et~al.}(2013)\citenamefont {Fernandez-Corbaton}, \citenamefont {Zambrana-Puyalto}, \citenamefont {Tischler}, \citenamefont {Vidal}, \citenamefont {Juan},\ and\ \citenamefont {Molina-Terriza}}]{Fernandez-Corbaton2013Aug}%
  \BibitemOpen
  \bibfield  {author} {\bibinfo {author} {\bibfnamefont {I.}~\bibnamefont {Fernandez-Corbaton}}, \bibinfo {author} {\bibfnamefont {X.}~\bibnamefont {Zambrana-Puyalto}}, \bibinfo {author} {\bibfnamefont {N.}~\bibnamefont {Tischler}}, \bibinfo {author} {\bibfnamefont {X.}~\bibnamefont {Vidal}}, \bibinfo {author} {\bibfnamefont {M.~L.}\ \bibnamefont {Juan}},\ and\ \bibinfo {author} {\bibfnamefont {G.}~\bibnamefont {Molina-Terriza}},\ }\bibfield  {title} {\bibinfo {title} {{Electromagnetic Duality Symmetry and Helicity Conservation for the Macroscopic Maxwell's Equations}},\ }\href {https://doi.org/10.1103/PhysRevLett.111.060401} {\bibfield  {journal} {\bibinfo  {journal} {Phys. Rev. Lett.}\ }\textbf {\bibinfo {volume} {111}},\ \bibinfo {pages} {060401} (\bibinfo {year} {2013})}\BibitemShut {NoStop}%
\bibitem [{\citenamefont {Kerker}\ \emph {et~al.}(1983)\citenamefont {Kerker}, \citenamefont {Wang},\ and\ \citenamefont {Giles}}]{Kerker1983Jun}%
  \BibitemOpen
  \bibfield  {author} {\bibinfo {author} {\bibfnamefont {M.}~\bibnamefont {Kerker}}, \bibinfo {author} {\bibfnamefont {D.-S.}\ \bibnamefont {Wang}},\ and\ \bibinfo {author} {\bibfnamefont {C.~L.}\ \bibnamefont {Giles}},\ }\bibfield  {title} {\bibinfo {title} {{Electromagnetic scattering by magnetic spheres}},\ }\href {https://doi.org/10.1364/JOSA.73.000765} {\bibfield  {journal} {\bibinfo  {journal} {J. Opt. Soc. Am. A}\ }\textbf {\bibinfo {volume} {73}},\ \bibinfo {pages} {765} (\bibinfo {year} {1983})}\BibitemShut {NoStop}%
\bibitem [{\citenamefont {Ndangali}\ and\ \citenamefont {Shabanov}(2010)}]{Ndangali2010Oct}%
  \BibitemOpen
  \bibfield  {author} {\bibinfo {author} {\bibfnamefont {R.~F.}\ \bibnamefont {Ndangali}}\ and\ \bibinfo {author} {\bibfnamefont {S.~V.}\ \bibnamefont {Shabanov}},\ }\bibfield  {title} {\bibinfo {title} {{Electromagnetic bound states in the radiation continuum for periodic double arrays of subwavelength dielectric cylinders}},\ }\href {https://doi.org/10.1063/1.3486358} {\bibfield  {journal} {\bibinfo  {journal} {J. Math. Phys.}\ }\textbf {\bibinfo {volume} {51}},\ \bibinfo {pages} {102901} (\bibinfo {year} {2010})}\BibitemShut {NoStop}%
\bibitem [{\citenamefont {Alagappan}\ \emph {et~al.}(2024)\citenamefont {Alagappan}, \citenamefont {Garc{\ifmmode\acute{\imath}\else\'{\i}\fi}a-Vidal},\ and\ \citenamefont {Png}}]{Alagappan2024Nov}%
  \BibitemOpen
  \bibfield  {author} {\bibinfo {author} {\bibfnamefont {G.}~\bibnamefont {Alagappan}}, \bibinfo {author} {\bibfnamefont {F.~J.}\ \bibnamefont {Garc{\ifmmode\acute{\imath}\else\'{\i}\fi}a-Vidal}},\ and\ \bibinfo {author} {\bibfnamefont {C.~E.}\ \bibnamefont {Png}},\ }\bibfield  {title} {\bibinfo {title} {{Fabry-Perot Resonances in Bilayer Metasurfaces}},\ }\href {https://doi.org/10.1103/PhysRevLett.133.226901} {\bibfield  {journal} {\bibinfo  {journal} {Phys. Rev. Lett.}\ }\textbf {\bibinfo {volume} {133}},\ \bibinfo {pages} {226901} (\bibinfo {year} {2024})}\BibitemShut {NoStop}%
\bibitem [{\citenamefont {Ustimenko}\ \emph {et~al.}(2026)\citenamefont {Ustimenko}, \citenamefont {Evlyukhin}, \citenamefont {Kyrimi}, \citenamefont {Kildishev},\ and\ \citenamefont {Rockstuhl}}]{Ustimenko2026Jan}%
  \BibitemOpen
  \bibfield  {author} {\bibinfo {author} {\bibfnamefont {N.}~\bibnamefont {Ustimenko}}, \bibinfo {author} {\bibfnamefont {A.~B.}\ \bibnamefont {Evlyukhin}}, \bibinfo {author} {\bibfnamefont {V.}~\bibnamefont {Kyrimi}}, \bibinfo {author} {\bibfnamefont {A.~V.}\ \bibnamefont {Kildishev}},\ and\ \bibinfo {author} {\bibfnamefont {C.}~\bibnamefont {Rockstuhl}},\ }\bibfield  {title} {\bibinfo {title} {{Lattice-induced sound trapping in biperiodic metasurfaces of acoustic resonators}},\ }\href {https://doi.org/10.1103/wnmk-zhrb} {\bibfield  {journal} {\bibinfo  {journal} {Phys. Rev. Res.}\ }\textbf {\bibinfo {volume} {8}},\ \bibinfo {pages} {013074} (\bibinfo {year} {2026})}\BibitemShut {NoStop}%
\bibitem [{\citenamefont {Morse}\ and\ \citenamefont {Feshbach}(1953)}]{Morse1953}%
  \BibitemOpen
  \bibfield  {author} {\bibinfo {author} {\bibfnamefont {P.~M.}\ \bibnamefont {Morse}}\ and\ \bibinfo {author} {\bibfnamefont {H.}~\bibnamefont {Feshbach}},\ }\href@noop {} {\emph {\bibinfo {title} {{Methods of Theoretical Physics}}}}\ (\bibinfo  {publisher} {McGraw-Hill},\ \bibinfo {address} {New York, NY, USA},\ \bibinfo {year} {1953})\BibitemShut {NoStop}%
\end{thebibliography}%
\end{document}